\documentclass[iop,apj,tighten]{emulateapj}
\pdfoutput=1 
\usepackage{apjfonts} 
\usepackage{amsmath,amstext}
\usepackage[breaklinks,colorlinks,citecolor=blue,linkcolor=magenta]{hyperref} 
\usepackage[all]{hypcap} 
\usepackage{ulem}
\usepackage{color}

\shorttitle{3D Jet Formation}
\shortauthors{Gonz\'alez-Avil\'es et al.}

\begin{document}

\title{I. Jet formation and evolution due to 3D magnetic reconnection}

\author{J.J. Gonz\'alez-Avil\'es \altaffilmark{1} , F.S. Guzm\'an  \altaffilmark{1}, V. Fedun \altaffilmark{2}, G. Verth \altaffilmark{3}, S. Shelyag \altaffilmark{4} and S. Regnier  \altaffilmark{4}}
\affil{$^1$Laboratorio de Inteligencia Artificial y Superc\'omputo. Instituto de F\'{\i}sica y Matem\'{a}ticas, Universidad
        Michoacana de San Nicol\'as de Hidalgo. Morelia, Michoac\'{a}n, M\'{e}xico.\\
        $^2$ Department of Automatic Control and Systems Engineering, University of Sheffield, Sheffield, S1 3JD, UK\\
        $^3$ School of Mathematics and Statistics, University of Sheffield, Sheffield, S3 7RH, UK\\
        $^4$ Department of Mathematics, Physics and Electrical Engineering, Northumbria University, Newcastle upon Tyne, NE1 8ST, UK}

\begin{abstract}
Using simulated data-driven three-dimensional resistive MHD simulations of the solar atmosphere, we show that 3D magnetic reconnection may be responsible of the formation of jets with characteristics of Type II spicules. For this, we numerically model the photosphere-corona region using the C7 equilibrium atmosphere model. The initial magnetic configuration is a 3D potential magnetic field, extrapolated up to the solar corona region from a dynamic realistic simulation of solar photospheric magnetoconvection model which is mimicking quiet-Sun. In this case we consider a uniform and constant value of the magnetic resistivity of $12.56 ~\Omega~{\rm m}$. We have found that formation of the jet depends on the Lorentz force, which helps to accelerate the plasma upwards. Analyzing various properties of the jet dynamics, we found that the jet structure shows Doppler shift near to regions with high vorticity.  
The morphology, upward velocity, covering a range up to 130 $\rm km$ $\rm s^{-1}$, and timescale formation of the estructure between 60 and 90 s, are similar to those expected for Type II spicules. 
\end{abstract}

\maketitle


\section{Introduction}
\label{sec:Introduction}

Jet-like emissions of plasma in the solar atmosphere have been extensively observed over a range of wavelengths, e.g. X-ray, EUV and H$\alpha$, that usually occur in active regions and polar coronal holes. It is believed that many plasma jets are produced directly by magnetic reconnection, when oppositely directed magnetic fields come in contact \citep[see e.g.][]{Shibata_et_al_2007}. The magnetic reconnection acts as a mechanism of  conversion of the magnetic field energy into thermal and kinetic energy of the ejected plasma and can occur from the convection zone to the solar corona. In particular, the observed chromospheric dynamics at the solar limb is dominated by spicules \citep{Beckers_1968}, which are ubiquitous, highly dynamic jets of plasma \citep{Secchi_1878,Tsiropoula_et_al_2012,De_Pontieu_et_al_2007b}. The improvement in the resolution of the observations by the Hinode satellite and Swedish 1 m Solar Telescope (SST) on La Palma \citep{Kosugi_et_al_2007, Scharmer_2003, Scharmer_et_al_2008} has suggested the existence of two classes of spicules.  

The first type of spicules are so-called Type I, which reach maximum heights of 4-8 Mm, maximum ascending velocities of 15-40 km s$^{-1}$, have a lifetime of 3-6.5 minutes \citep{Pereira_et_al_2012}, and show up and downward motions \citep{Beckers_1968, Suematsu_et_al_1995}. These Type I spicules are probably the counterpart of the dynamic fibrils on the disk. They follow a parabolic (ballistic) path in space and time. In general the dynamics of these spicules is produced by mangneto-acoustic shock wave passing or wave-driving through the chromosphere \citep{Shibata_et_al_1982, De_Pontieu_et_al_2004, Hansteen_et_al_2006, Martinez-Sykora_et_al_2009, Matsumoto&Shibata_2010,Scullion_et_al_2011}. The second type of spicules (Type II) reach maximum heights of 3-9 Mm (longer in coronal holes) and have shorter lifetimes of 50-150 s than Type I spicules \citep{De_Pontieu_et_al_2007a, Pereira_et_al_2012}. These Type II spicules show apparent upward motions with speeds of order 30-110 km s$^{-1}$. At the end of their life they usually exhibit rapid fading in chromospheric lines \citep{De_Pontieu_et_al_2007b}. However the timescale of both types of spicules depends on the temperature, i.e., Ca II observations show short spicules, whereas Mg II or transition region lines show lifetimes of the order of ten minutes \citep{Pereira_et_al_2014,Skogsrud_et_al_2015}. Also in \cite{Zhang_et_al_2012}, the authors stablished the complexity of differentiating between Type I and Type II, so in general we can say that the Spicules are not well understood. It has been suggested from observations that Type II spicules are continuously accelerated while being heated to at least transition region temperatures \citep{De_Pontieu_et_al_2009,De_Pontieu_et_al_2011}. Another observations indicate that some Type II spicules also show an increase or a more complex velocity dependence with height \citep{Sekse_et_al_2012}. 

Apart from the upward motion, Type II spicules show swaying or transverse motions at the limb with velocity amplitudes of the order 10-30 km s$^{-1}$ and periods of 100-500 s \citep{De_Pontieu_et_al_2007b,Tomczyk_et_al_2007,Zaqarashvili&Erdelyi_2009,McIntosh_et_al_2011, Sharma_2017}, suggesting generation of upward, downwards and standing Alfvén waves \citep{Okamoto&De_Pontieu_2011,Tavabi_et_al_2015}, the generation of MHD kink mode waves or Alfv\'en waves due to magnetic reconnection \citep{Nishizuka_et_al_2008,He_et_al_2009, McLaughlin_et_al_2012,Kuridze_et_al_2012}. Also, \cite{Suematsu_et_al_2008}  suggest that some spicules show multi-thread structure as result of possible rotation. Another possible motions that Type II spicules show are the torsional motions as suggested in \citep{Beckers_1972}, and established using high-resolution spectroscopy at the limb \citep{De_Pontieu_et_al_2012}. According to the latter, Type II spicules show torsional motions with 25-30 km s$^{-1}$ speeds. 

There are observational results and theoretical models for the Type II spicules, however our understating of their physical origins remains limited. Some possibilities are that Type II spicules are due to magnetic reconnection \citep{Isobe_et_al_2008,De_Pontieu_et_al_2007b,Archontis_et_al_2010,Gonzalez-Aviles_et_al_2017a}, oscillatory reconnection processes \citep{Heggland_et_al_2009,McLaughlin_et_al_2012}, strong Lorentz force \citep{Martinez-Sykora_et_al_2011,Goodman_2012} or propagation of $p$-modes \citep{de_Wijn_et_al_2009}. More recently, \citep{Martinez-Sykora_et_al_2017} showed that spicules occur when magnetic tension is amplified and transported upward through interaction between ions and neutrals or ambipolar diffusion. The tension is impulsively released to drive flows, heat plasma, and generate Alfv\'enic waves.  

In this paper, we show that 3D magnetic reconnection may be responsible for formation of a jet with characteristics of a Type II spicule. For that (i) we assume a completely ionized solar atmosphere which is governed by the resistive MHD equations subject to a constant gravitational field, (ii) we model the solar atmosphere based on the C7 model in combination with a 3D potential magnetic field configuration extrapolated from a realistic photospheric quiet-Sun model. 

The system of equations, the magnetic field configuration, the numerical methods and the model of the solar atmosphere are described in detail in Section \ref{sec:model_numerical_methods}. The results of the numerical simulations are presented in Section \ref{sec:Results}. Finally in the Section \ref{sec:conclusions}, we present the final comments and conclusions.


\section{Model and Numerical Methods}
\label{sec:model_numerical_methods}


\subsection{The system of Resistive MHD equations}
\label{sub_sec:eglm_resistive_mhd_equations}

We solve the dimensionless Extended Generalized Lagrange Multiplier (EGLM) resistive MHD \citep{Jiang_et_al_2012} equations that include gravity:

\begin{eqnarray}
\frac{\partial\rho}{\partial t} +\nabla\cdot(\rho{\bf{v}})=0, \label{eglm_cont_equation} \\
\frac{\partial(\rho{\bf v})}{\partial t} + \nabla\cdot\left(\left(p+\frac{1}{2}{\bf B}^{2}\right){\bf I}+\rho{\bf vv}-{\bf BB}\right)\nonumber\\
=-(\nabla\cdot{\bf B}){\bf B}+\rho{\bf g}, \label{eglm_mom_equation} \\
\frac{\partial E}{\partial t}+\nabla\cdot\left({\bf v}\left(E+\frac{1}{2}{\bf B}^{2}+p \right)-{\bf B}({\bf B}\cdot{\bf v})\right)\nonumber\\
=-{\bf B}\cdot(\nabla\psi)-\nabla\cdot((\eta{\bf J})\times{\bf B})+\rho{\bf g}\cdot{\bf v}, \label{eglm_energy_equation} \\[0.3 cm]
\frac{\partial{\bf B}}{\partial t}+\nabla\cdot({\bf Bv}-{\bf vB}+\psi{\bf I})=-\nabla\times(\eta{\bf J}), \label{eglm_induction_equation} \\
\frac{\partial\psi}{\partial t}+c_{h}^{2}\nabla\cdot{\bf B}=-\frac{c_{h}^{2}}{c_{p}^{2}}\psi, \label{eglm_psi_equation} \\ 
{\bf J}=\nabla\times{\bf B},  \label{eq:corriente} \\
E=\frac{p}{(\gamma-1)}+\frac{\rho{\bf v}^{2}}{2}+\frac{{\bf B}^{2}}{2},  \label{eq:energiatotal}
\end{eqnarray}

\noindent where $\rho$ is the mass density, ${\bf v}$ is the velocity vector field, ${\bf B}$ is the magnetic vector field, $E$ is the total energy density and $\gamma=5/3$ is the adiabatic index. The plasma pressure $p$ is described by the equation of state of an ideal gas. ${\bf g}$ is the gravitational field, ${\bf J}$ is the current density, $\eta$ is the magnetic resistivity tensor and $\psi$ is a scalar potential that aims at damping out the violation of the constraint $\nabla\cdot{\bf B}=0$. Here $c_h$ is the wave speed and $c_p$ is the damping rate of the wave of the characteristic mode associated with $\psi$. In this study we consider uniform and constant magnetic resistivity for simplicity. The system of Equations (\ref{eglm_cont_equation})-(\ref{eq:energiatotal}) was normalized by the quantities given in Table \ref{table:1}, which are typical scales in the solar atmosphere.   

In the EGLM-MHD formulation, Equation (\ref{eglm_psi_equation}) is the magnetic field divergence free constraint. As suggested in \cite{Dedner_2002}, the expressions for $c_h$ and $c_p$ are
 
\begin{equation} 
c_h = \frac{c_{cfl}}{\Delta t}min(\Delta x,\Delta y,\Delta z), 
~~~c_p = \sqrt{\left|\frac{-\Delta t}{\ln c_{d}}\right|c_{h}^{2}}, \nonumber 
\end{equation}

\noindent where $\Delta t$ is the time step, $\Delta x$, $\Delta y$ and $\Delta z$ are the spatial resolutions, $c_{cfl}<1$ is the Courant factor, $c_d$ is a problem dependent coefficient between 0 and 1, this constant determines the damping rate of divergence errors. The parameters $c_h$ and $c_p$ are not independent of the grid resolution and  the numerical scheme used, for that reason one should adjust their values. In our simulations we use $c_p=\sqrt{c_r}c_h$, with $c_r=0.18$ and $c_h=0.1$. In this work we solve the 3D resistive MHD equations with resolutions $\Delta x$, $\Delta y$ and $\Delta z$. 

The gas pressure is computed using the thermal energy, which is obtained by subtracting the kinetic and magnetic energy from the total energy, defined by the total energy Equation (\ref{eq:energiatotal}). In the solar corona region, the plasma-$\beta$ can become very small, and the thermal energy could be many orders of magnitude smaller than magnetic energy. Therein, small discretization errors in the total energy can produce unphysical negative pressure. We fix this problem by replacing the total energy density Equation (\ref{eglm_energy_equation}) in low-beta regions ($\beta\leq10^{-2}$) with the entropy density equation.

\begin{equation}
\frac{\partial S}{\partial t}  + \nabla\cdot(S{\bf v}) = (\gamma-1)\rho^{1-\gamma}\eta{\bf J}^{2}, \label{entropy_equation}
\end{equation}

\noindent where $S=\frac{p}{\rho^{\gamma-1}}$ is the entropy density and ${\bf J}^{2}=J_{x}^{2}+J_{y}^{2}+J_{z}^{2}$. In this way, we calculate the pressure directly using the entropy, which, by definition, is a positive quantity. The entropy density equation is  used to maintain the positivity of gas pressure in the context of the ideal MHD simulations \citep{Balsara&Spicer_1999,Li_2008,Derings_et_al_2016}, and is also used in some resistive MHD simulations of the solar corona \citep{Takasao_et_al_2015}. In the ideal MHD limit, equation (\ref{entropy_equation}) is an advection type of equation, whereas in the case of the resistive MHD equations the Ohmic dissipation is added as a source term. This entropy equation is consistent with the second law of thermodynamics in the continuum limit \citep{Derings_et_al_2016}.

\begin{table}
\caption{Normalization units}
\centering
\begin{tabular}{c c c c}
\hline\hline
Variable & Quantity & Unit & Value \\ [0.5ex]
\hline
x,y,z & Length & $l_0$ & $10^{6}$ m \\
$\rho$ & Density & $\rho_{0}$ & $10^{-12}$ kg m$^{-3}$\\
${\bf B}$& Magnetic field & $B_{0}$ & 11.21 G \\
${\bf v}$ & Velocity & $v_{0}=B_0/\sqrt{\mu_0\rho_0}$ & $10^{6}$ m s$^{-1}$ \\
$t$& Time & $t_{0}=l_0/v_0$ & 1 s \\
$\eta$ & Resistivity & $\eta_{0}=l_0\mu_0 v_0$ & 1.256$\times10^{6}$ m$^{2}$ s$^{-1}$ N A$^{-2}$ \\ [1ex]
\hline
\end{tabular}
\label{table:1}
\end{table}


\subsection{The magnetic field}
\label{sub_sec:initial_conditions}

As an initial magnetic configuration, we use a 3D potential (current-free) magnetic field extrapolated from a simulated quiet-Sun photospheric field. The latter has been obtained from a large-scale, high-resolution self-consistent simulation of solar magnetoconvection in a bipolar photospheric region with MURaM code \citep{Shelyag_et_al_2012, Vogler_et_al_2005}. The original computational box had a size of 480$\times$480$\times$400 pixels with a spatial resolution of 25 km in the horizontal directions and 10 km in the vertical direction. The initial magnetic field was created as a checkerboard (positive-negative) pattern with the unsigned vertical magnetic field strength of $200~\mathrm{G}$. This field configuration was inserted into a well-developed non-magnetic photospheric convection model and evolved for 20 minutes of physical time. During this simulation phase the magnetic field partially cancelled and partially concentrated in the intergranular lanes forming the intergranular magnetic field concentrations with random polarities and with the strength of $\sim 1.5~\mathrm{kG}$ \citep{Shelyag_et_al_2012}. 

\begin{figure*}
\centering
\includegraphics[scale=0.25]{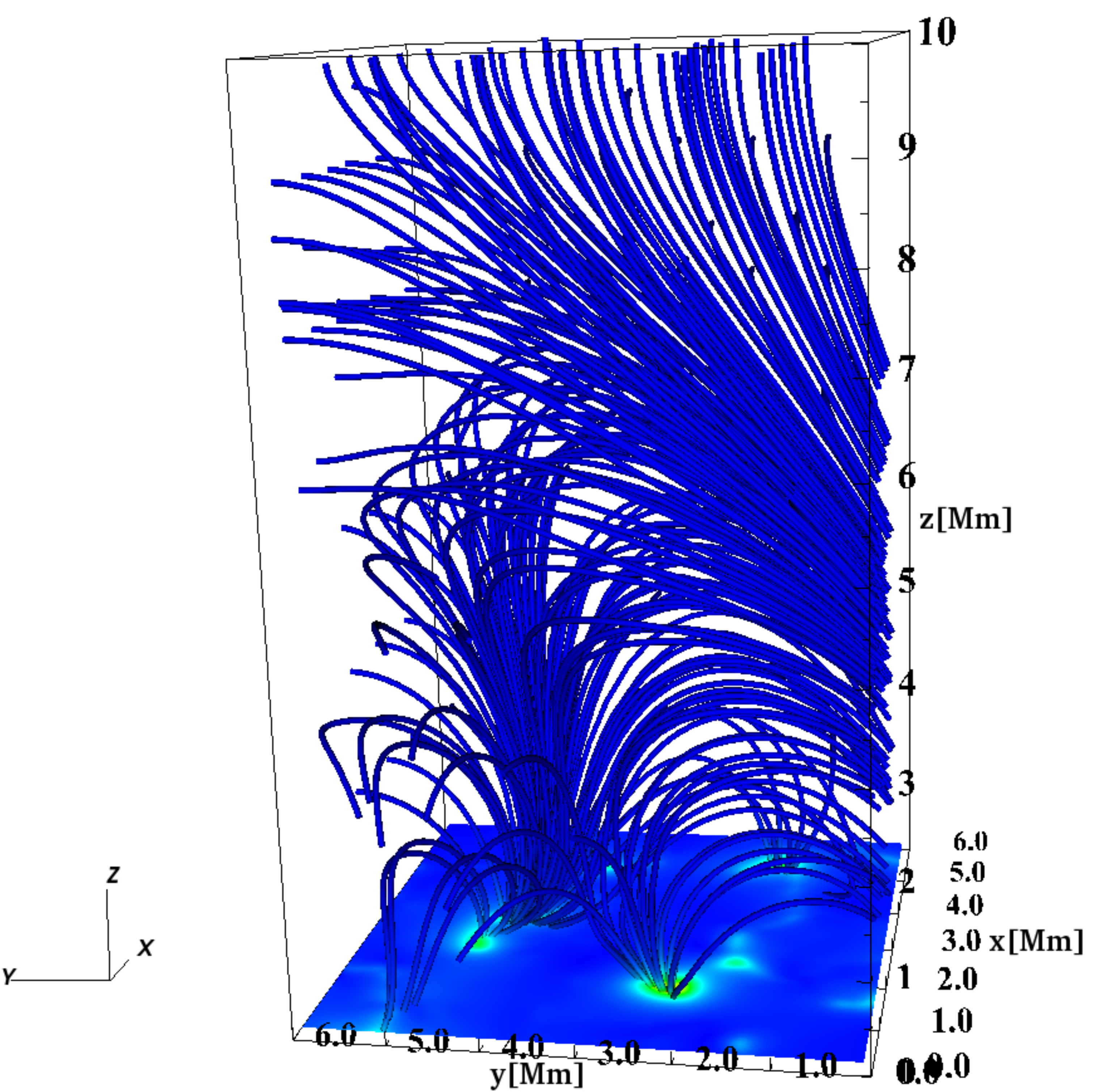}
\includegraphics[scale=0.22]{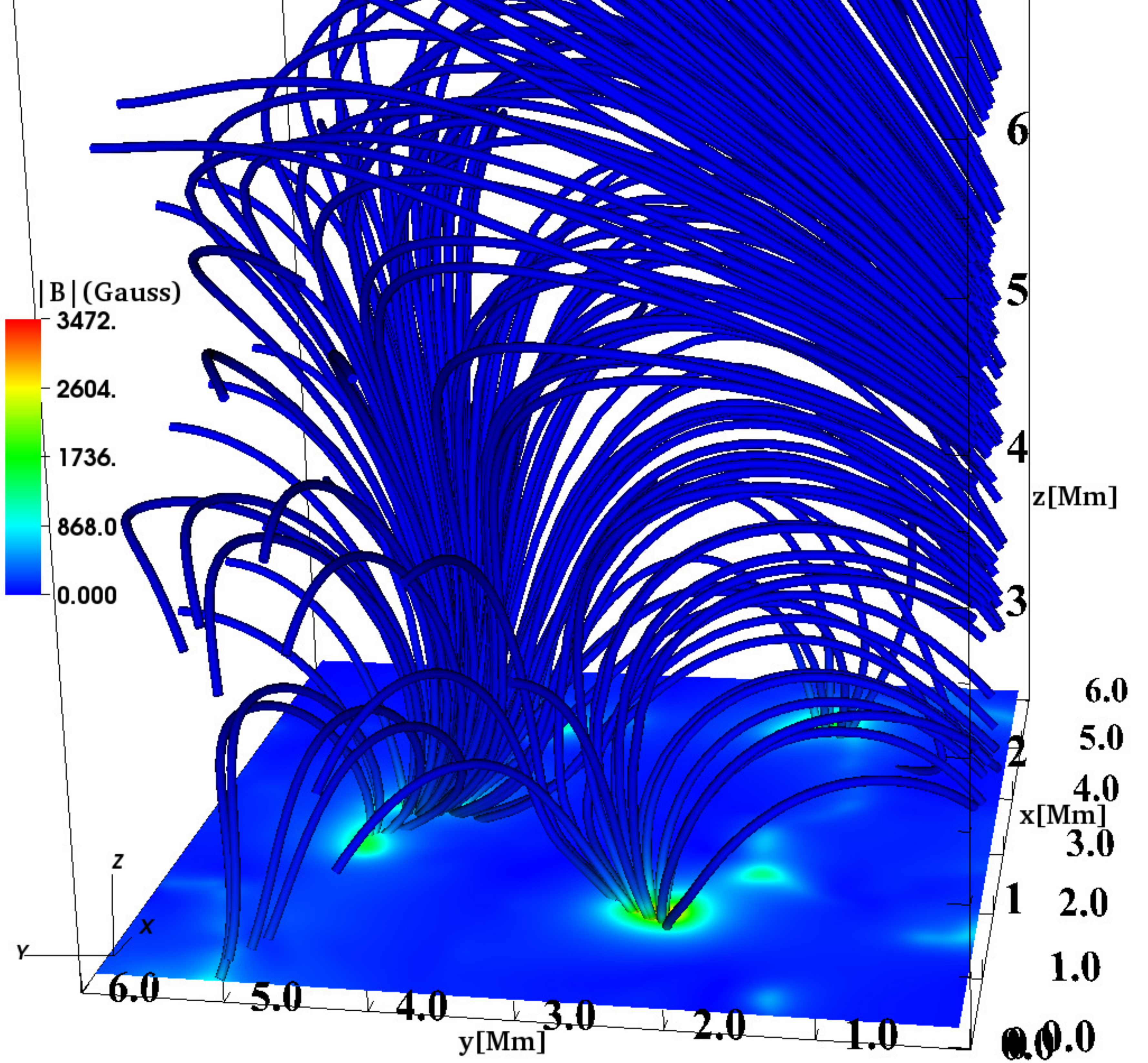}\\
\includegraphics[scale=0.25]{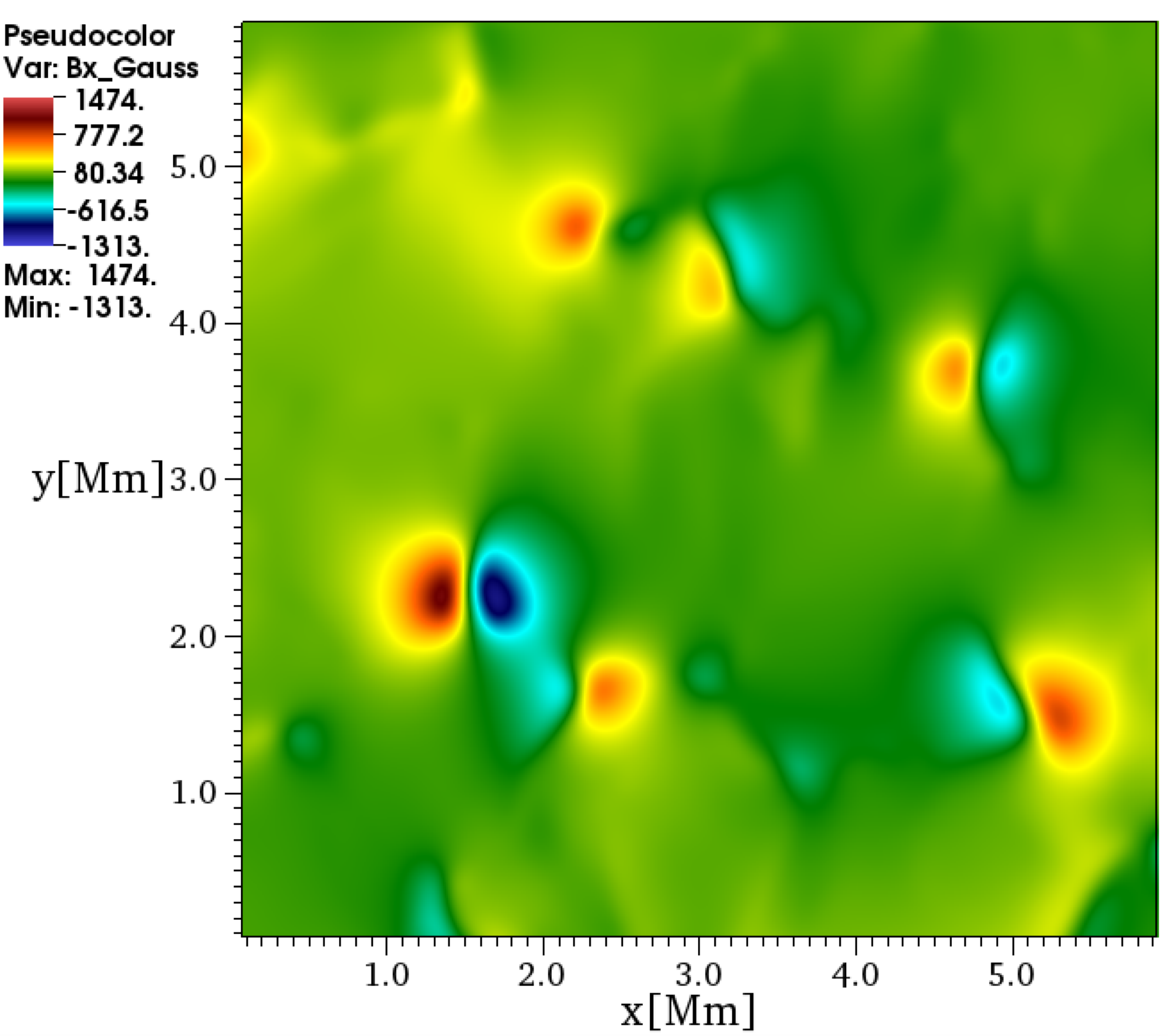}
\includegraphics[scale=0.25]{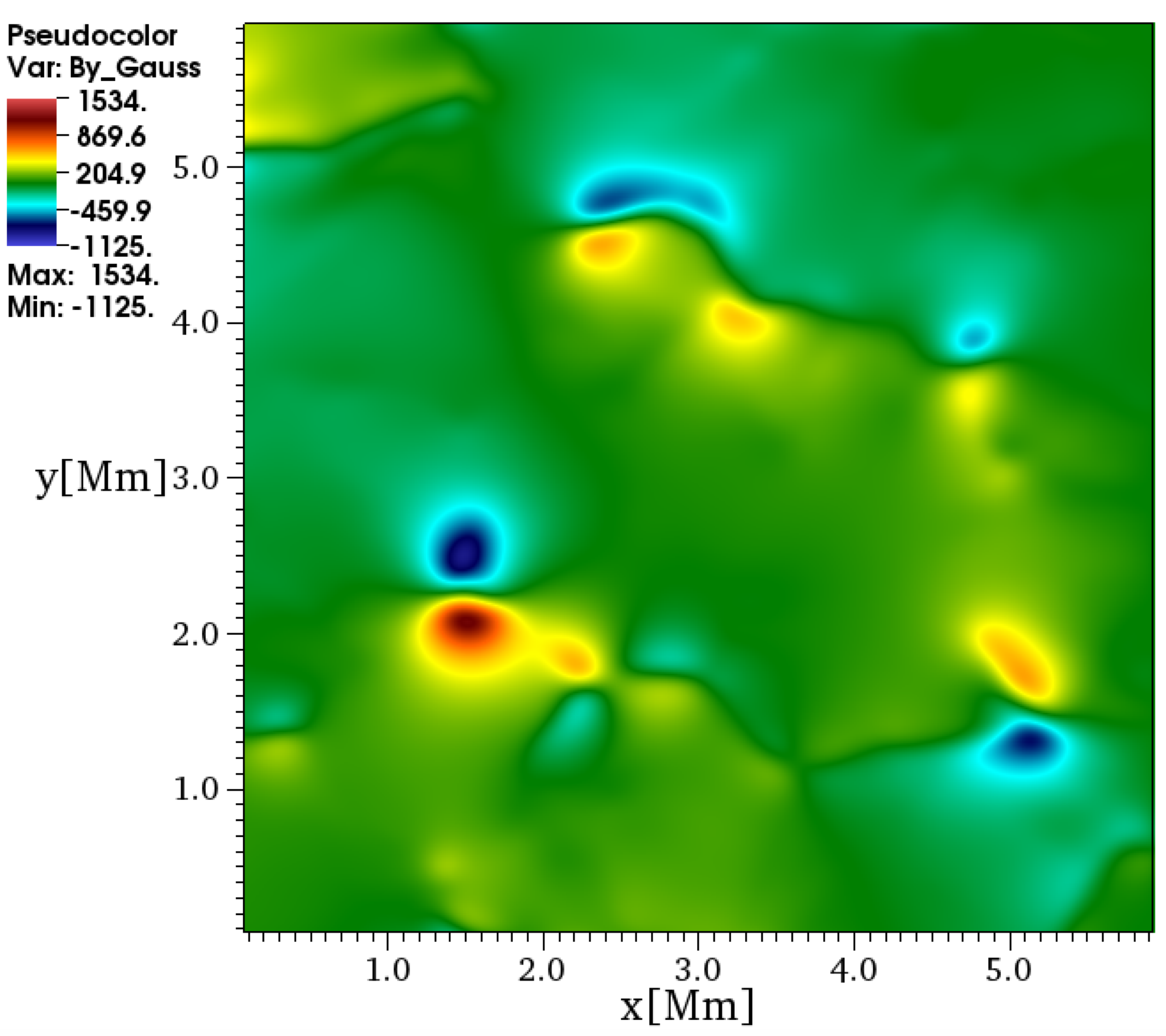}
\includegraphics[scale=0.25]{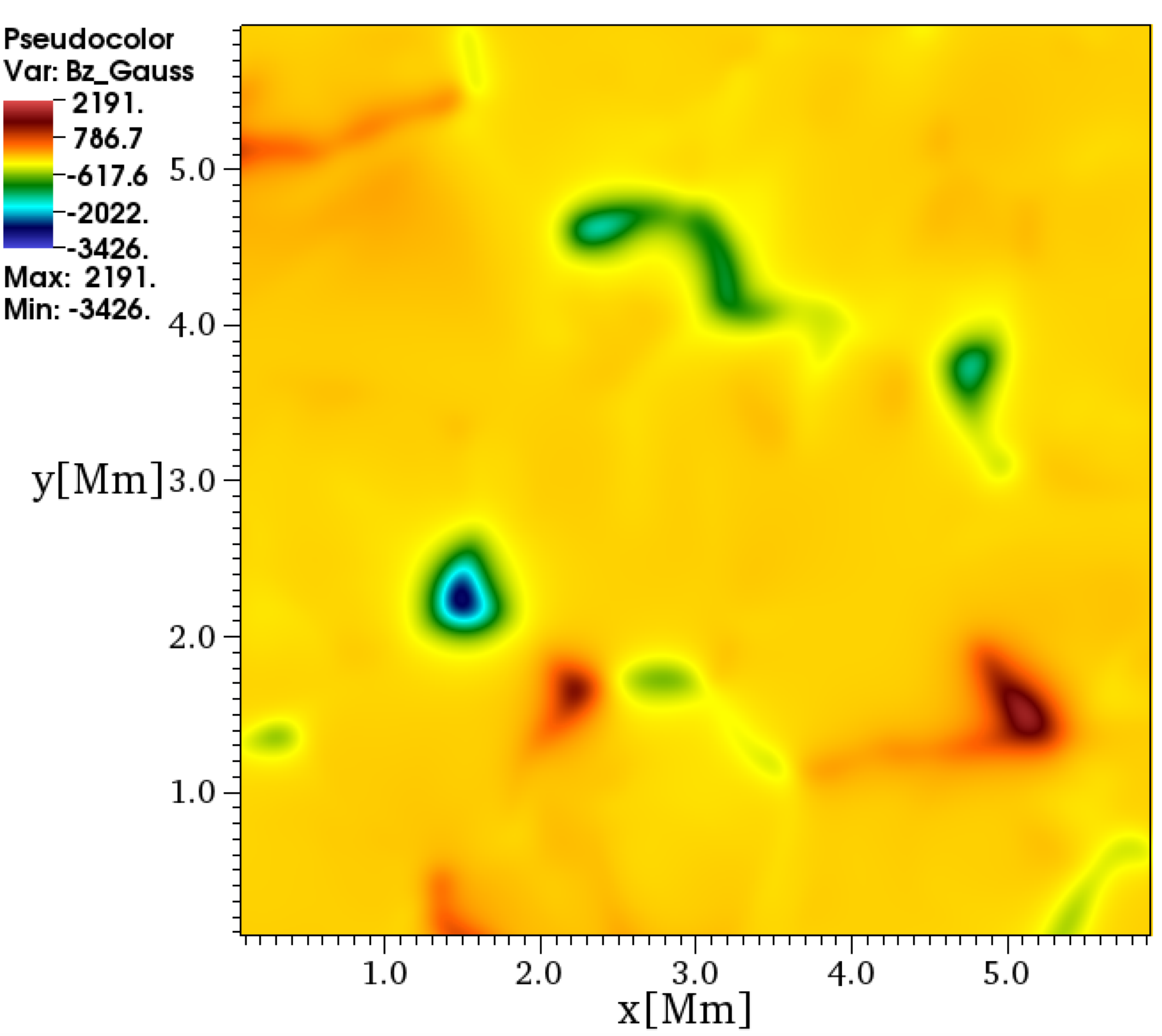}
\caption{\label{3D_magnetic_field} (Top) Magnetic field lines and zoom of strong bipolar regions in the 3D domain at initial time. At the bottom plane we show the magnitude of the magnetic field. (Bottom) Three components of the magnetic field $B_x$ $B_{y}$ and $B_{z}$ at the plane $z=0.1$Mm. The color bars represent the magnitude of the magnetic field in Gauss.}
\end{figure*}

The potential field extrapolation is based on a vector-potential Grad-Rubin-like method as described in \cite{Amari_et _al_1997}. The potential field extrapolation uses open boundary conditions on the side and top of the computational box: the first derivative of the magnetic field component normal to the surface of the box vanishes. We select a 3D domain of 6$\times$6$\times$10 Mm containing a topology of interest to perform our numerical simulations, the reason is that with such structure it is likely that reconnection may happen and lead to jet generation. The magnetic field lines of the 3D configuration and the magnitude of the magnetic field at $z=0.1$Mm are shown on the top panel of Figure \ref{3D_magnetic_field}. The all three components of the magnetic field $B_x$, $B_y$, $B_z$ in the plane  $z=$0.1 Mm are shown in the bottom panels, where dipolar structures can be observed. In our convention the $xy$ plane is horizontal and $z$ labels height. These plots show the region used to simulate the evolution of the system, which contains magnetic dipoles at around the location $(x,y,z)$$\sim$$(1.4, 2.3, 0.1)$ Mm.
  

\subsection{Numerical methods}
\label{sub_sec:numerical_methods}

The implementation is the same High Resolution Shock Capturing method as used in \citep{Gonzalez-Aviles_et_al_2017a}, based on finite volume approximation. However, in the present paper we exploit the full three-dimensional capabilities of the Newtonian CAFE code \citep{Gonzalez-Aviles_et_al_2015}. A summary of the specific numerical methods is as follows. We solve numerically the system of Equations (\ref{eglm_cont_equation})-(\ref{entropy_equation}) on a uniform cell centered grid, using the method of lines with a third order  Runge-Kutta time integrator (RK3) \citep{Shu&Osher_1989}. The discretization of the resistive MHD equations above is based on finite volume approximation. We use the MINMOD and MC limiters for the flux reconstruction, and a combination of the HLLE and HLLC approximate flux formulas \citep{Einfeldt_1988,Harten_et_al_1983,Li_2005}. The combinations of limiters and flux formulas is adaptive and depends on the magnitude of the discontinuities and shocks formed during the evolution, using the maximum dissipative combination MINMOD-HLLE in zones where $\beta < 10^{-2}$ and the least dissipative combination MC-HLLC otherwise. 


\subsection{Model of the solar atmosphere}
\label{sub_sec:solar_atmosphere}

\begin{figure}
\centering
\includegraphics[width=8.25cm]{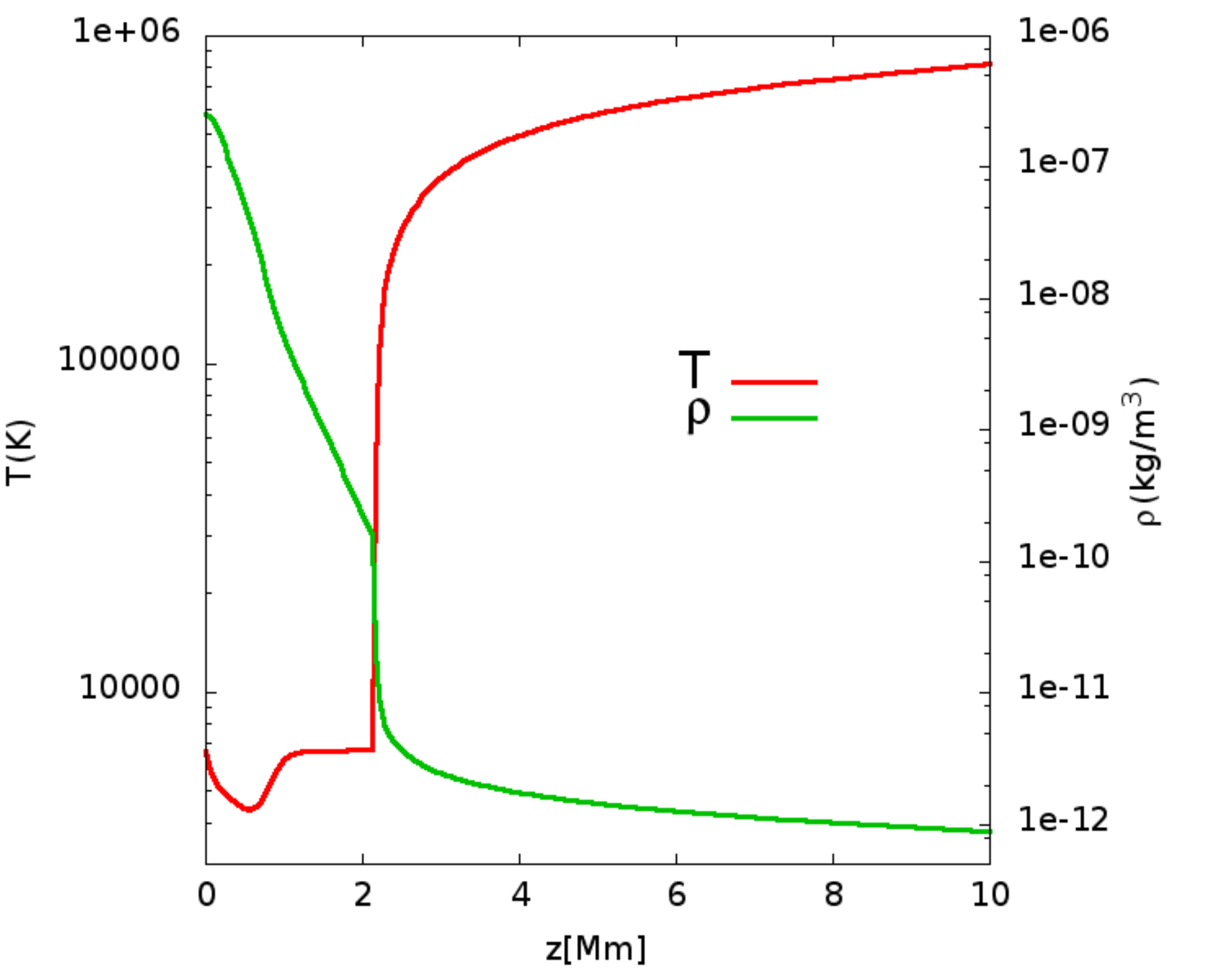}
\caption{Temperature and mass density as a function of height for the C7 equilibrium solar atmosphere model. }
\label{fig:atmosphere}
\end{figure} 

We choose the numerical domain to cover part of the interconnected solar photosphere, chromosphere and corona (see top left panel of Figure \ref{3D_magnetic_field} and Figure~\ref{fig:atmosphere}). For this the atmosphere is initially assumed to be in hydrostatic equilibrium. The temperature field is considered to obey the semi-empirical C7 model of the chromosphere-transition region \citep{Avrett&Loeser2008} and is distributed consistently with observed line intensities and  profiles from the SUMER atlas of the extreme ultraviolet spectrum \citep{Curdt_et_al_1999}. The photosphere is extended to the solar corona as described by \cite{Fontela_et_al_1990} and \cite{Griffiths_et_al_1999}. The temperature $T(z)$ and density $\rho(z)$ as functions of height $z$ are shown in Figure \ref{fig:atmosphere}, where the transition region is characterized by the steep gradients. 


\section{Results of Numerical Simulations}
\label{sec:Results}

We carried out a numerical simulation within a specific domain with magnetic fields constructed with the MURaM code, which contained a region with a high magnetic field strength dipoles. We define the numerical domain to be $x\in[0,6]$, $y\in[0,6]$, $z\in[0,10]$ Mm, covered with 240$\times$240$\times$400 grid cells, i.e., the effective resolution is 25 km in each direction. In the faces of the numerical box we set fixed in time boundary conditions, which keep the value of the variables set to their initial condition value at a ghost boundary three ghost cells out from the six faces of the physical boundary.

Once we set the magnetic field and the atmosphere model described above, we start evolving the plasma according to the Equations (\ref{eglm_cont_equation}-\ref{entropy_equation}). We do not apply any explicit perturbation to the system, instead, the round-off errors suffice to trigger the instability of the whole system, including the magnetic field and hydrodynamic equilibria, that later on traduces into the burst of material upwards. The reconnection happens and is accompanied by the introduction of a finite magnetic resistivity $\eta=12.56~\Omega  ~{\rm m}$. 

\begin{figure*}
\centering
\includegraphics[scale=0.21]{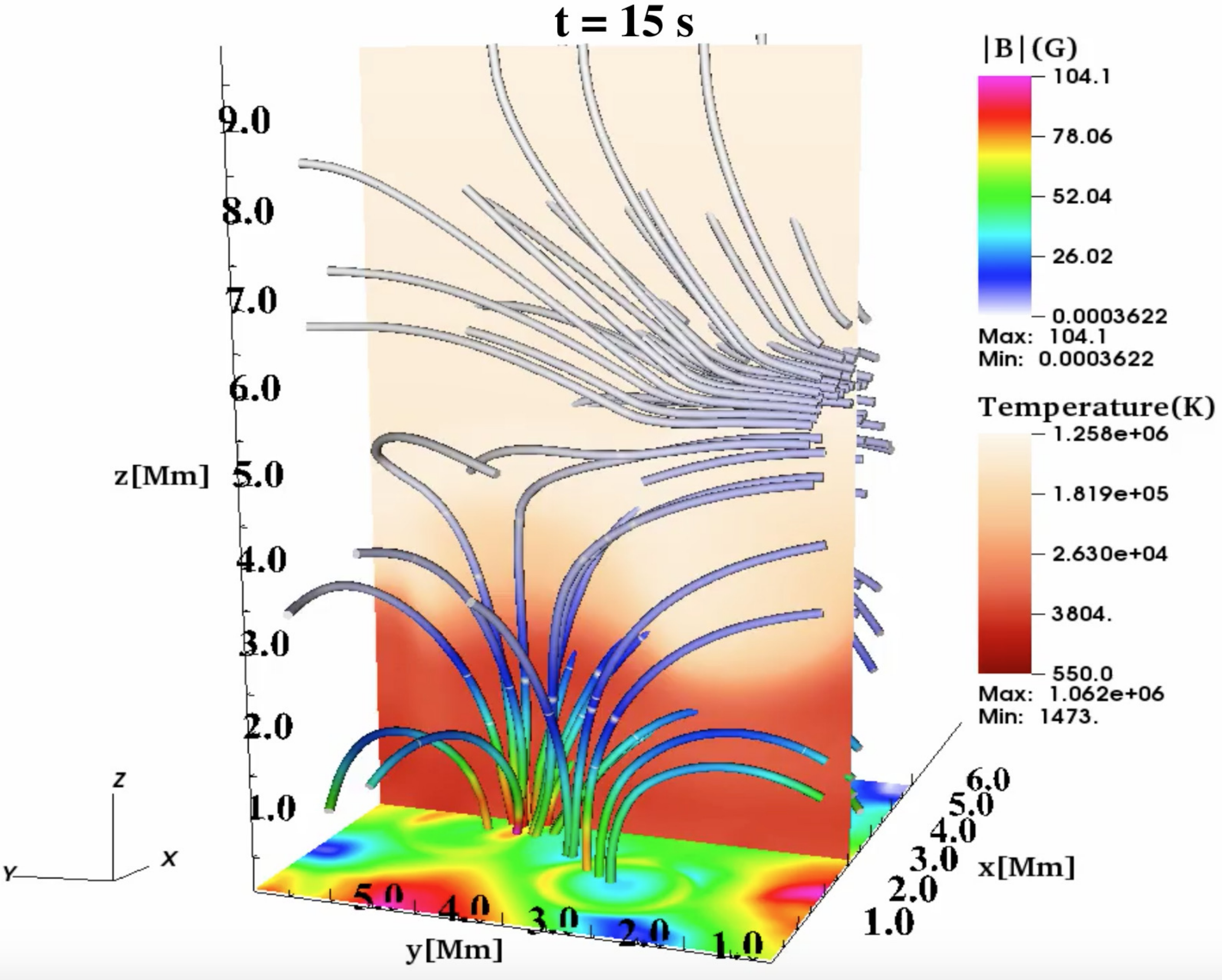}
\includegraphics[scale=0.21]{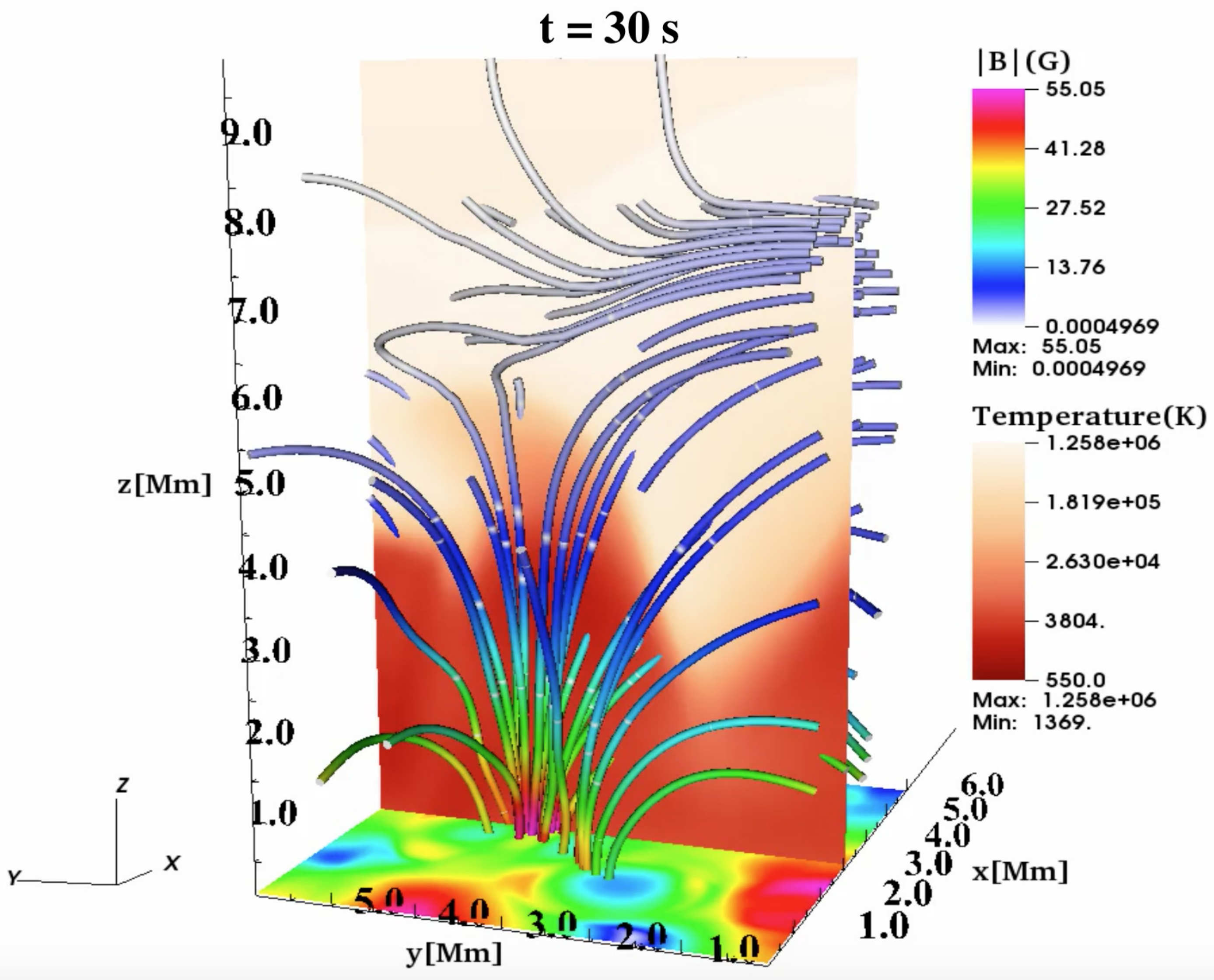}
\includegraphics[scale=0.21]{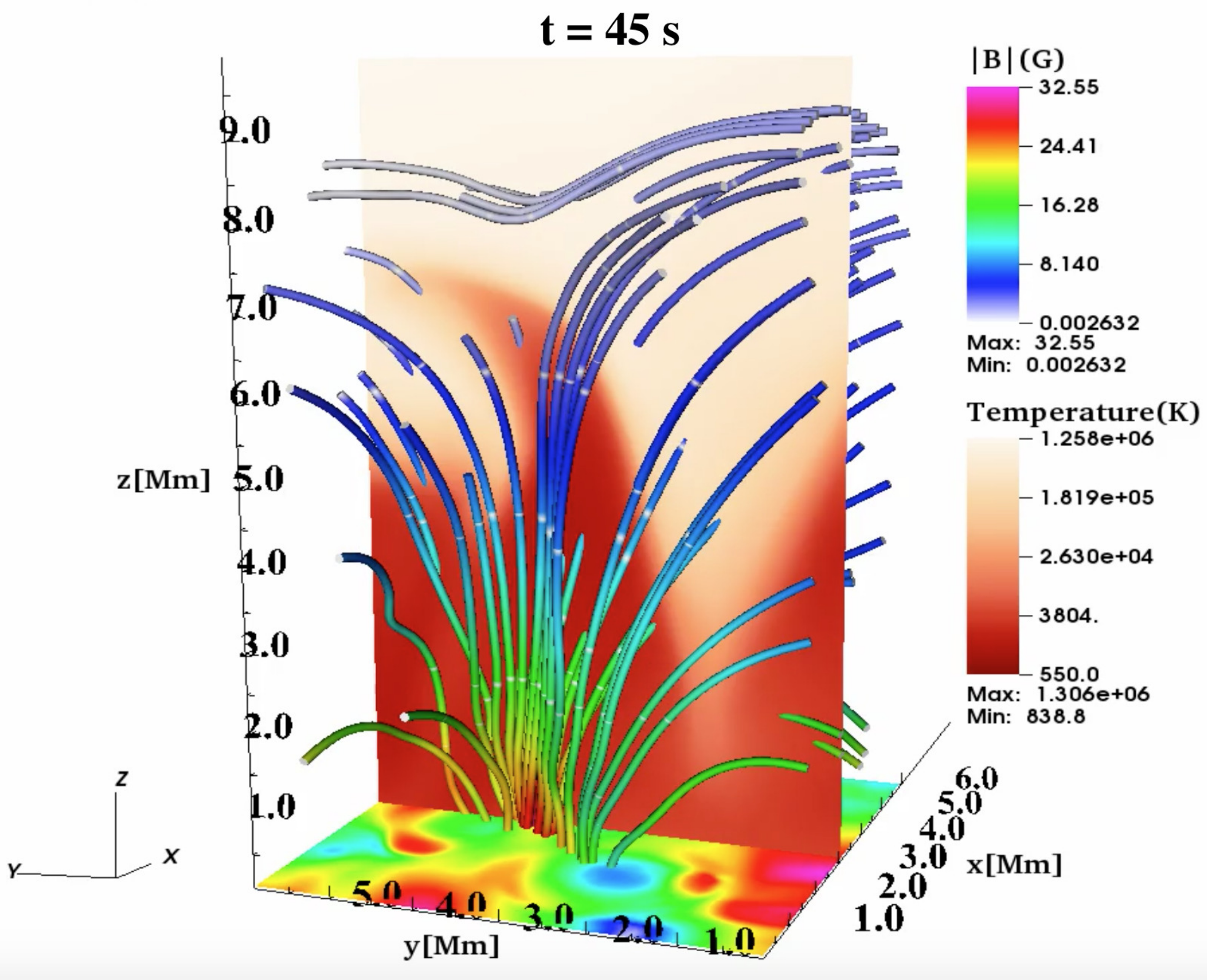}\\
\includegraphics[scale=0.21]{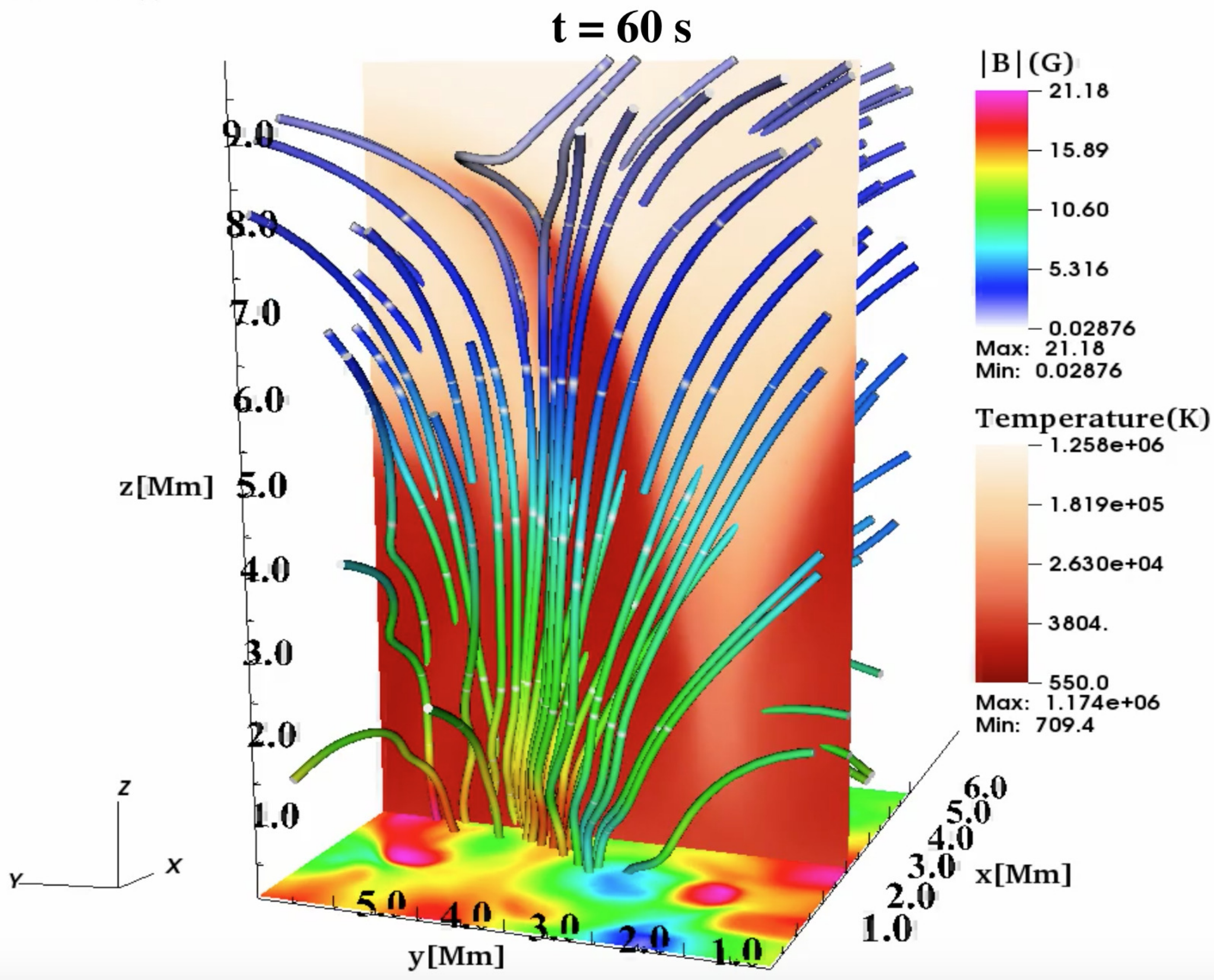}
\includegraphics[scale=0.21]{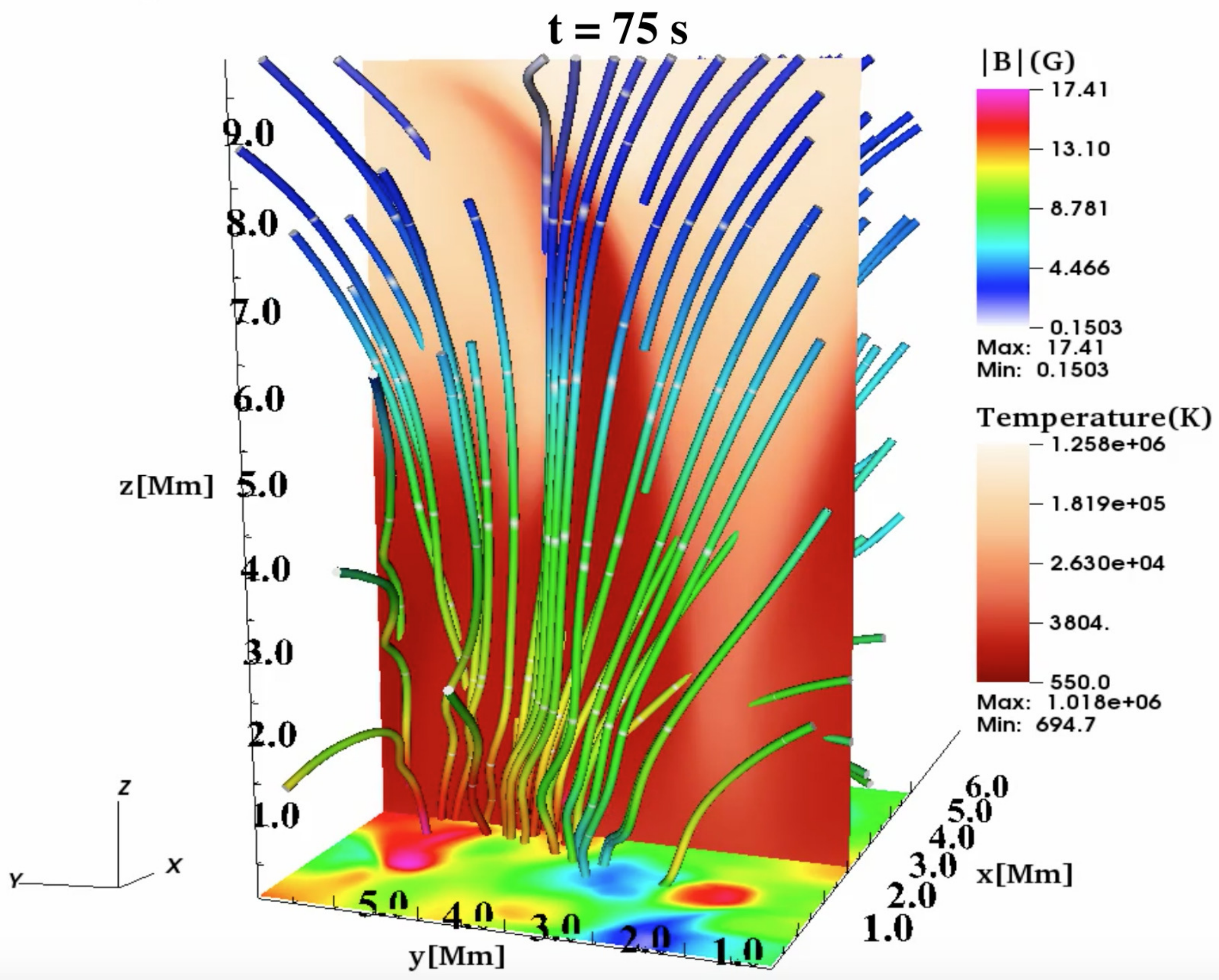}
\includegraphics[scale=0.21]{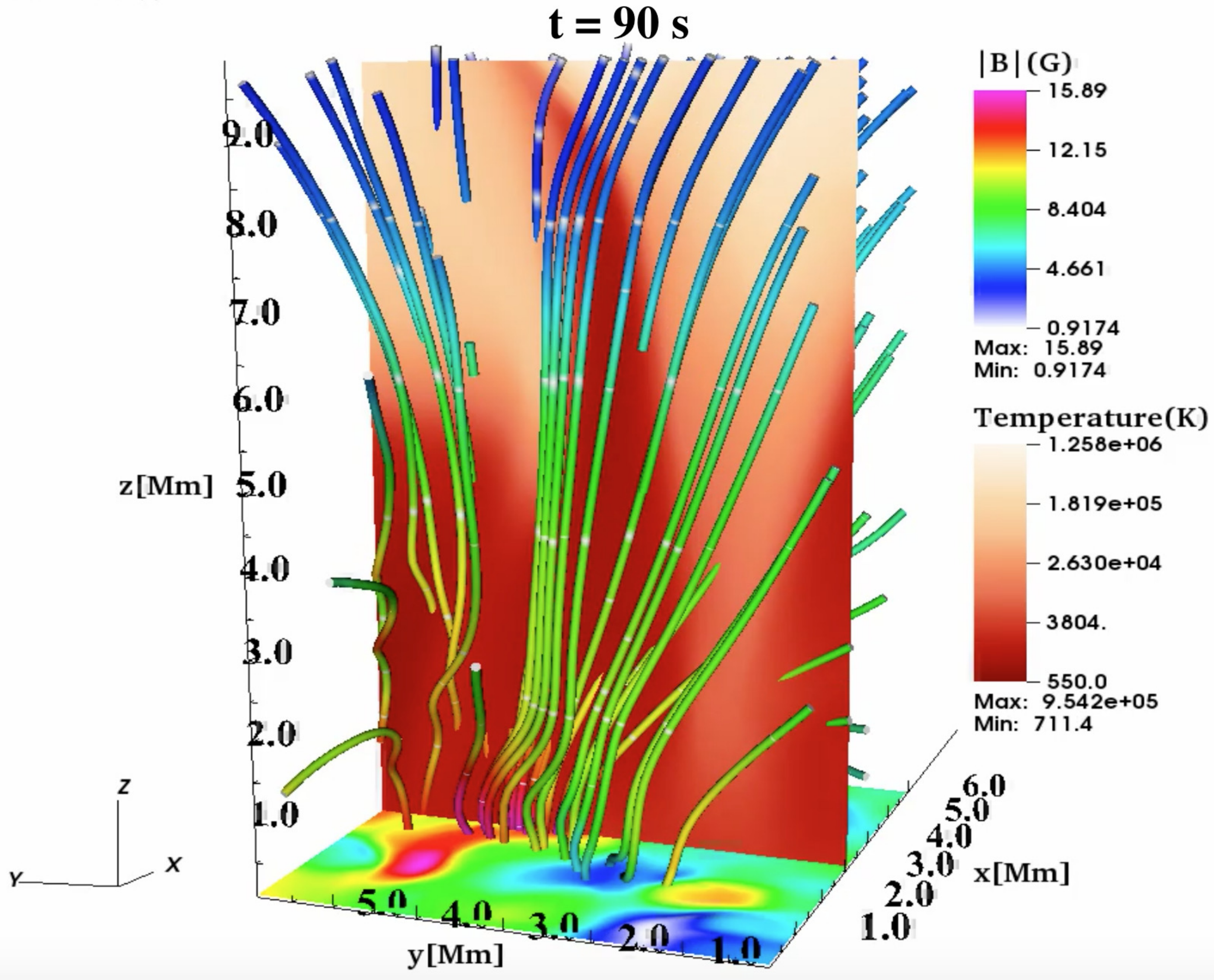}
\caption{\label{logTemp_3D_fieldlines} The 3D representation of jet formation. Snapshots of the logarithm of temperature in Kelvin and magnetic field lines in 3D at times 15, 30, 45, 60, 75 and 90 s. At the bottom we display the magnitude of the magnetic field in the $xy$ plane at $z=$0.1 Mm. The color bars represent the magnitude of the magnetic field in Gauss and Temperature.}
\end{figure*}

We focus on the process of jet formation and track the temperature evolution that helps understanding the dynamics of the system. In Figure~\ref{logTemp_3D_fieldlines} we show snapshots of temperature on the plane $x=2.5$ Mm and the magnetic field lines in 3D at different times. For instance, at time $t=15$ s the jet-like structure starts to develop in the region of magnetic reconnection which accelerates the plasma. Between $t=30$ s and $t=45$ s, the jet continues to develop and moves upwards. The most representative time of the jet formation is $t=60$ s, at this time we can see a structure with a similar morphology of a Type II spicule, which reaches a height of about $z\approx$7 Mm measured from the transition region \citep{Tavabi_et_al_2015} and vertical velocity of about $v_z\approx 130$ km s$^{-1}$ as shown in Figure \ref{vz_temp_contours_vfield_yz_plane_times}, these characteristics are similar to those of a Type II spicule \citep{De_Pontieu_et_al_2007a}. At time $t=90$ s, the spicule-like structure reaches the top of the domain located at  $z=10$ Mm. 

\begin{figure*}
\centering
\includegraphics[scale=0.26]{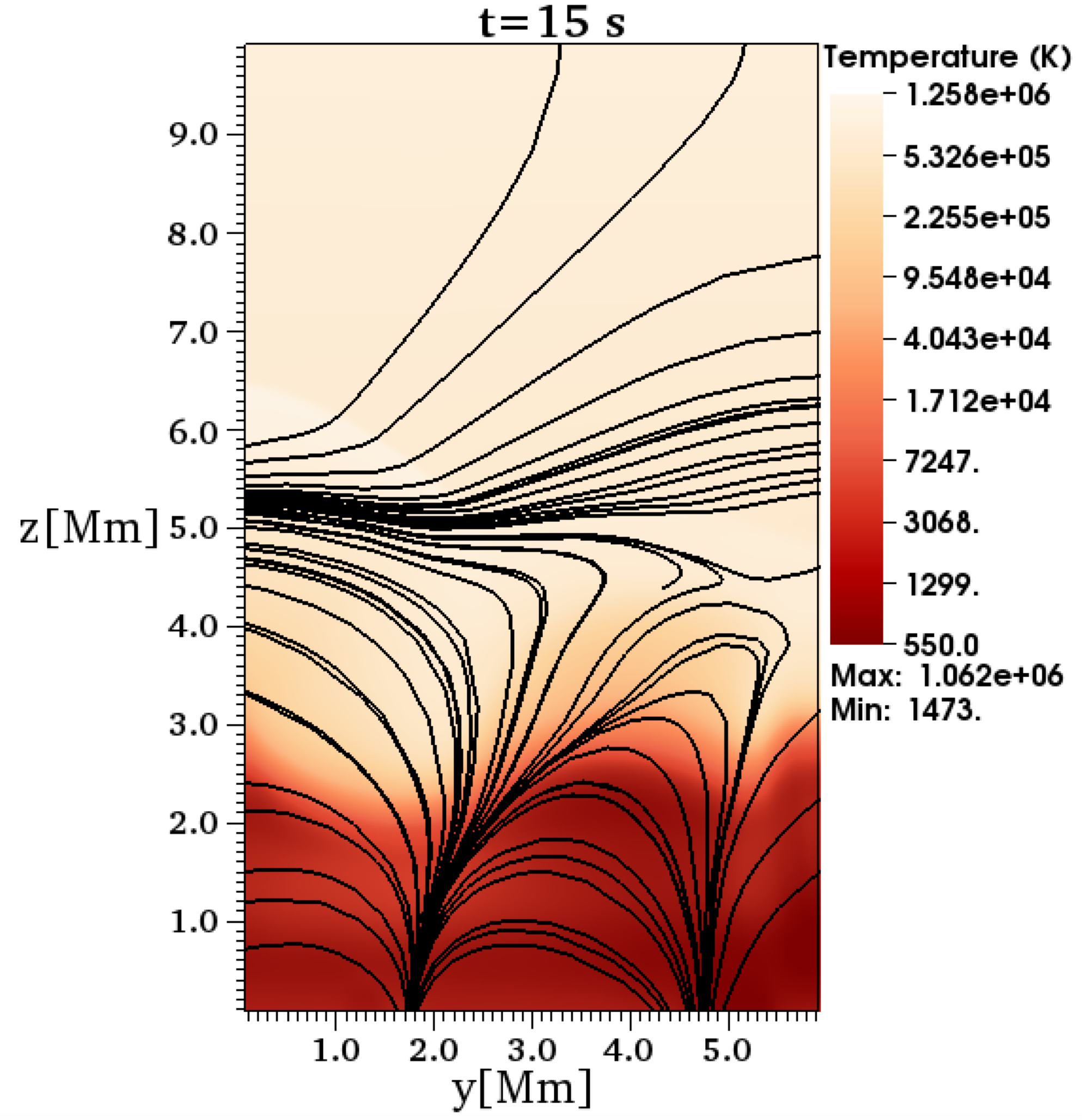}
\includegraphics[scale=0.26]{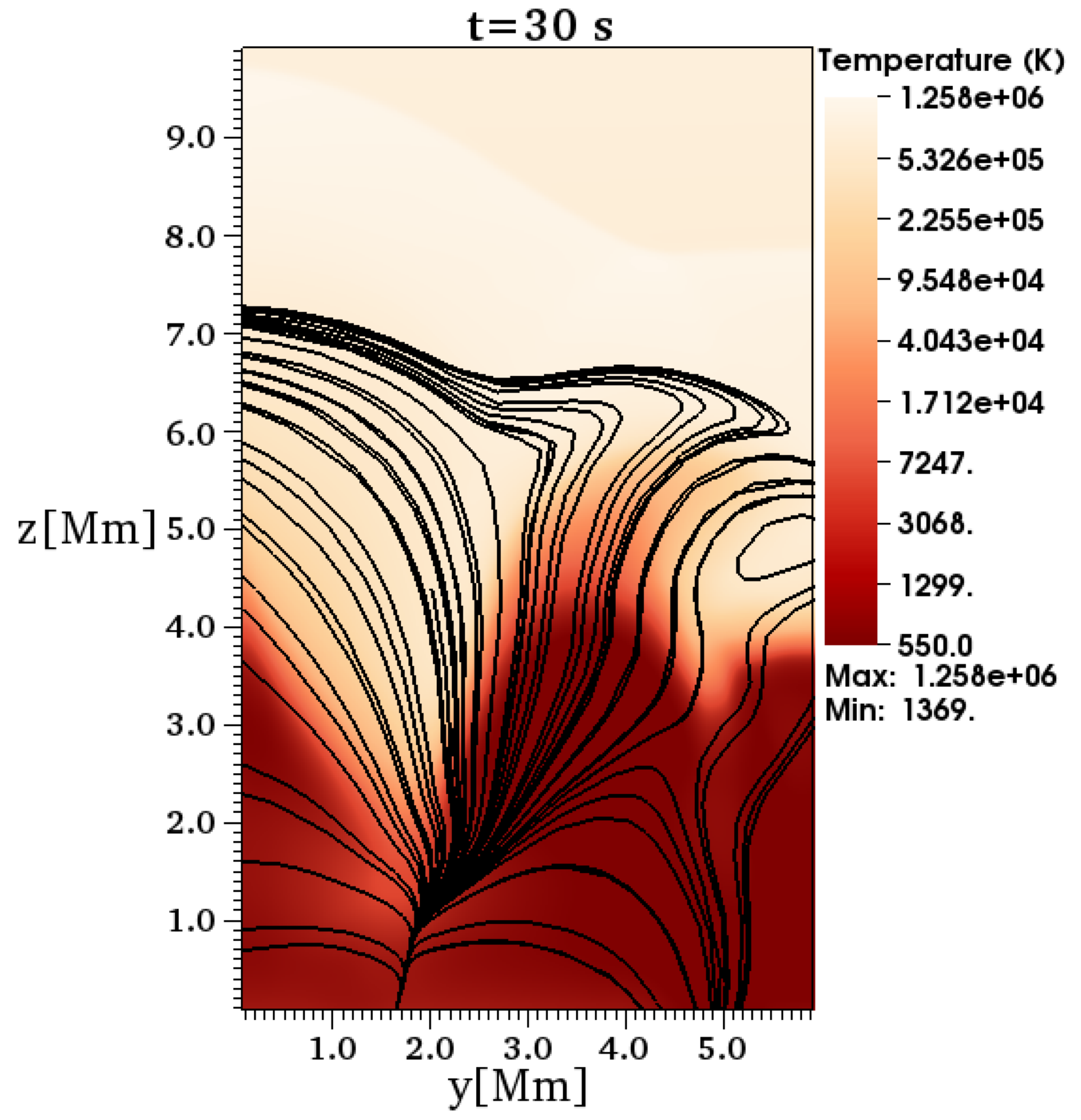}
\includegraphics[scale=0.26]{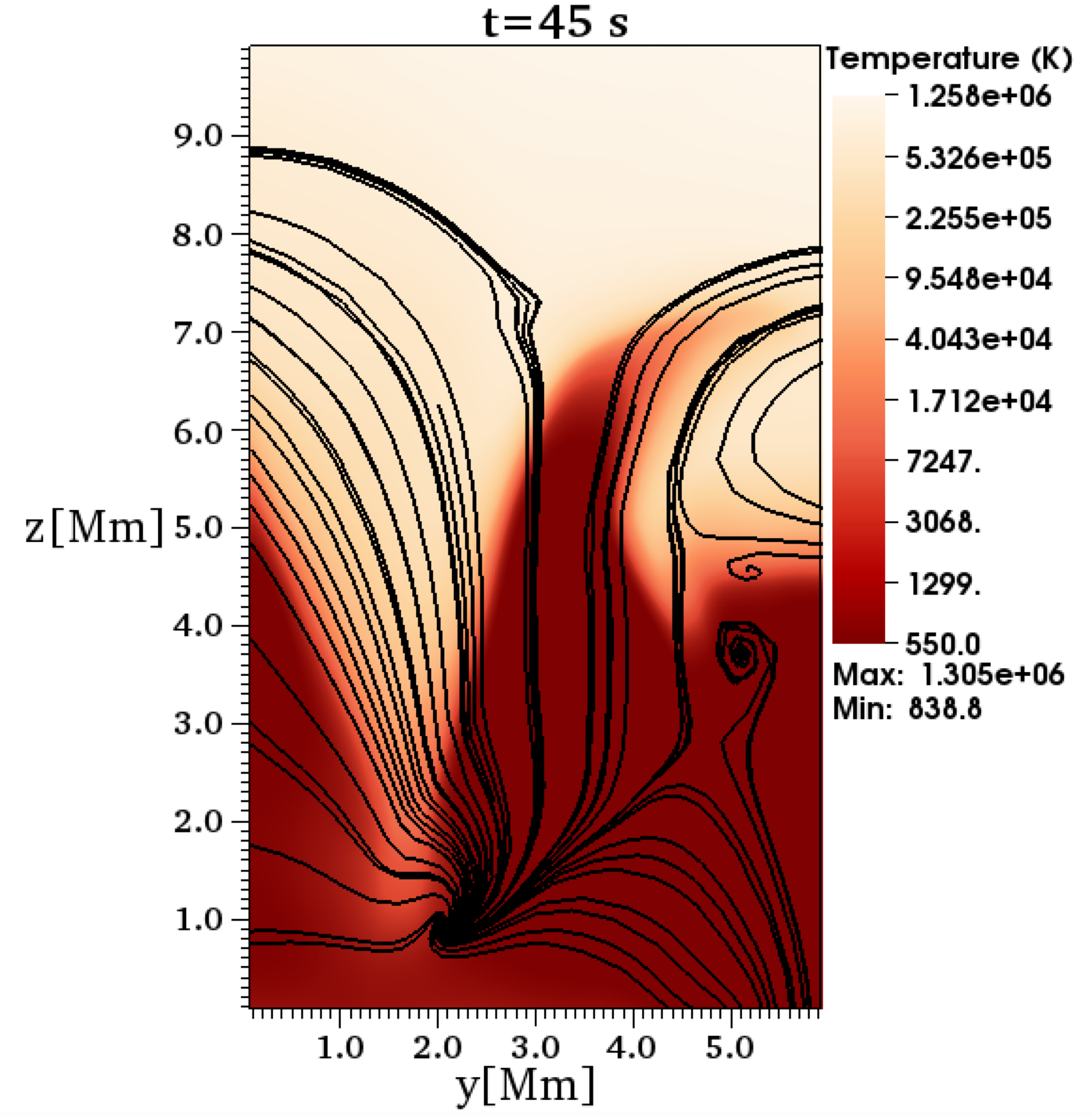}
\includegraphics[scale=0.26]{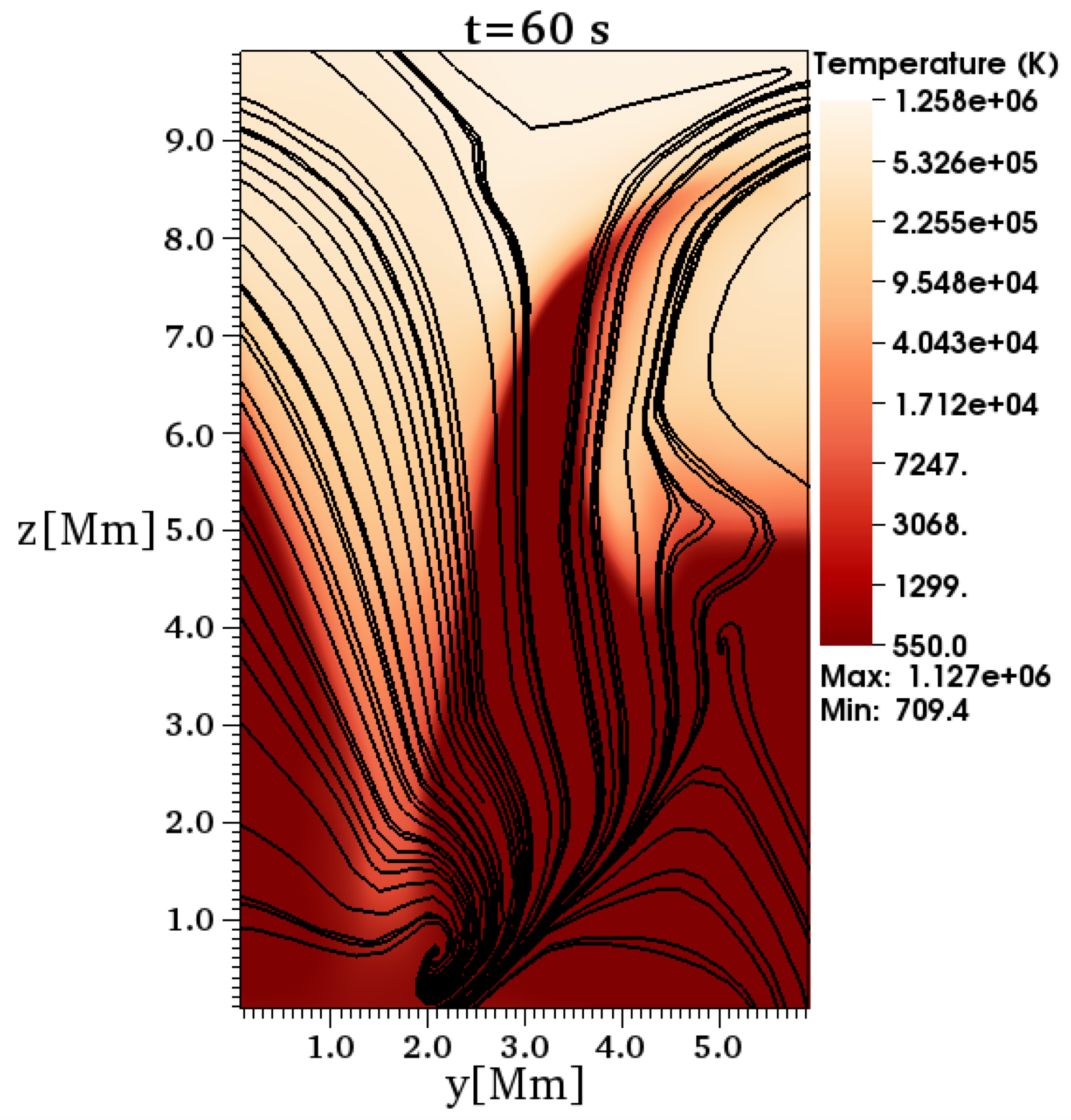}
\includegraphics[scale=0.26]{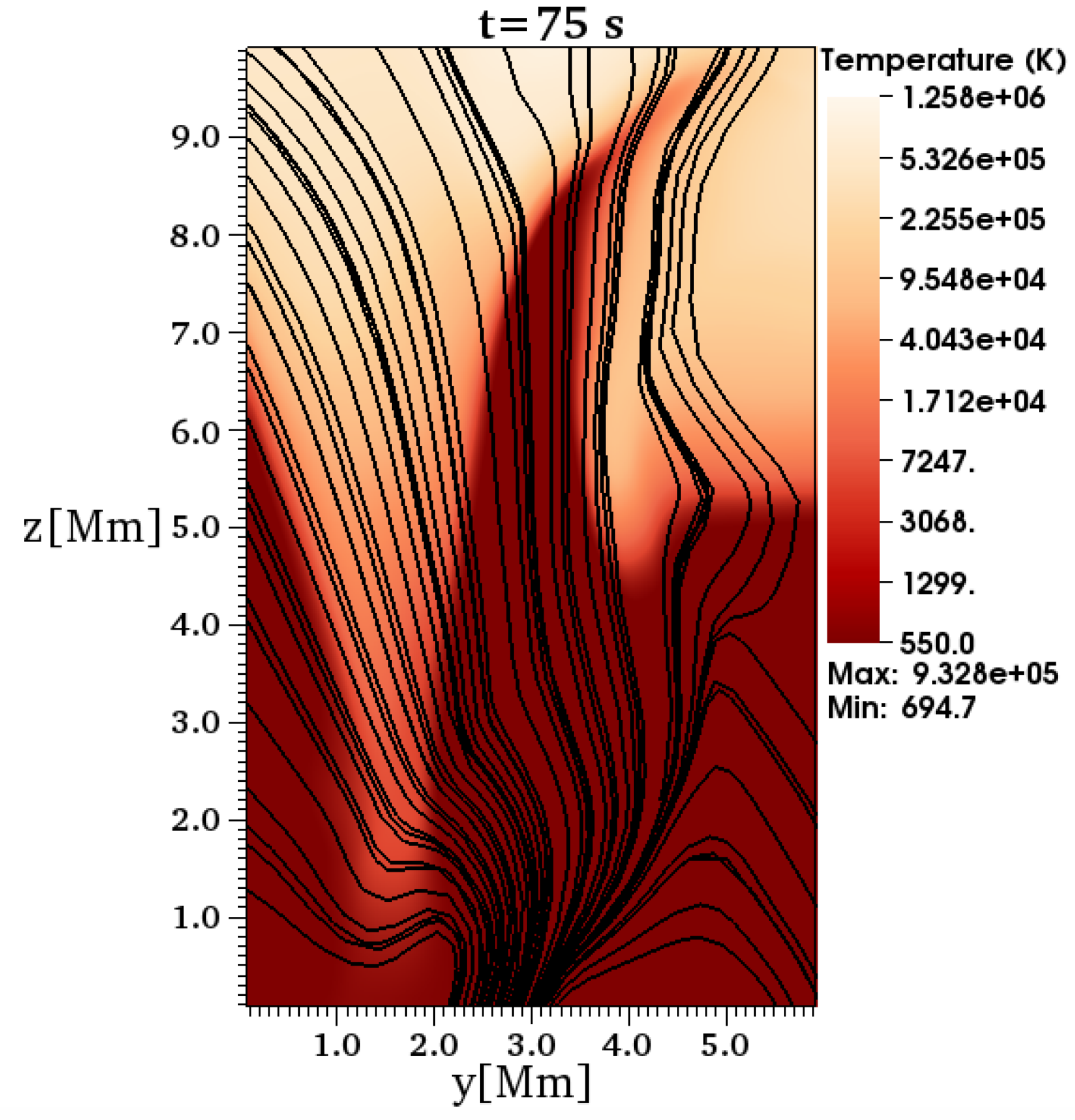}
\includegraphics[scale=0.26]{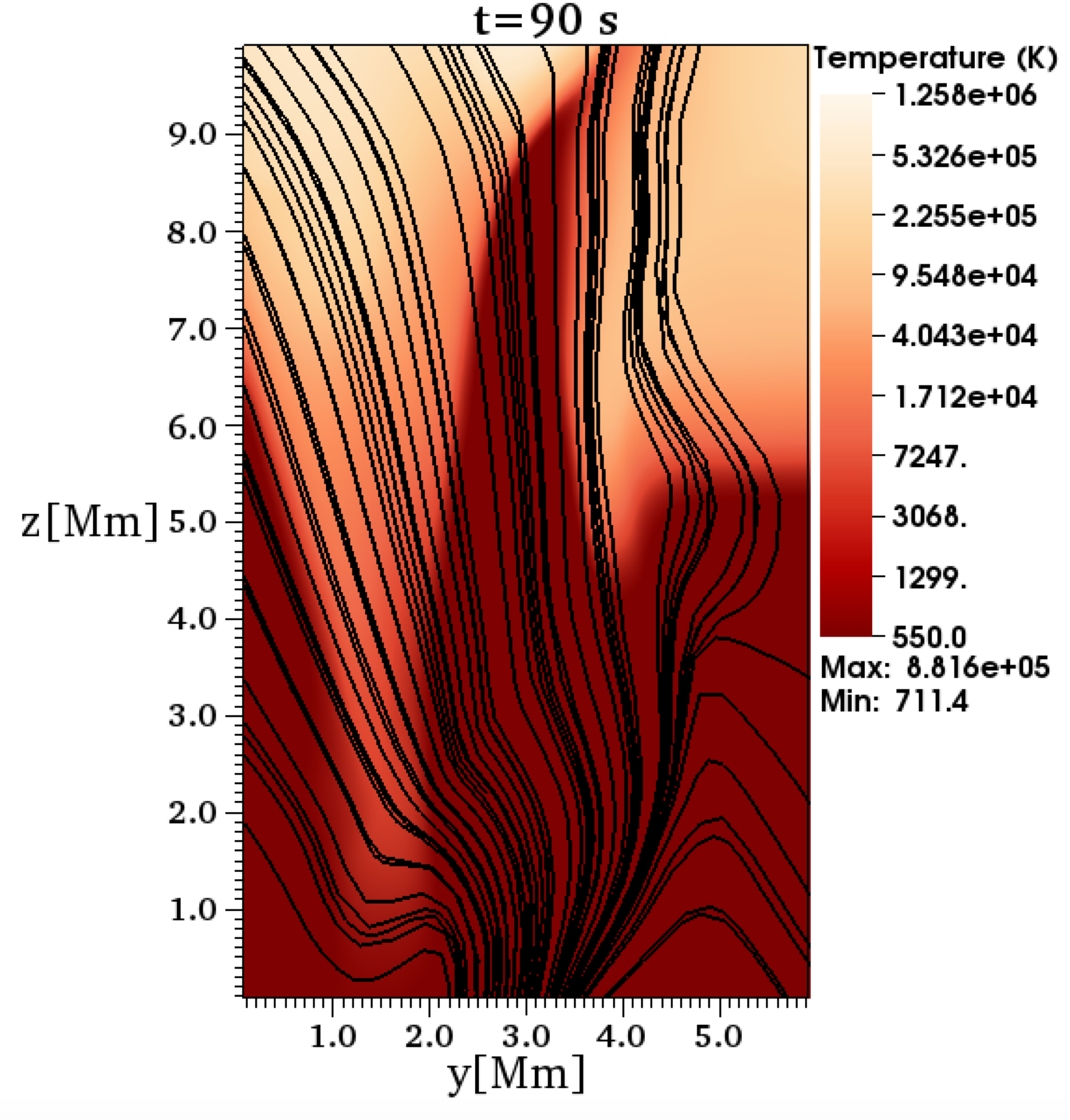}
\caption{\label{log_Temp_yz_plane_times} Snapshots of the logarithm of temperature (K) and magnetic field lines in the cross cut at the plane $x=$0.1 Mm at times 15, 30, 45, 60, 75, and 90 s are shown.}
\end{figure*}

We show a  2D perspective of the process with a cut of the 3D domain at the plane $x=0.1$ Mm in Figure \ref{log_Temp_yz_plane_times}, where various snapshots of the evolution of the temperature (in Kelvin) and the magnetic field lines are shown. For instance, at time $t=15$ s the jet starts to develop at the transition region level $z\approx$ 2.1 Mm where there is a strong current density, which may be an indication of  reconnection happening. The location of the exact reconnection process turns out to be crucially different (see e. g. \citep{Pontin2012}).  Between $t=30$ s and $t=45$ s the jet continues to form. At time $t=60$ s a jet with features of a Type II spicule appears with a basis located at $z\approx$2 Mm and reaches a height of about $z\approx$ 7 Mm measured from transition region (see Figure \ref{log_Temp_yz_plane_times}), which is in agreement with the observed heights of the Type II spicules, between 3-9 Mm \citep{Pereira_et_al_2012,Tavabi_et_al_2015}. The structure of the spicule obtained at time $t=60$ s is similar to the obtained in Figure 5 in \cite{Martinez-Sykora_et_al_2011}. At time $t=90$ s the spicule reaches the top of the domain and the magnetic field lines tend to be uniform.

\begin{figure*}
\centering
\includegraphics[scale=0.26]{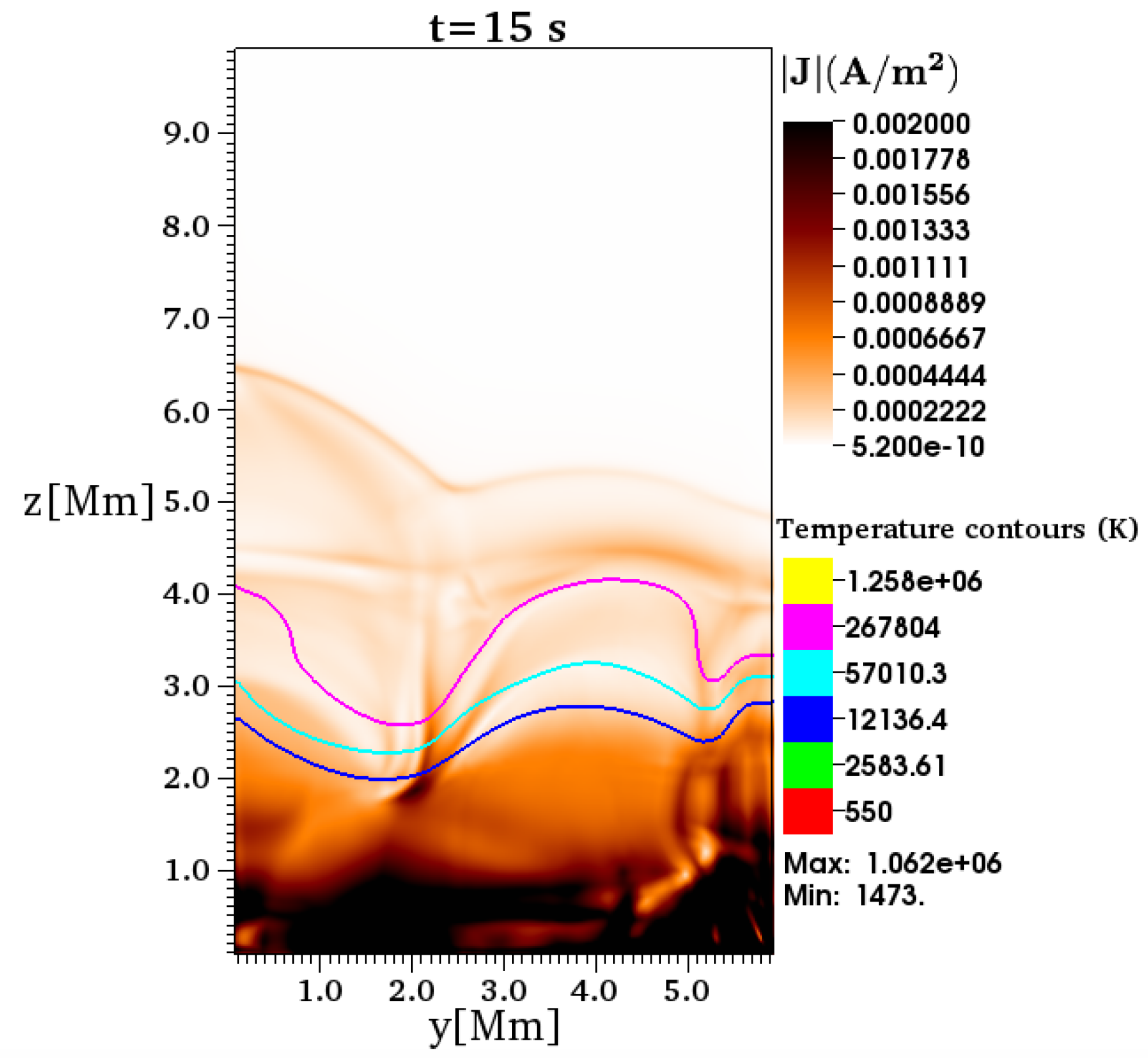}
\includegraphics[scale=0.26]{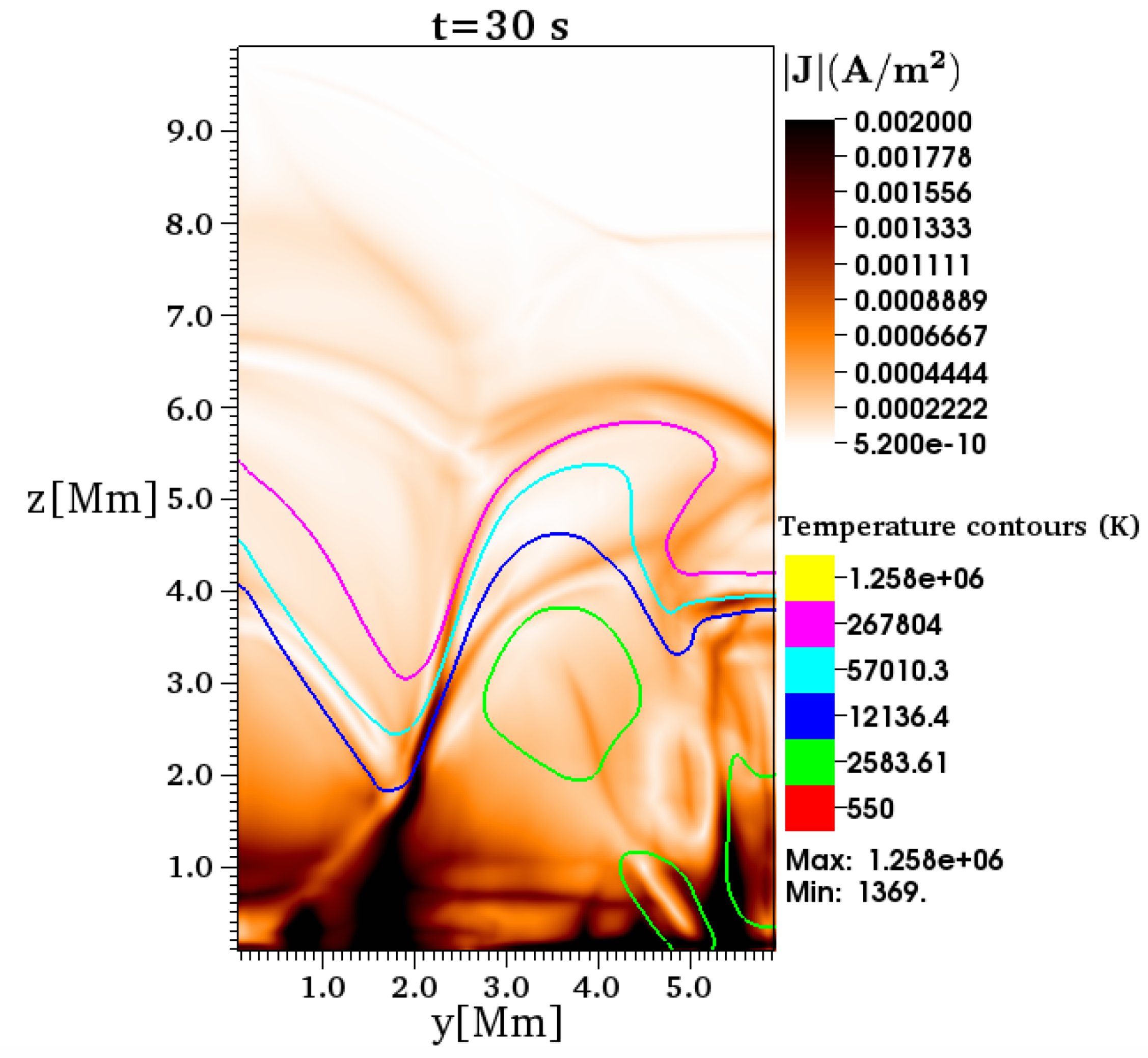}
\includegraphics[scale=0.26]{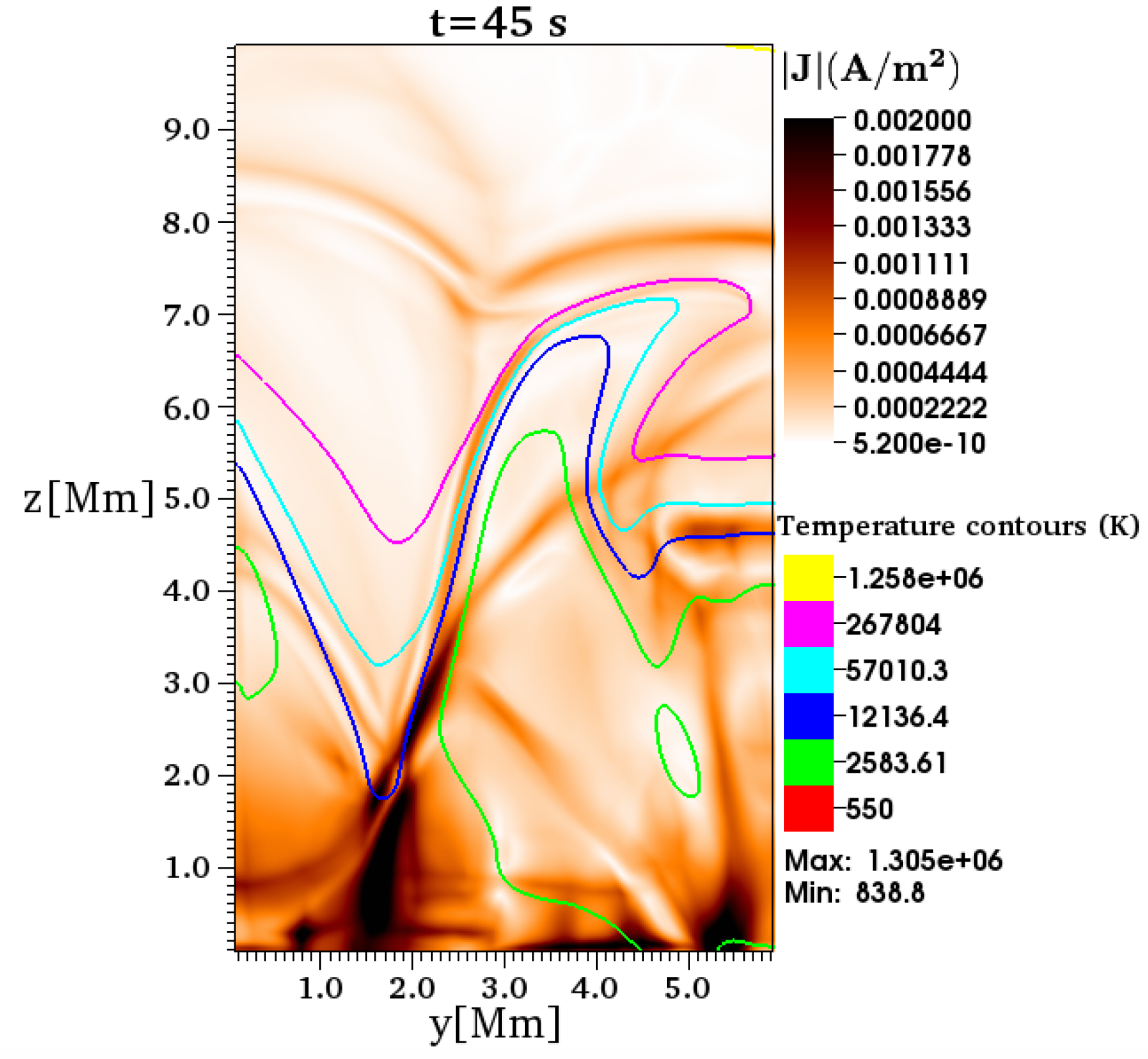}
\includegraphics[scale=0.26]{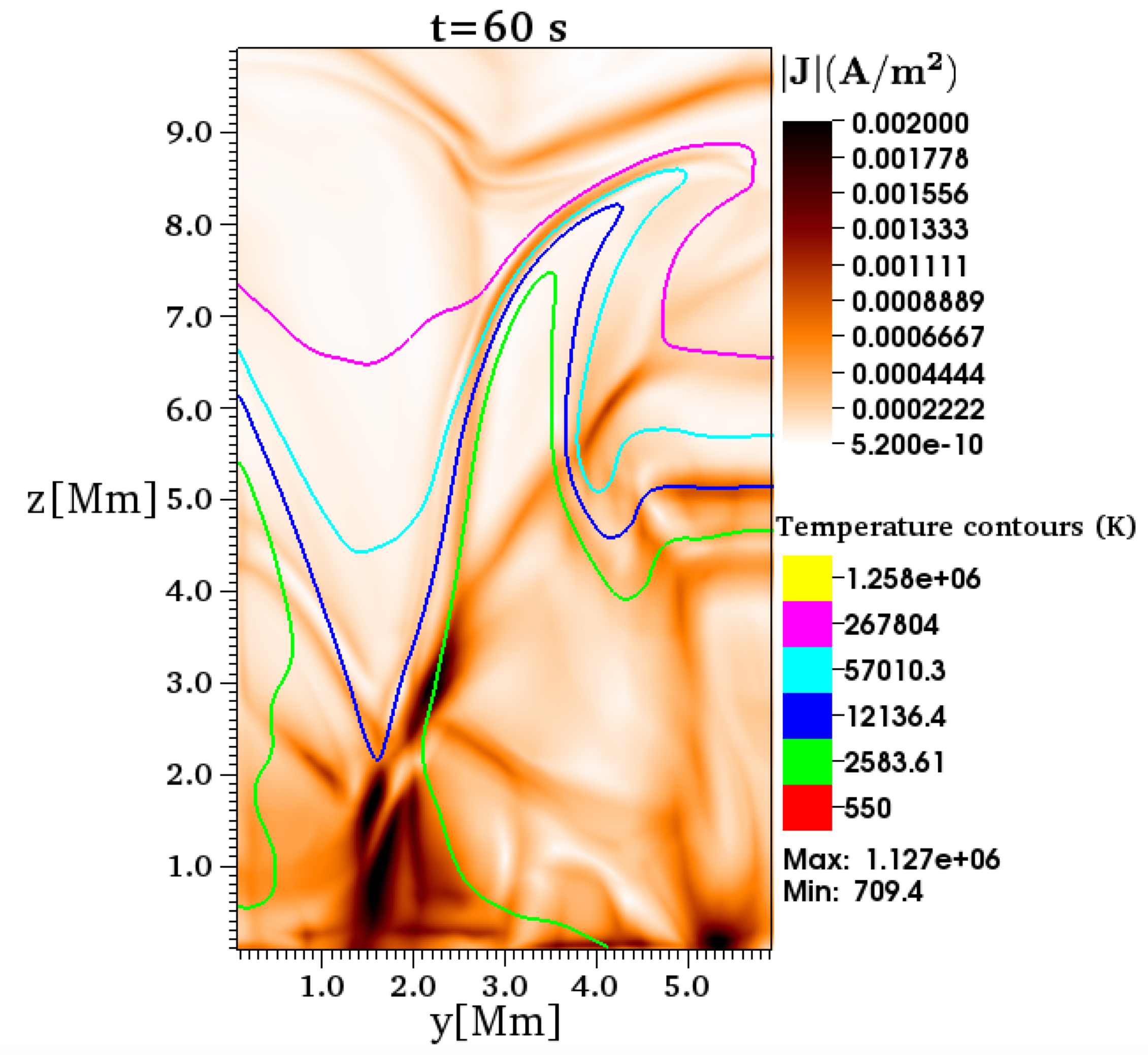}
\includegraphics[scale=0.26]{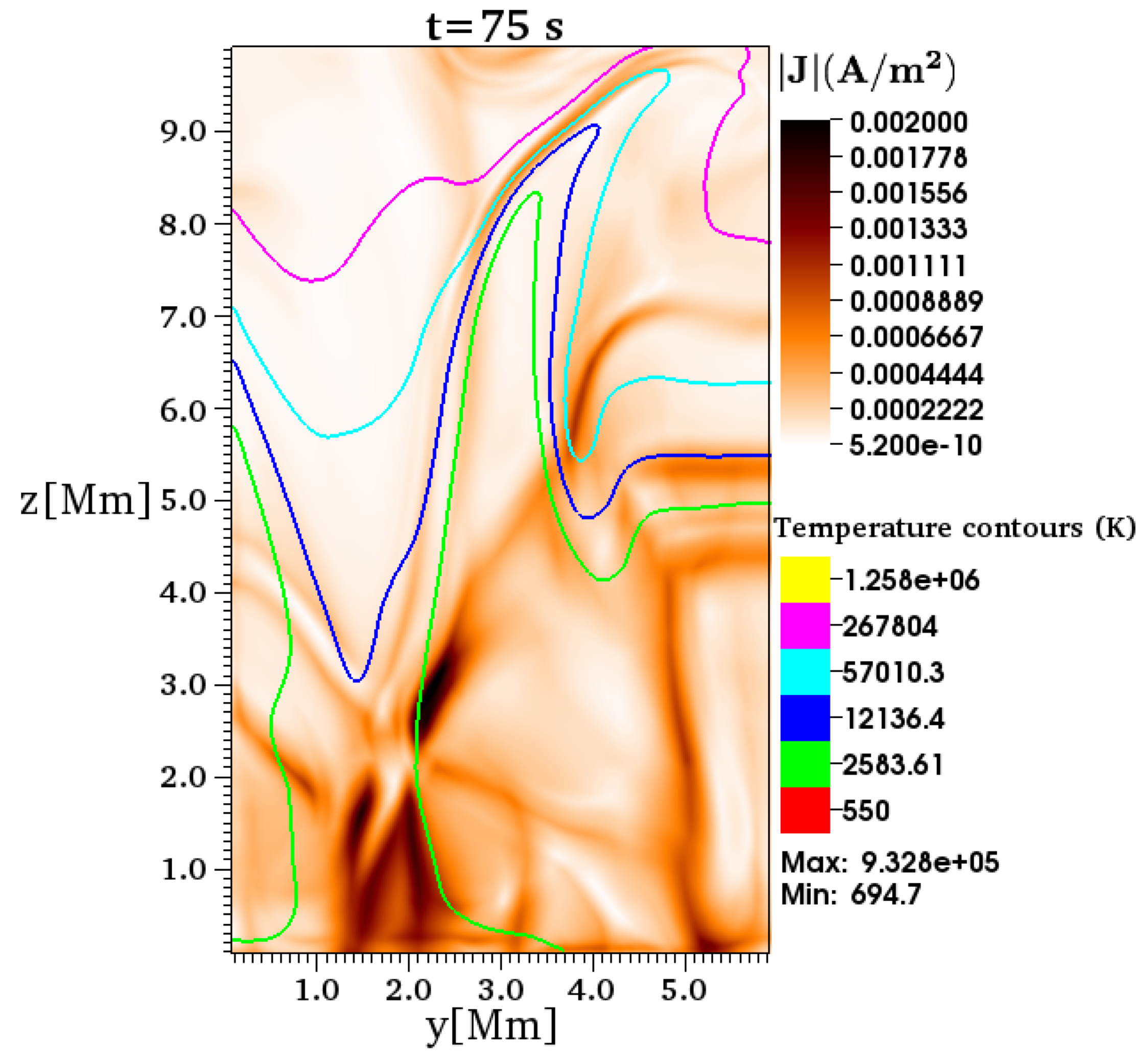}
\includegraphics[scale=0.26]{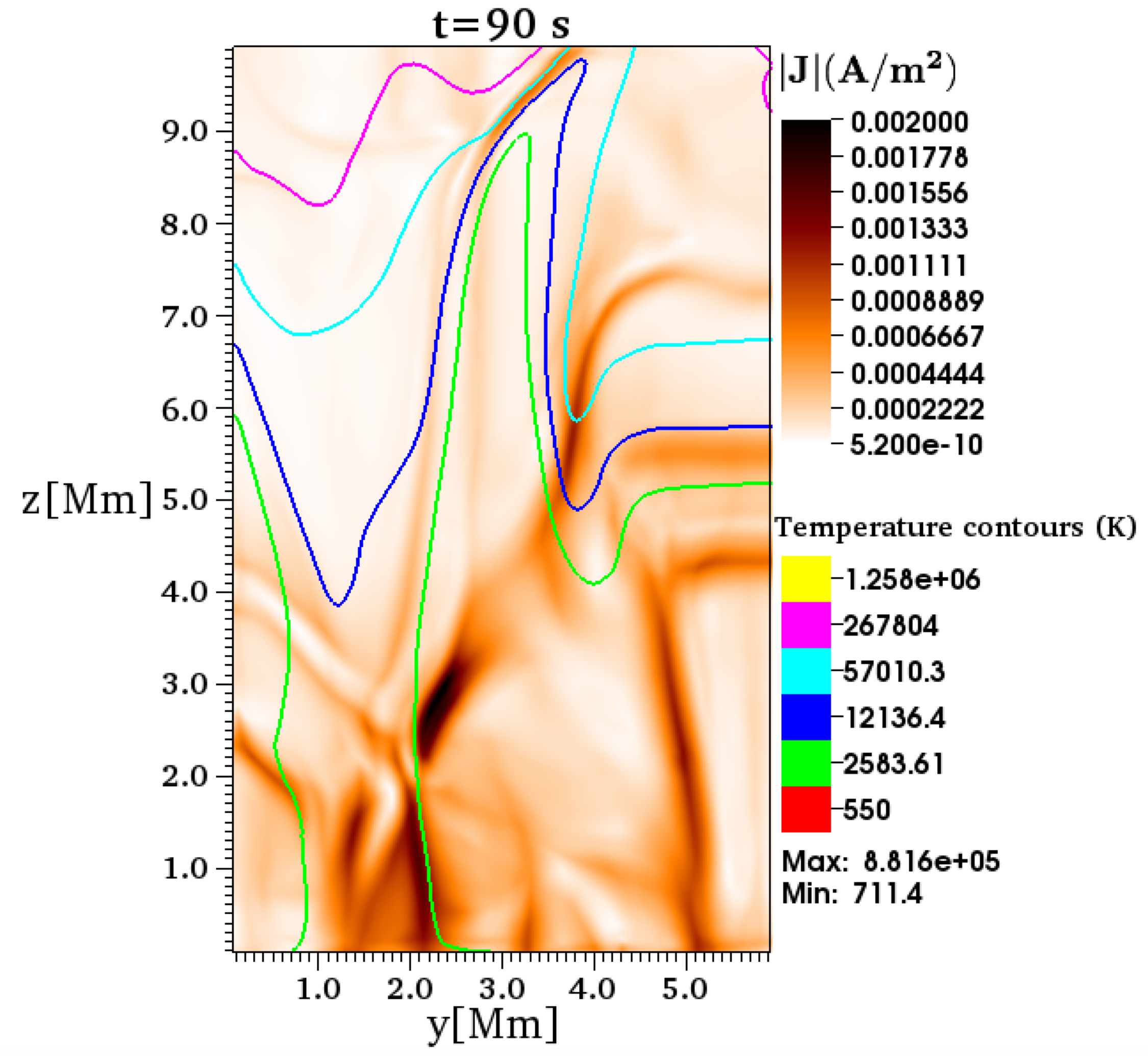}
\caption{\label{Jmag_yz_plane_times} Snapshots of $|{\bf J}|$ (A m$^{-2}$) and temperature contours (K) in the cross cut at the plane $x=0.1$Mm at various times.}
\end{figure*}

\begin{figure*}
\centering
\includegraphics[scale=0.26]{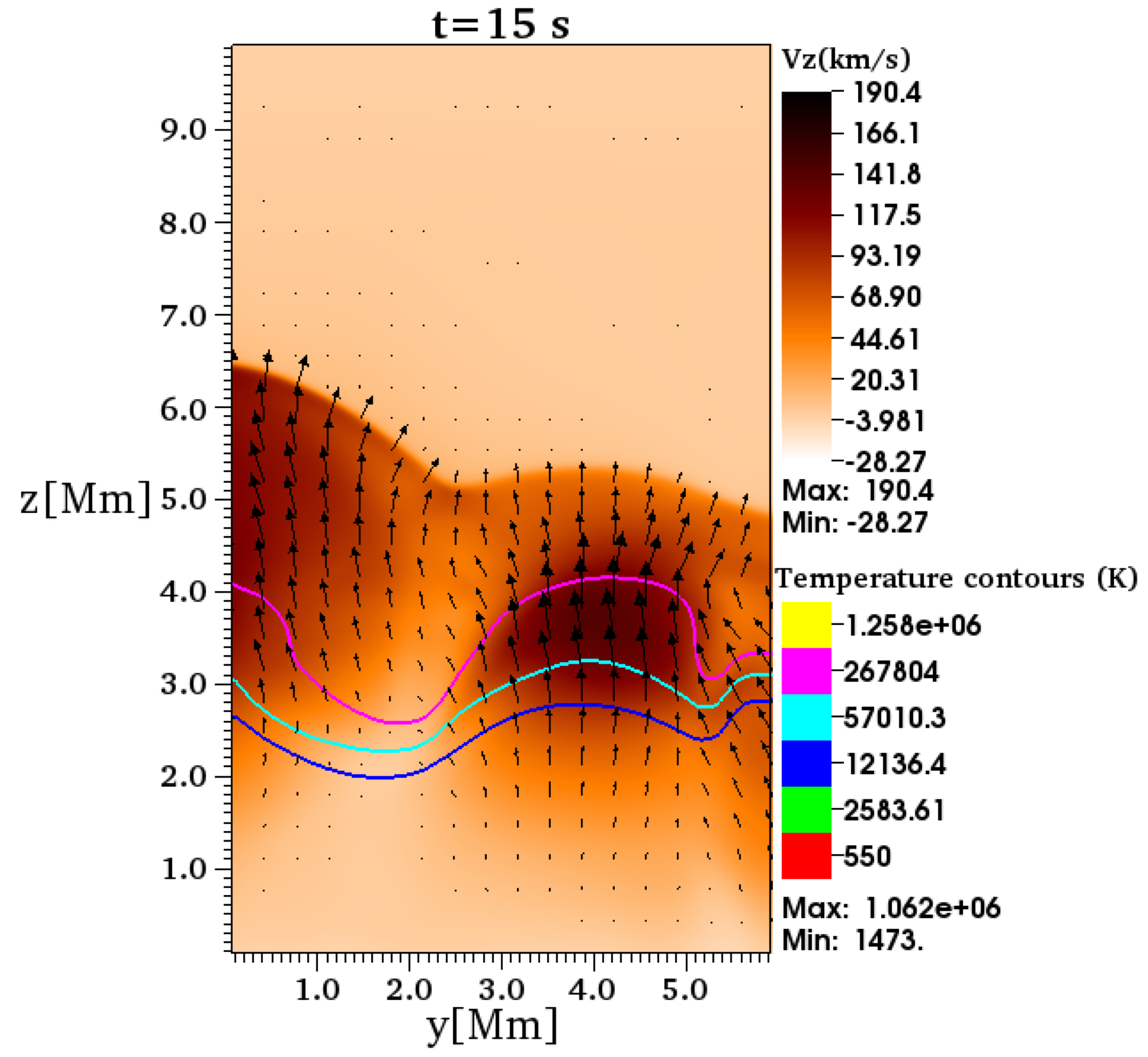}
\includegraphics[scale=0.26]{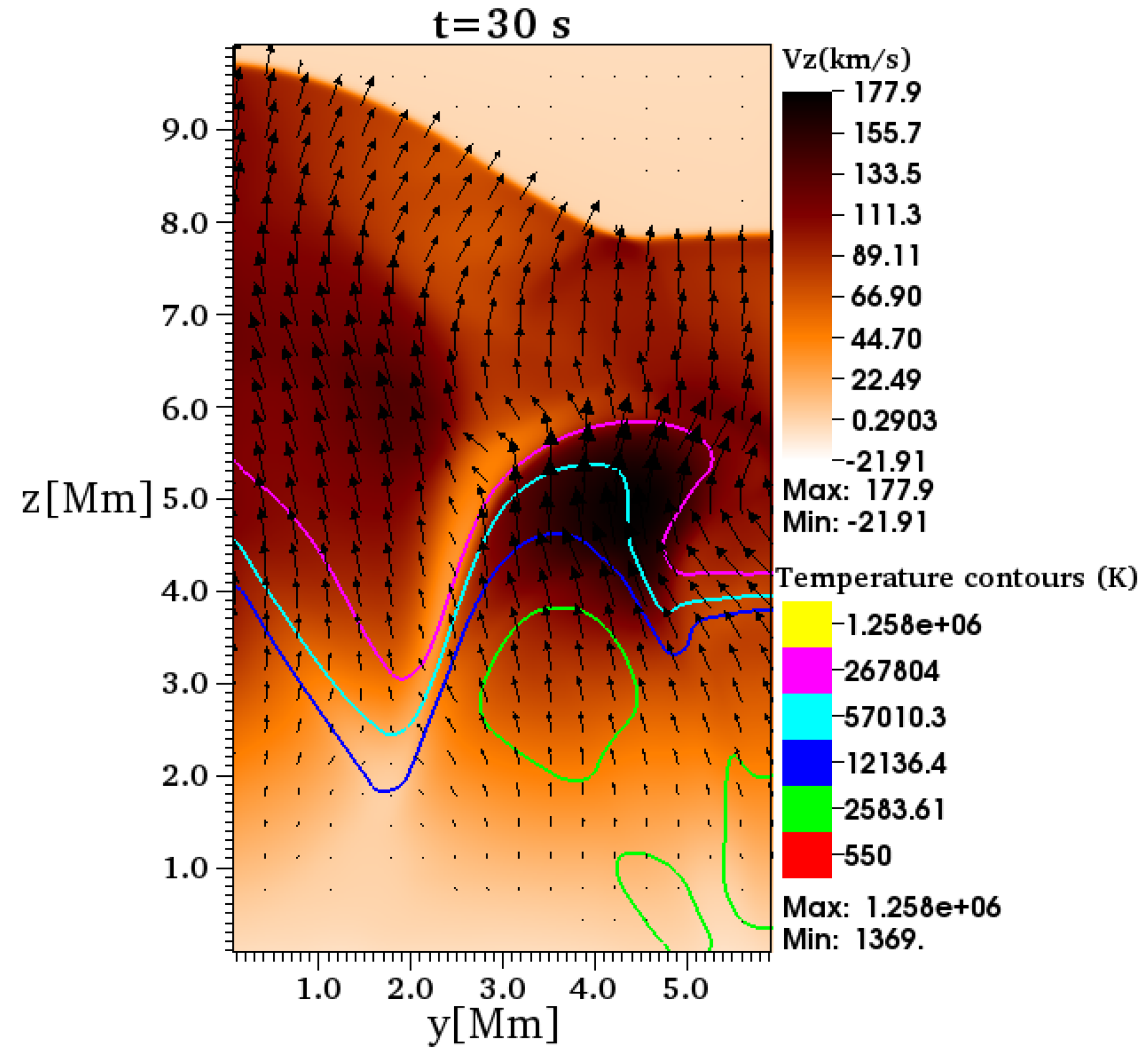}
\includegraphics[scale=0.26]{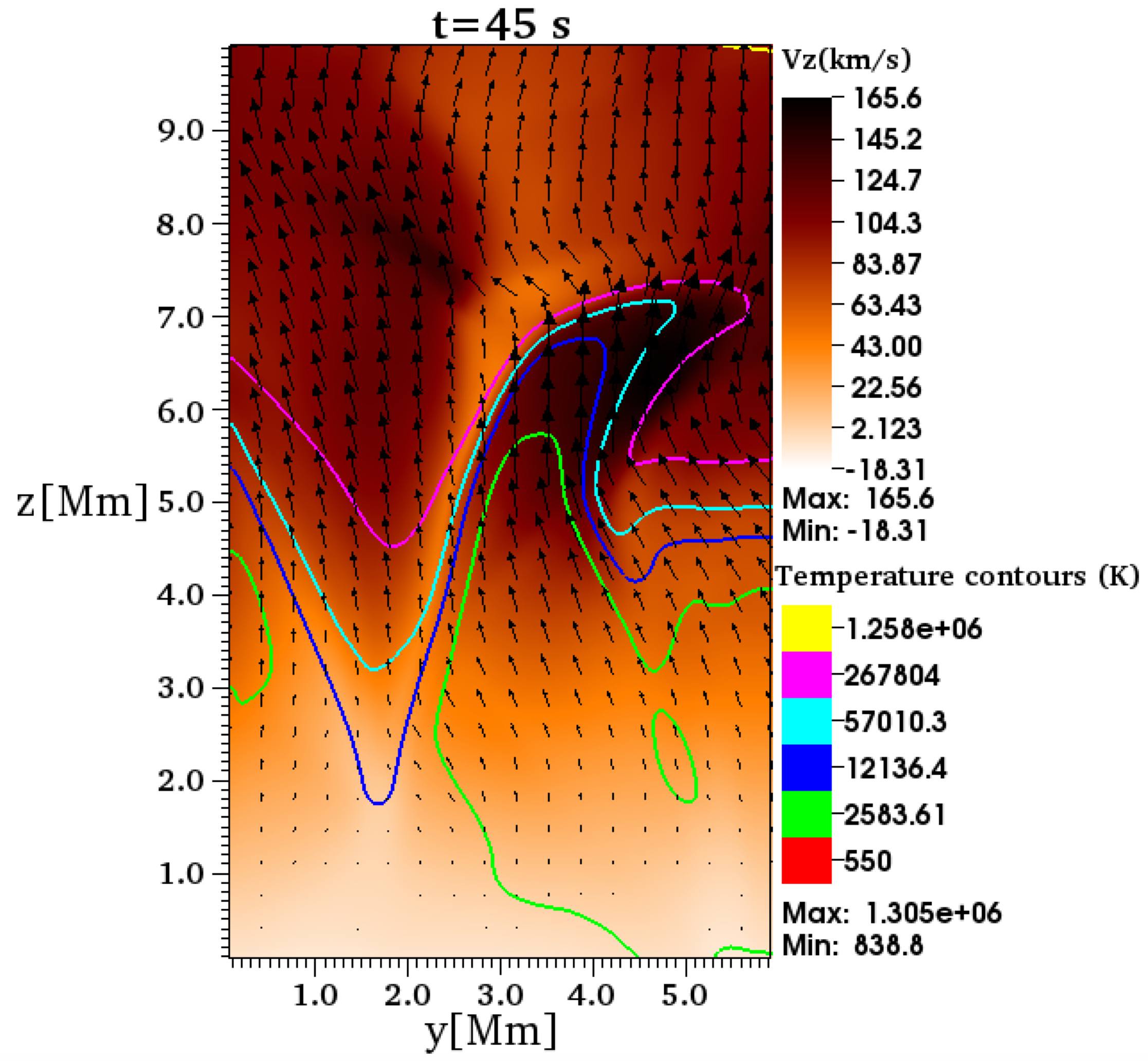}
\includegraphics[scale=0.26]{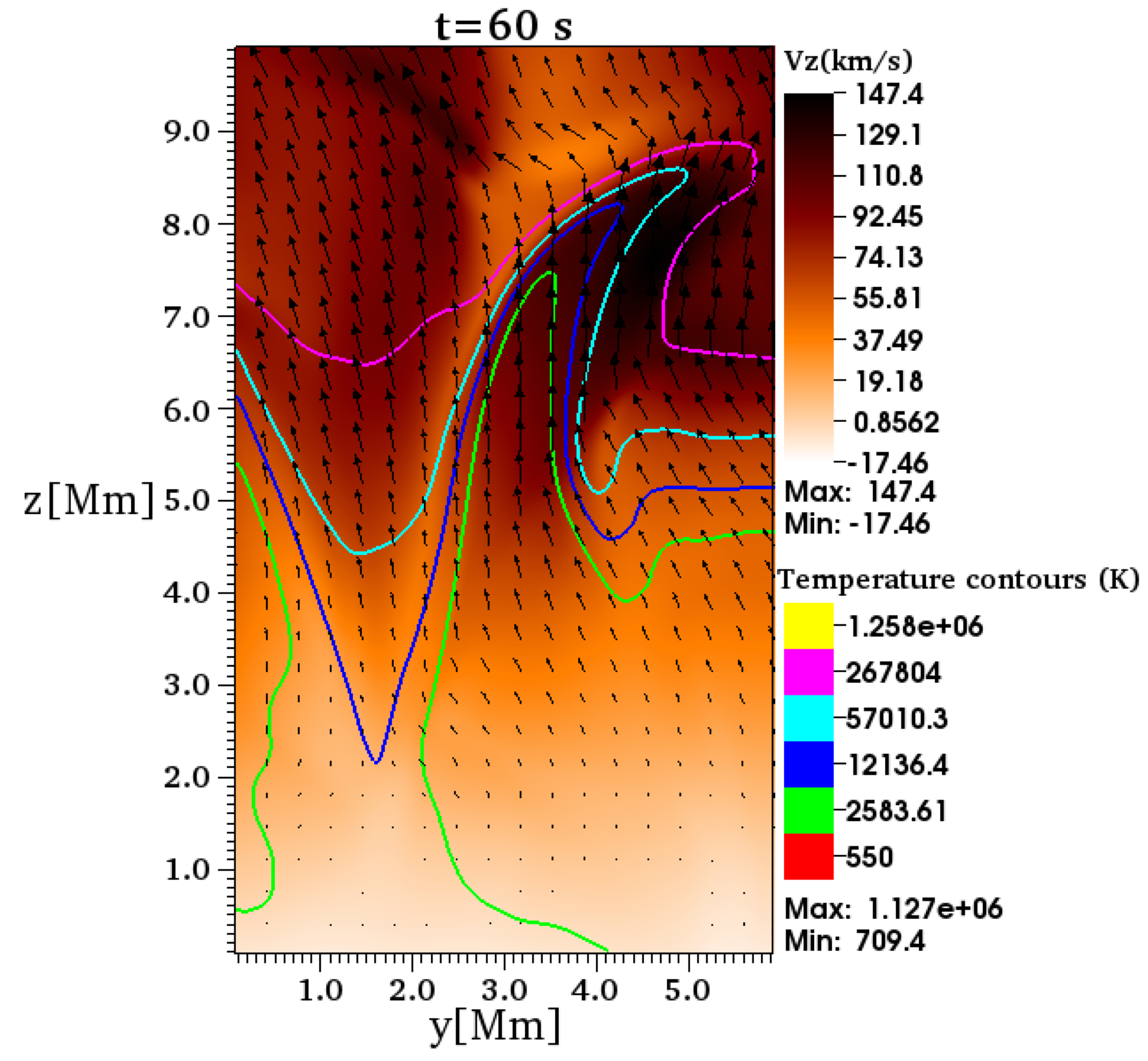}
\includegraphics[scale=0.26]{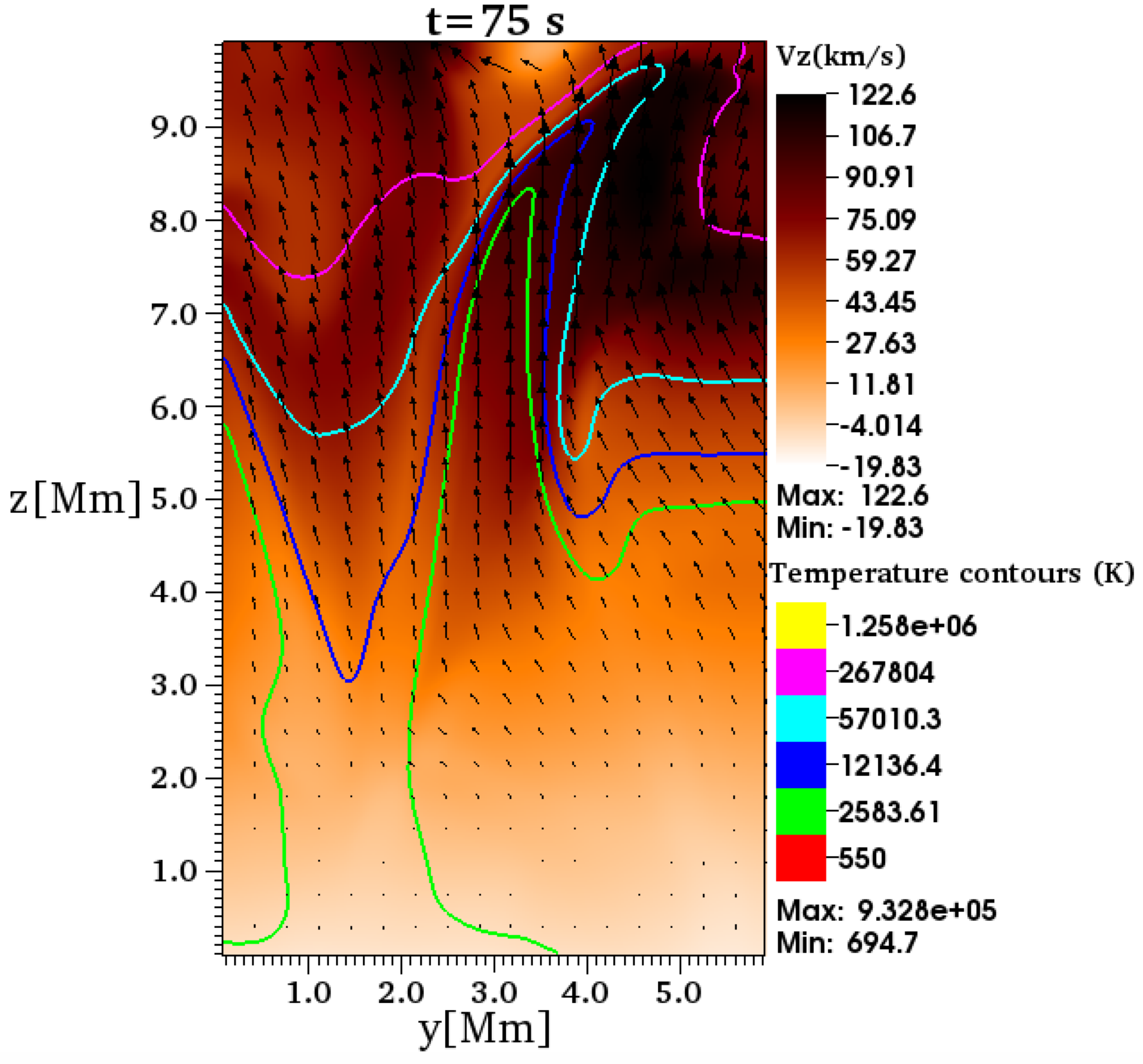}
\includegraphics[scale=0.26]{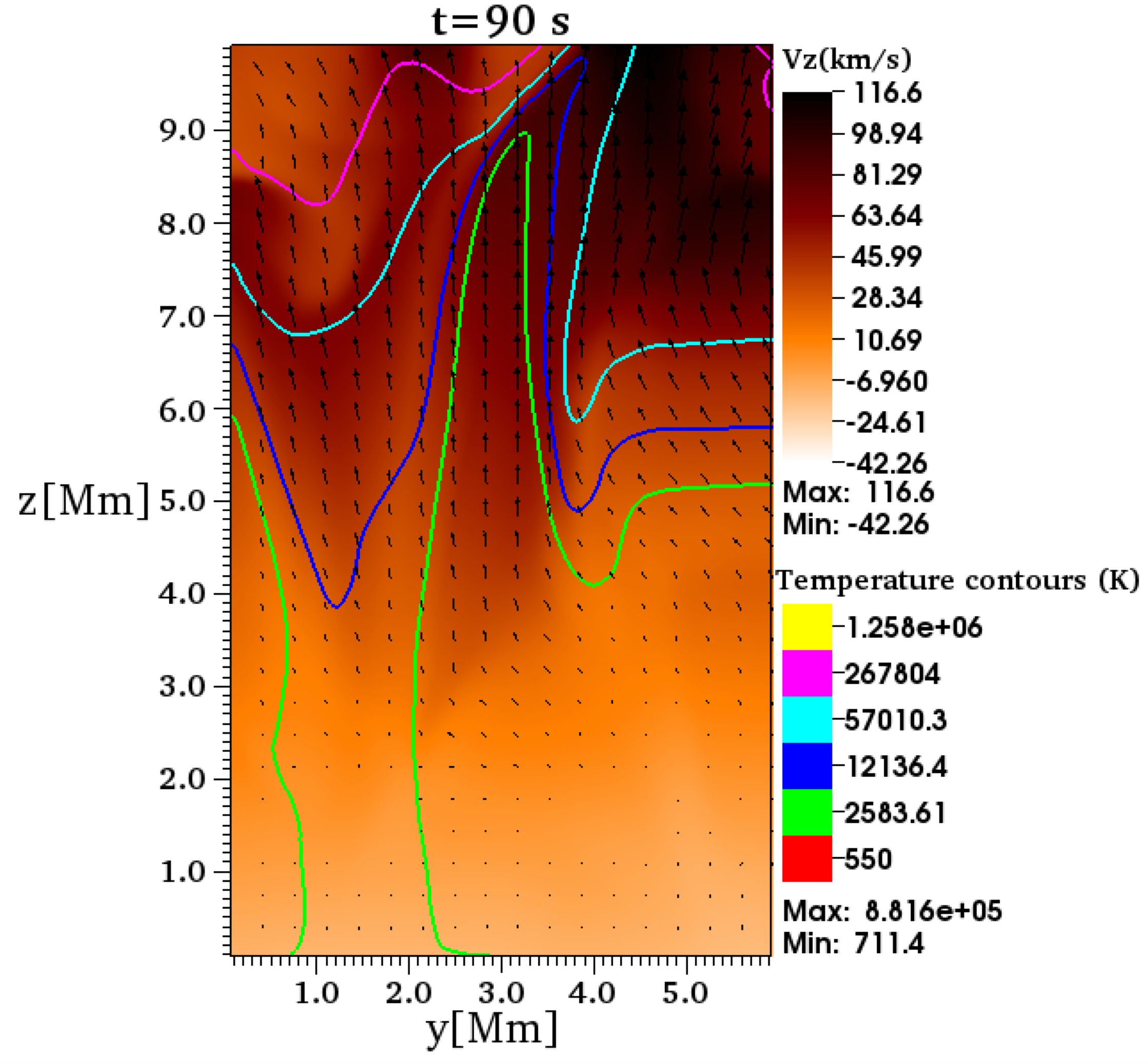}
\caption{\label{vz_temp_contours_vfield_yz_plane_times} Snapshots of the vertical component of velocity $v_z$ (km s$^{-1}$), temperature contours (K) and vector velocity field (black arrows) in the cross cut at the plane $x=0.1$Mm at various times.}
\end{figure*}

In order to locate regions where magnetic reconnection can take place, we show 2D perspectives of the evolution of $|\bf J|$ (A m$^{-2}$) and temperature contours (K) in Figure \ref{Jmag_yz_plane_times}. For instance, at time $t=15$ s, which is the time when the spicule starts to develop, we can see regions of strong current density located the transition region and chromosphere. Between $t=30$ s and $t=45$ s the stronger current density regions are located at the basis of the spicule, which can accelerate the plasma upwards. At time $t=60$ s, when the spicule is well formed, the stronger current locates around $(y,z)\sim$(2,2) Mm, at the basis of the spicule, which is consistent with the results shown in Figure \ref{log_Temp_yz_plane_times} at $t=60$ s. At the next two snapshots $t=75$ s and $t=90$ s, the regions of stronger current are still located at the bottom of the spicule. This analysis shows that magnetic reconnection mainly happens at the chromosphere and transition region. 

As it has been reported in a number of observational papers \citep{De_Pontieu_et_al_2007a, Anan_et_al_2010,Pereira_et_al_2012,Zhang_et_al_2012}, the upward velocity of spicules is important, therefore we monitor this quantity in our simulation. For the analysis, we show 2D maps of the vertical velocity $v_z$ (km s$^{-1}$), the vector velocity field and temperature contours (K) in Figure \ref{vz_temp_contours_vfield_yz_plane_times}. At time $t=15$ s, the spicule is moving upwards with a maximum vertical velocity $v_z\sim 190$ km s$^{-1}$. At time $t=30$ s, the spicule continues to move upwards with a velocity of the order $v_z\sim 178$ km s$^{-1}$. At time $t=60$ s, the maximum vertical velocity of the spicule is of the order $v_z\sim 148$ km s$^{-1}$, which is slightly above the range of observed upward velocities of a Type II spicule. At times $t=75$ s and $t=90$ s, the velocity reaches a value of $v_z\sim 116$ km s$^{-1}$ at the top of the domain.

\begin{figure*}
\centering
\includegraphics[scale=0.25]{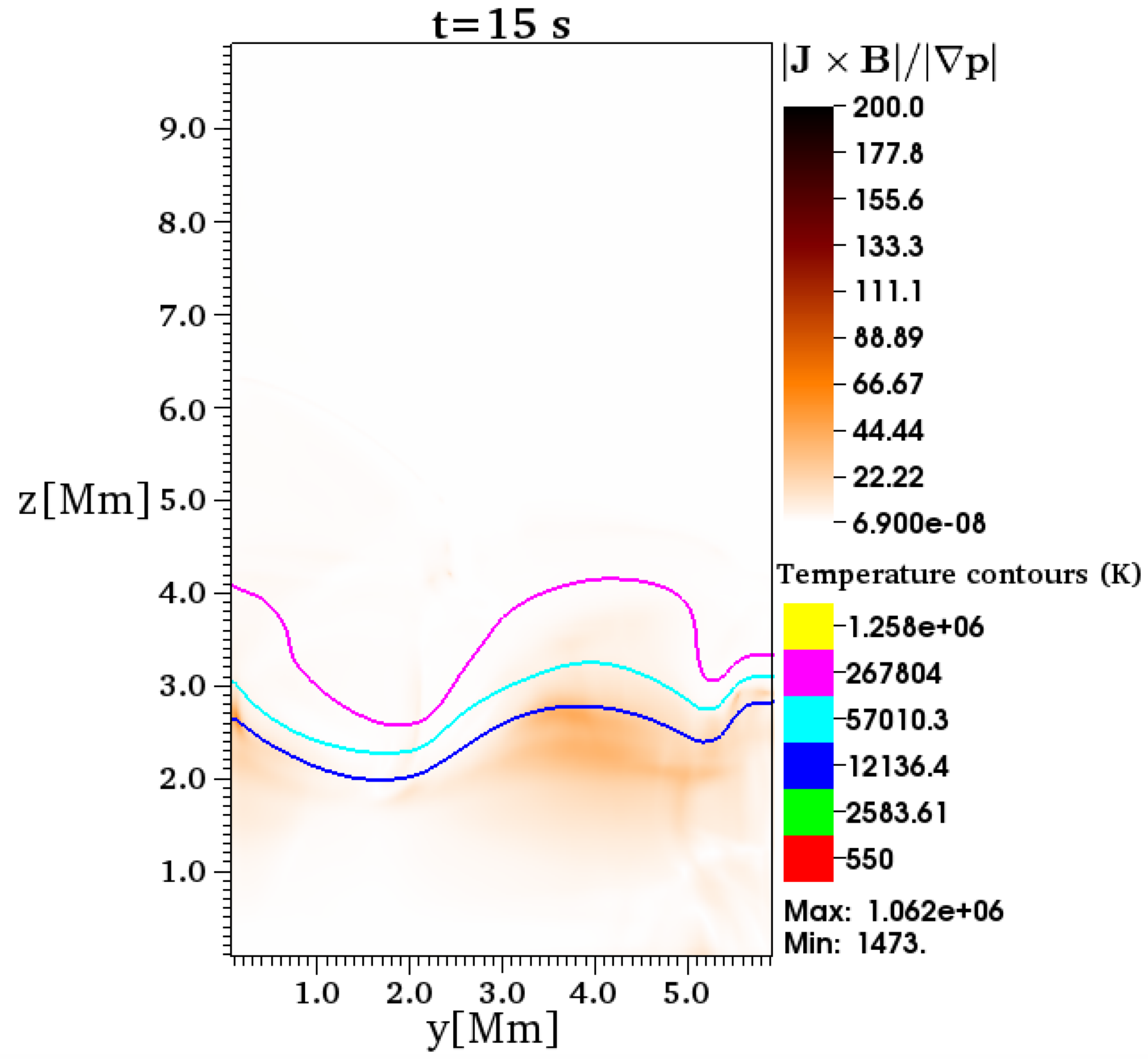}
\includegraphics[scale=0.25]{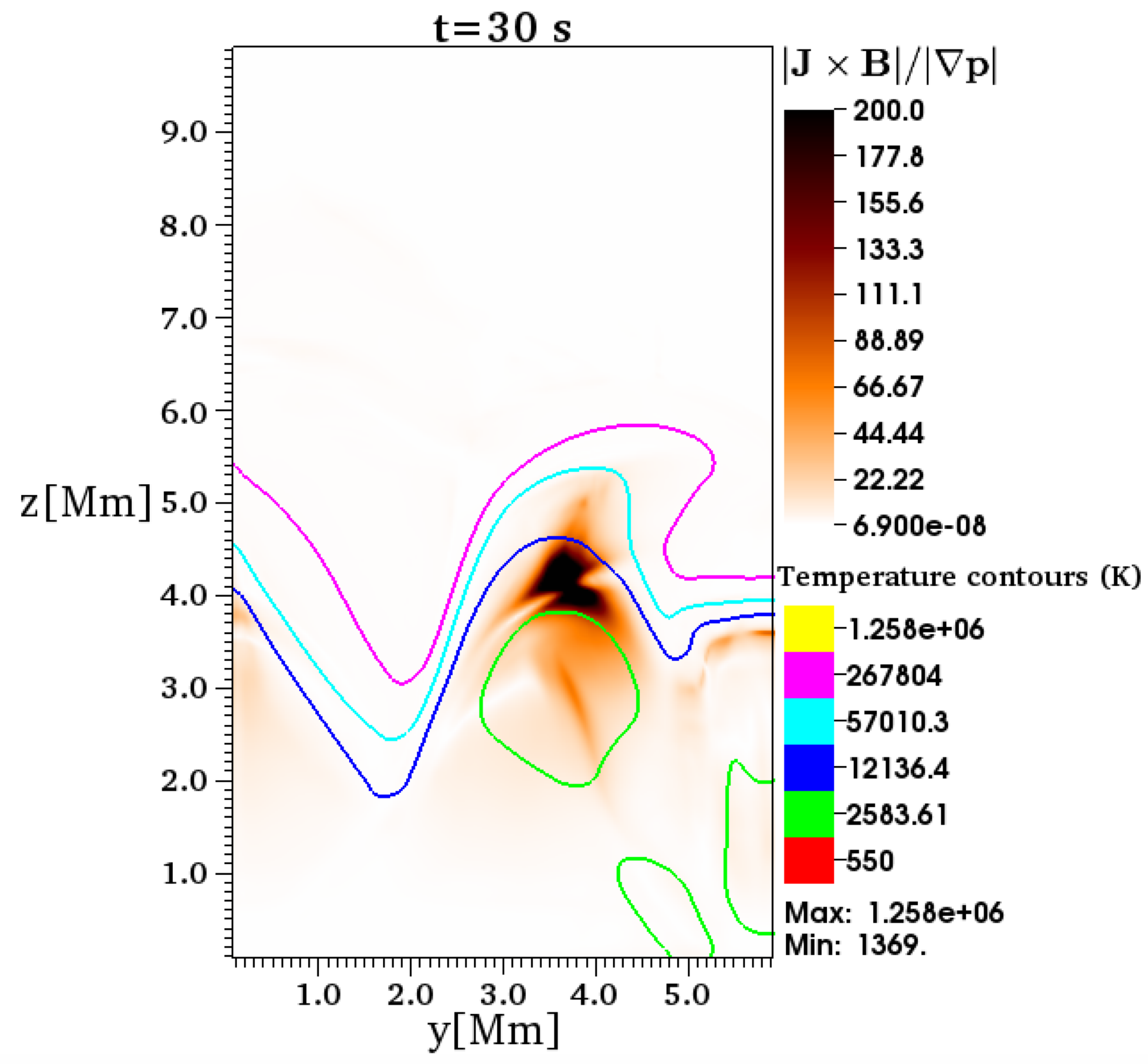}
\includegraphics[scale=0.25]{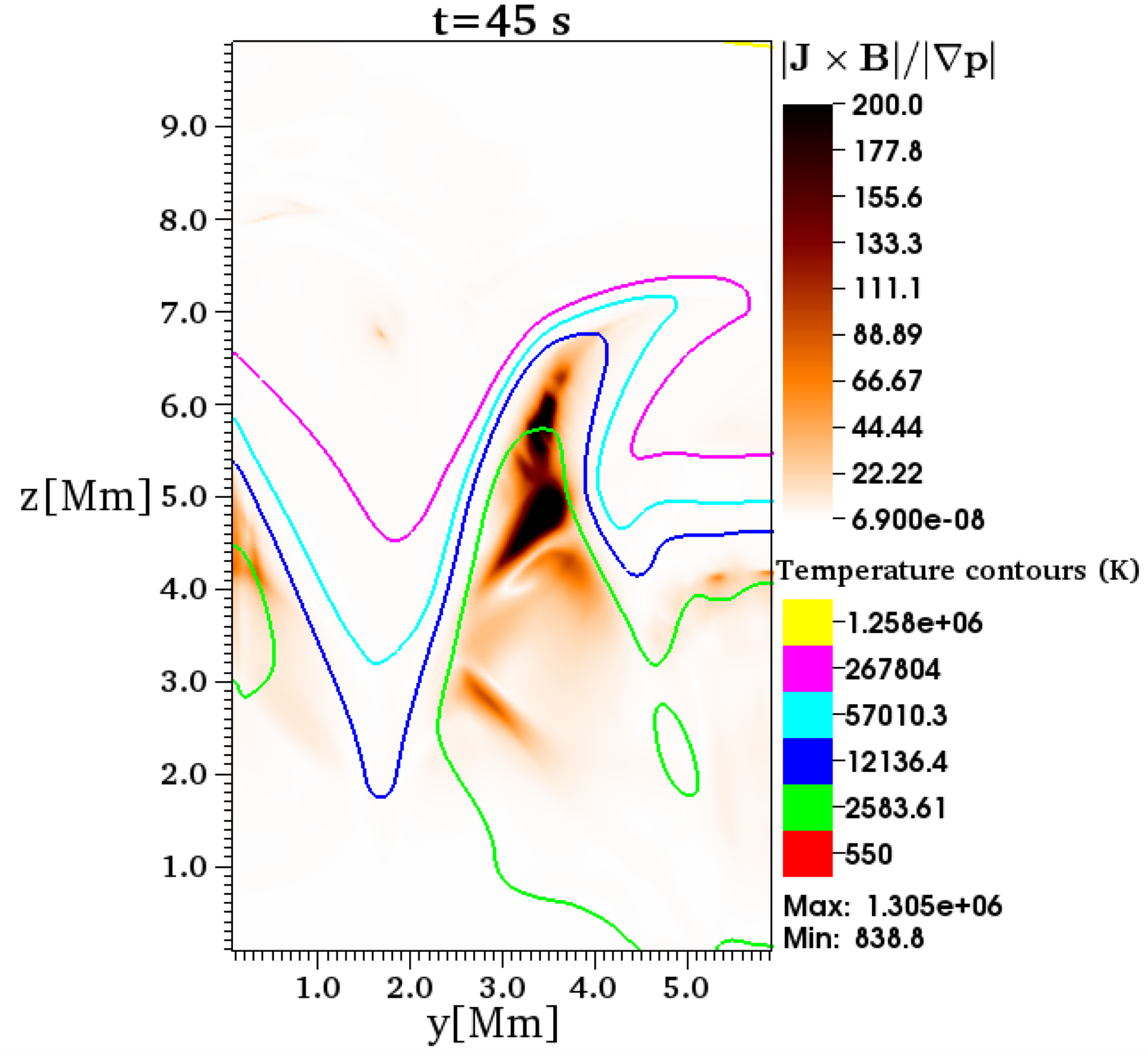}
\includegraphics[scale=0.25]{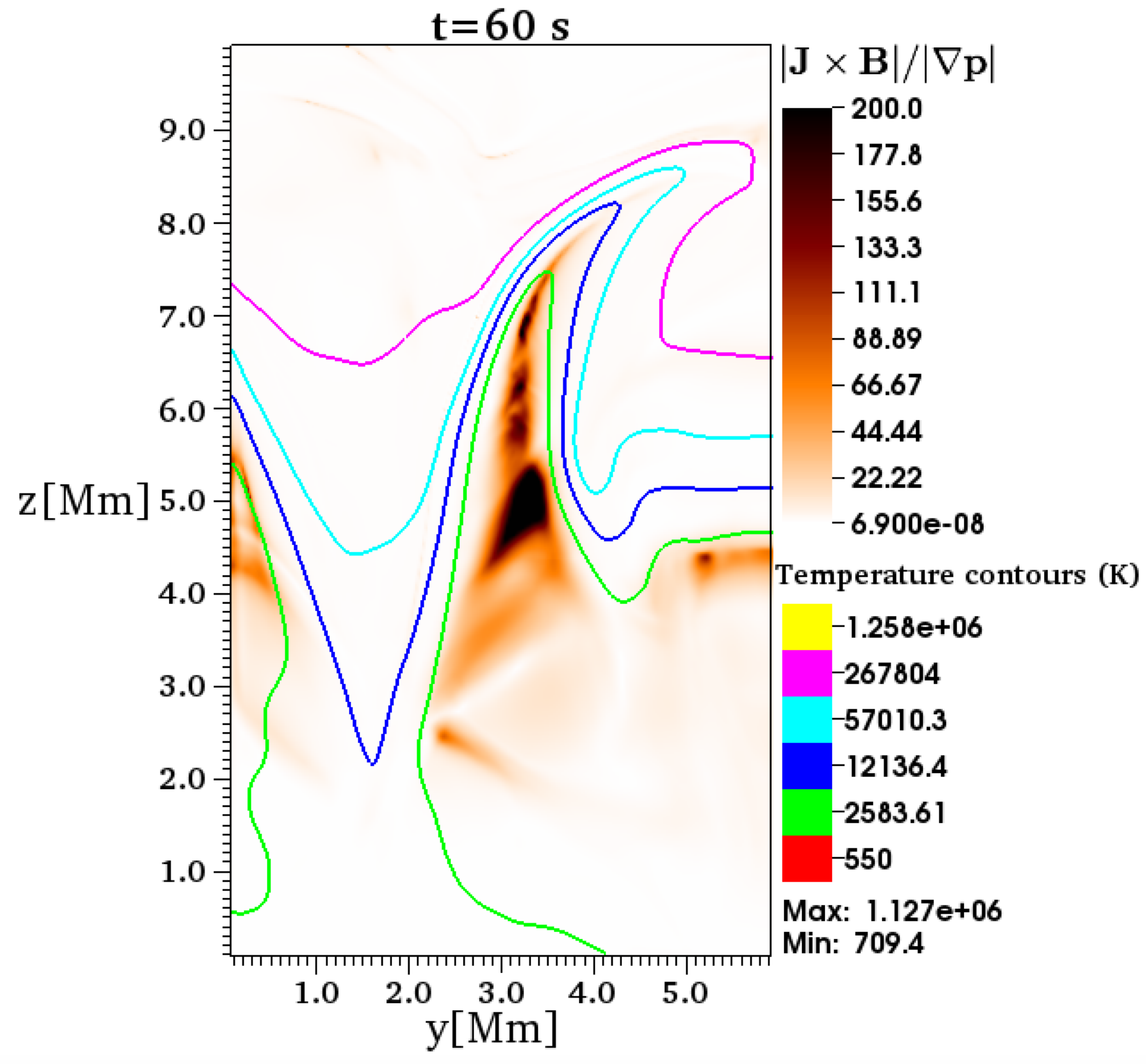}
\includegraphics[scale=0.25]{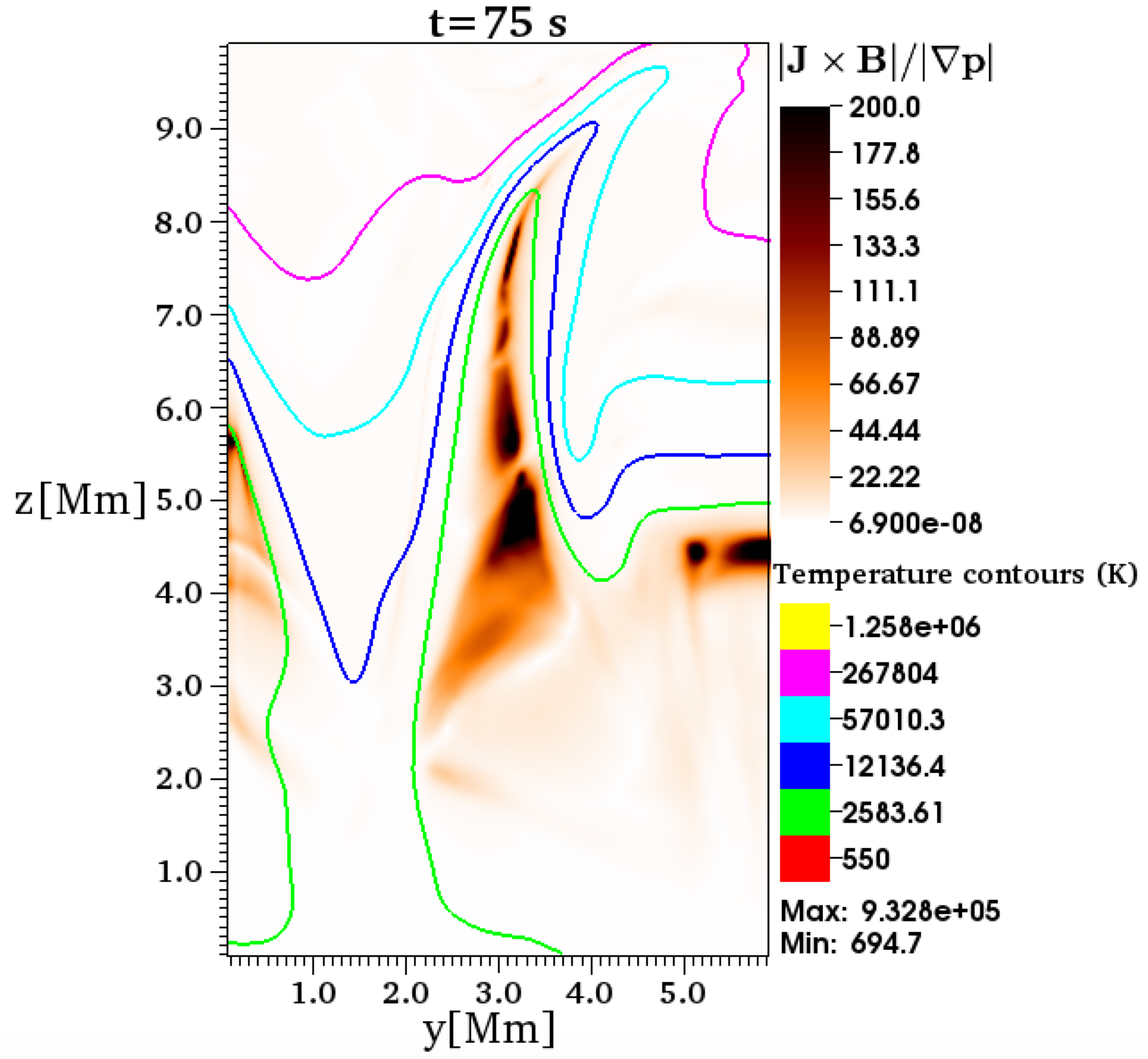}
\includegraphics[scale=0.175]{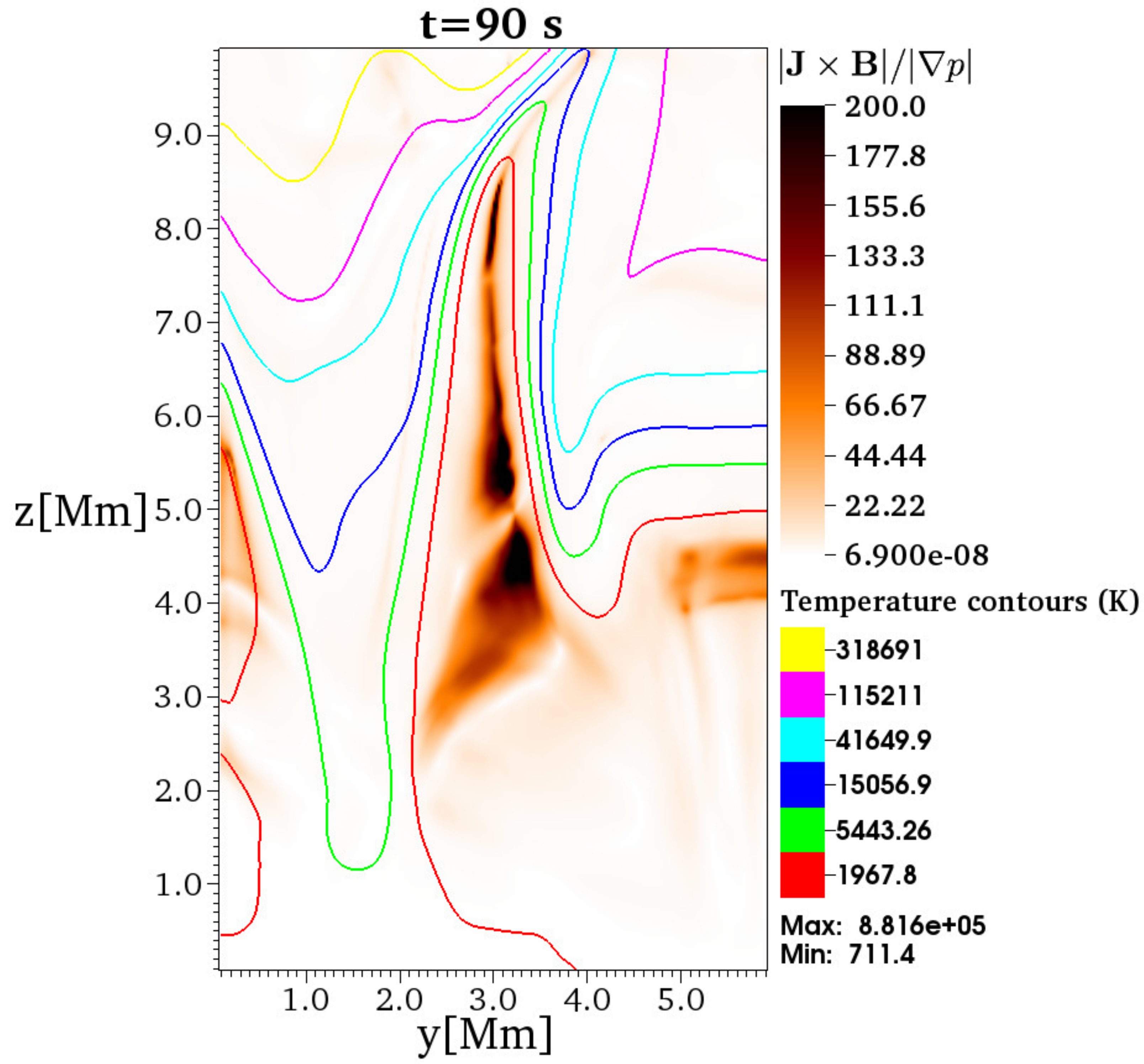}
\caption{\label{ratio_Lorentz_Force_grad_press_yz_plane} Snapshots of the ratio $|{\bf J}\times{\bf B}|/|\nabla p|$ and temperature contours (K) in the cross cut at the plane $x=0.1$Mm at various times. A comparison with  \ref{logTemp_3D_fieldlines} and \ref{log_Temp_yz_plane_times} indicates that the Lorentz force dominates in the region where the spicule is formed.}
\end{figure*}

In order to understand the physics behind the modeled spicule formation, it is important to identify the dominant force(s) acting during the formation and development of the spicule. For this we compare the forces due to the magnetic field and hydrodynamics, thus we calculate the ratio between the magnitude of the Lorentz force and the magnitude of pressure gradient $|{\bf J}\times{\bf B}|/|\nabla p|$. The results of the evolution of the ratio $|{\bf J}\times{\bf B}|/|\nabla p|$ and temperature contours (K) are shown in Figure \ref{ratio_Lorentz_Force_grad_press_yz_plane} in the cross cut at the plane $x=$0.1 Mm of the 3D domain. Notice that at time $t=15$ s, which is the time when the spicule starts to form, that Lorentz force dominates. At times $t=30$ s and $t=45$ s the dominance of the Lorentz force helps to the spicule moving upwards. At time $t=60$ s, the Lorentz force is stronger exactly where the spicule forms. This dominance is still clear at times $t=75$ s and $t=90$ s. This analysis shows that Lorentz force is an important ingredient of the jet formation.

In order to identify more clearly the behavior of the magnetic reconnection process, we calculate the velocity and magnetic field components, the gas pressure and mass density as functions of  $y$ along the constant line $z=2.1$ Mm, which is the location of the base of the jet (see Figure \ref{log_Temp_yz_plane_times}) and where the current density is strong. For instance in Figure \ref{reconnection_plots} we show the velocity and magnetic field components along this line at time $t=45$ s, we can see that $v_x$ and $v_y$ change sign indicating a bidirectional flow, which is characteristic of a current sheet region. We can see at the bottom of Figure \ref{reconnection_plots} that magnetic field components $B_x$, $B_y$ and $B_z$ also change sign, in particular the vertical magnetic field component $B_z$ indicates a current sheet region. We also analyze the behavior of the gas pressure $p$ and mass density $\rho$ at time $45$ s in Figure \ref{pressure_mass_rey_reconnection}, measured along the  line $z=$2.1 Mm. These plots show an increase in density and pressure near the reconnection region.

\begin{figure*}
\centering
\includegraphics[scale=0.2]{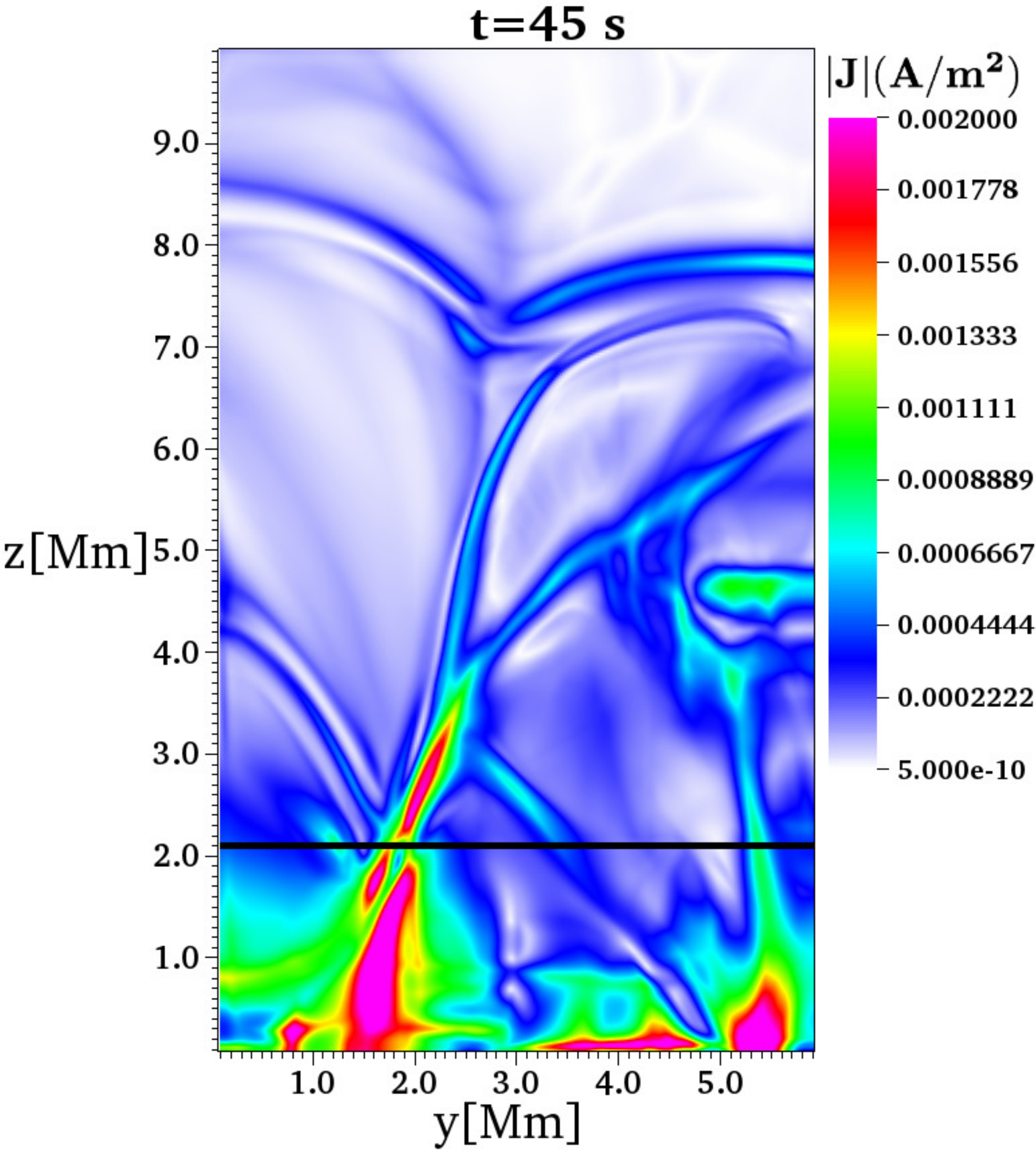}
\includegraphics[scale=0.2]{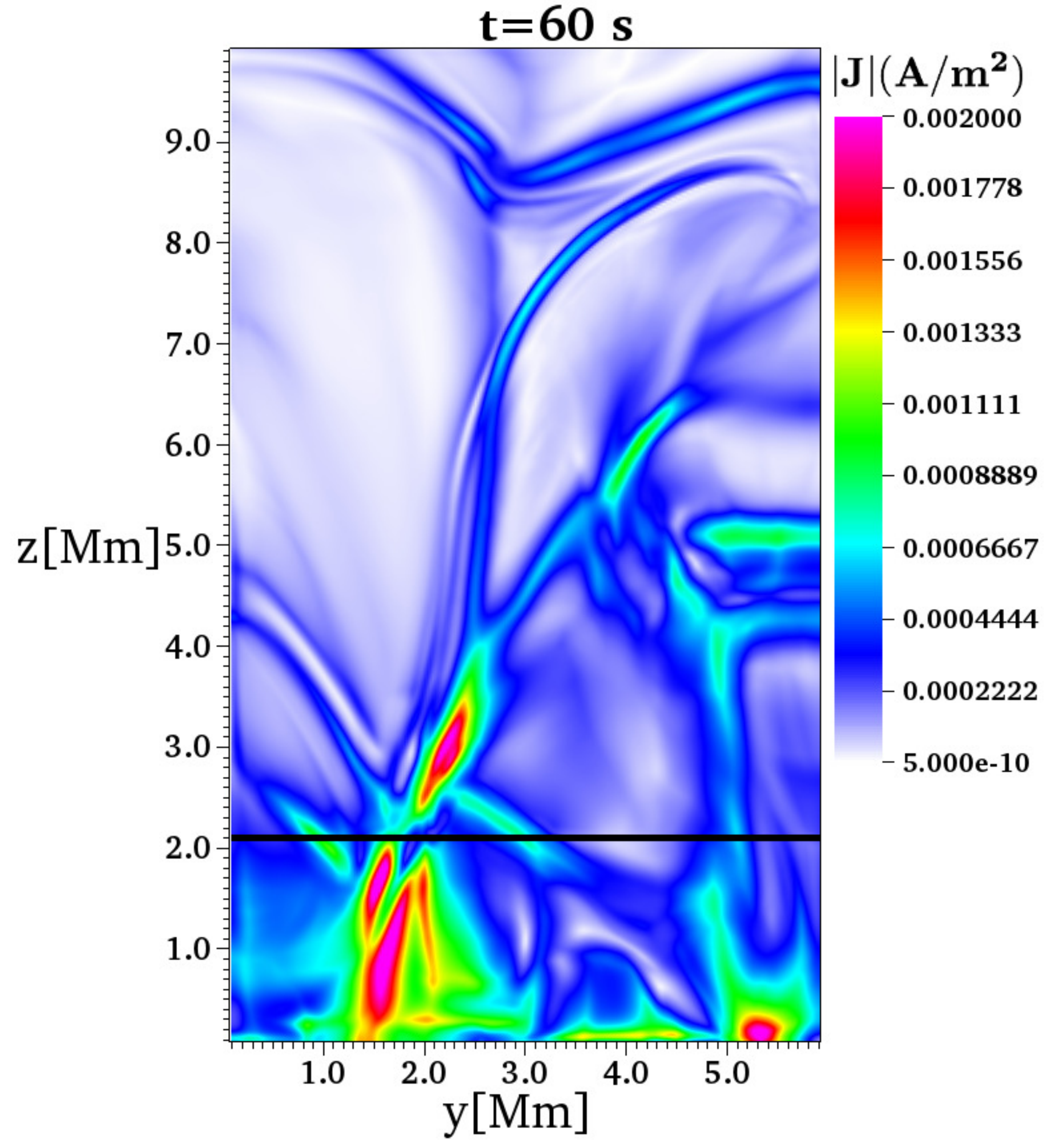}
\includegraphics[scale=0.2]{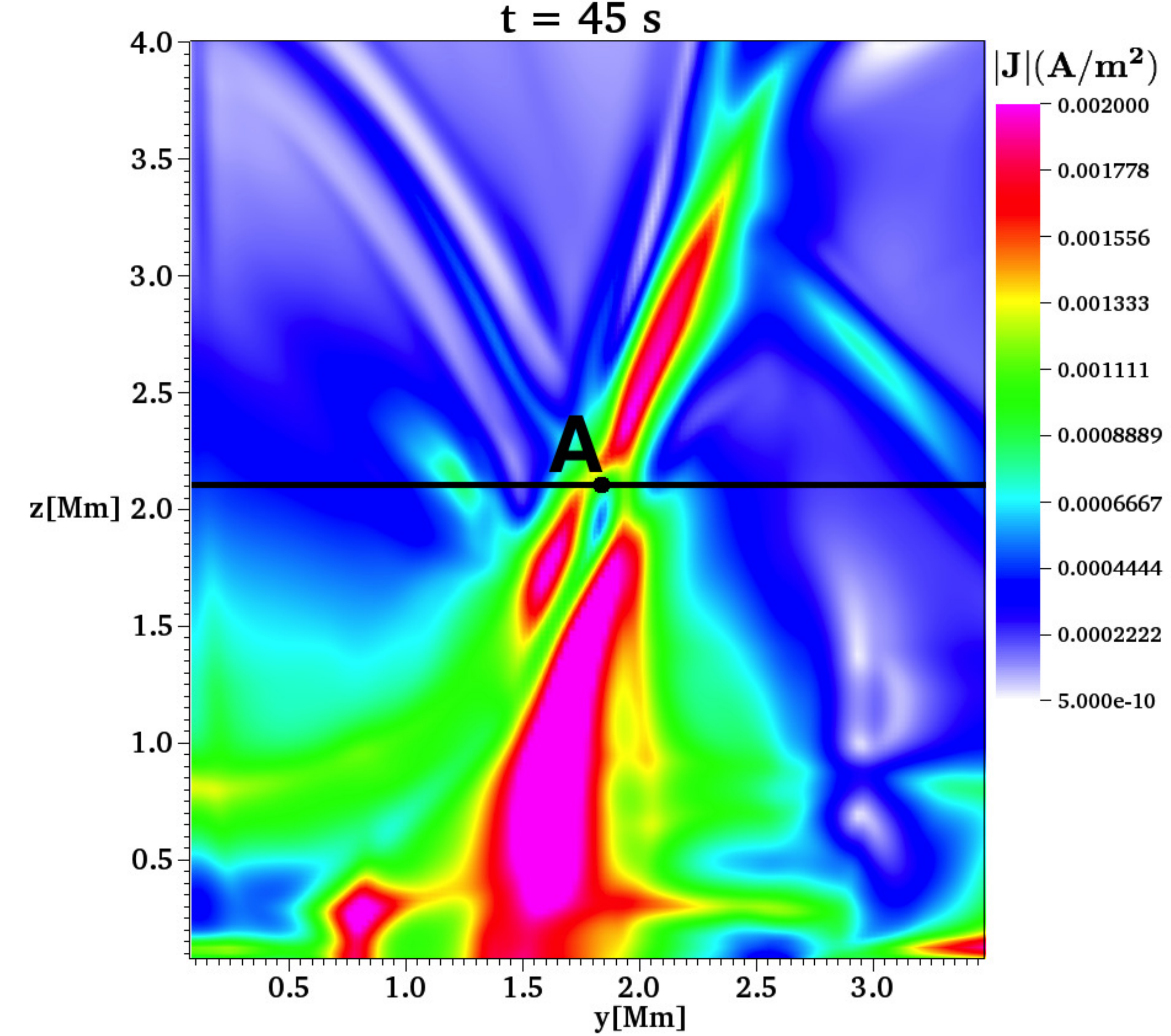}
\caption{\label{current_density_line} (Left) Snapshots of $|{\bf J}|$ (A m$^{-2}$) at two times and line $z=$2.1 Mm (black) to identify the region of strong current density in the cross cut at the plane $x=0.1$Mm. (Right) Zoom of the $|{\bf J}|$ (A m$^{-2}$) at time $t=45$ s and the point A where the ratio $E_{mag}/E_{kin}$ is estimated}.
\end{figure*}

\begin{figure*}
\centering
\includegraphics[scale=0.18]{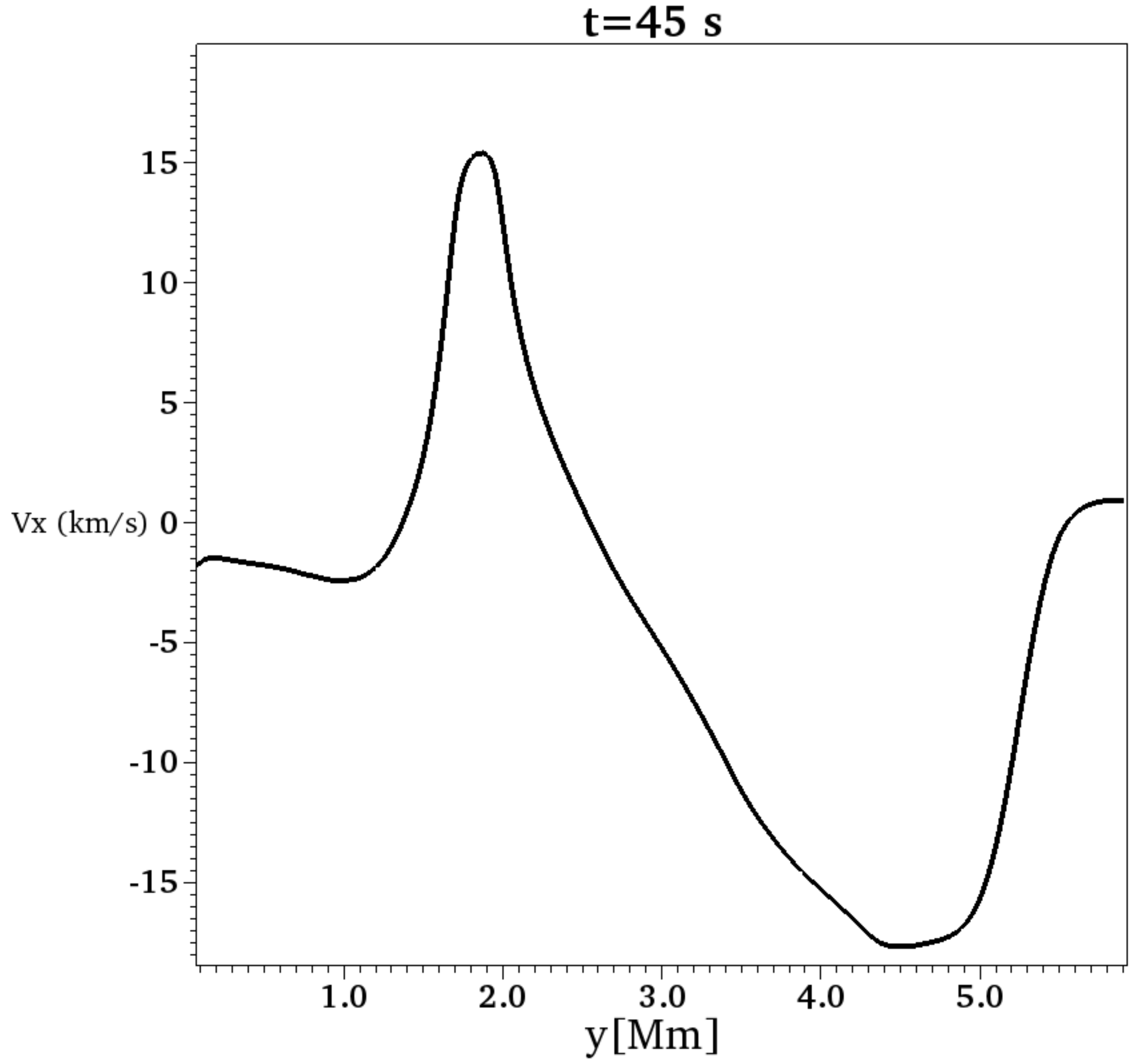}
\includegraphics[scale=0.18]{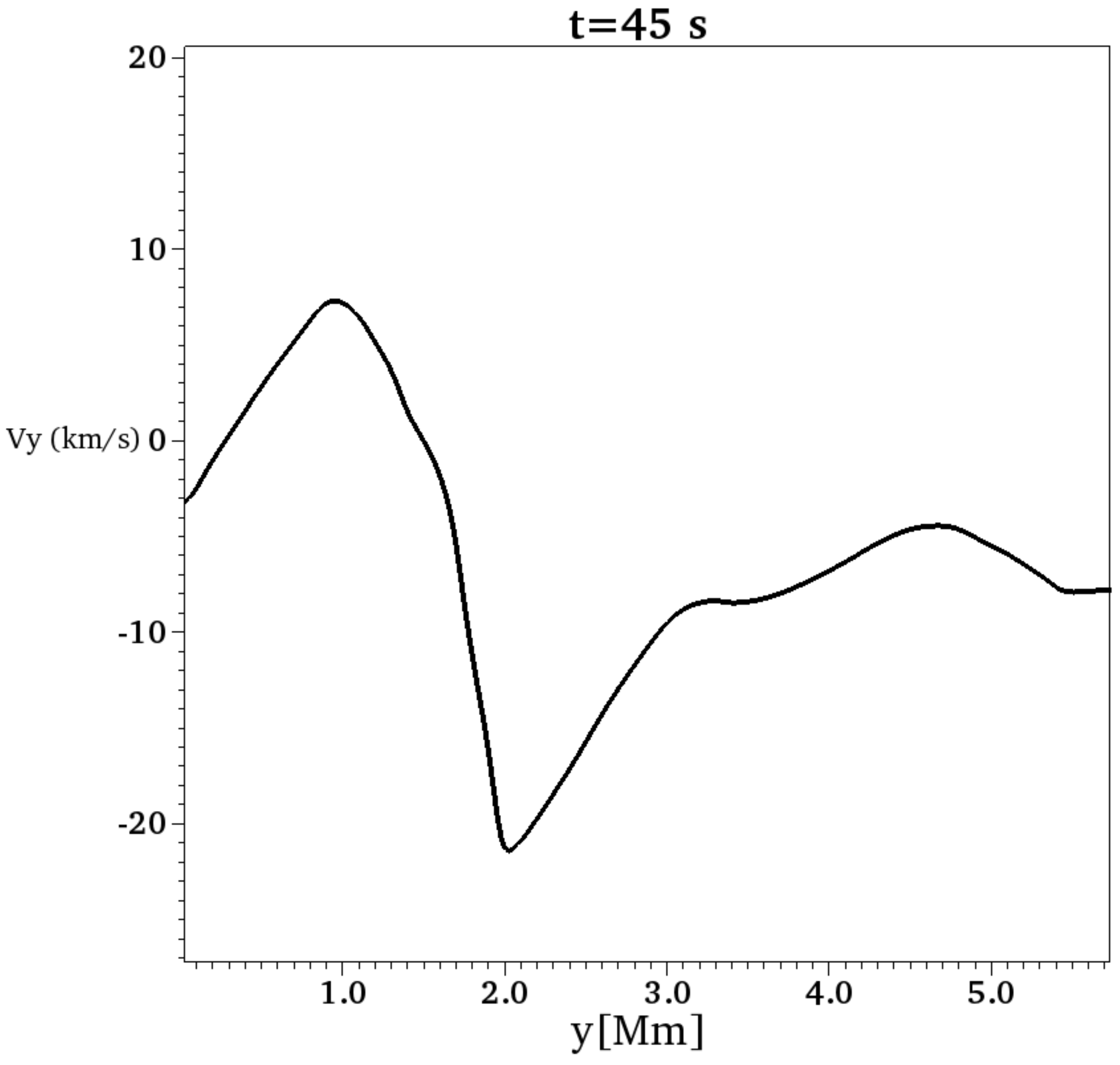}
\includegraphics[scale=0.18]{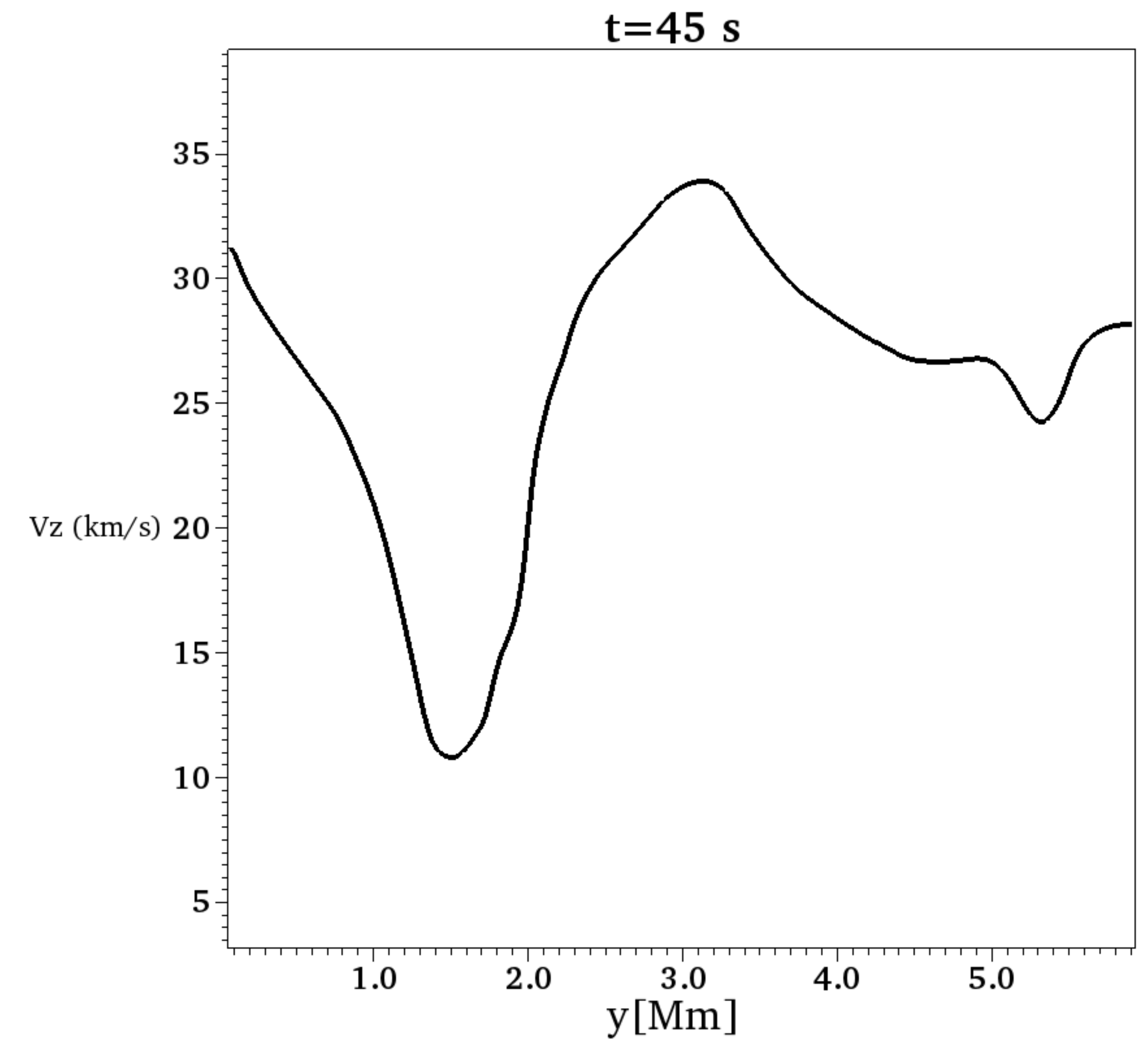}\\
\includegraphics[scale=0.18]{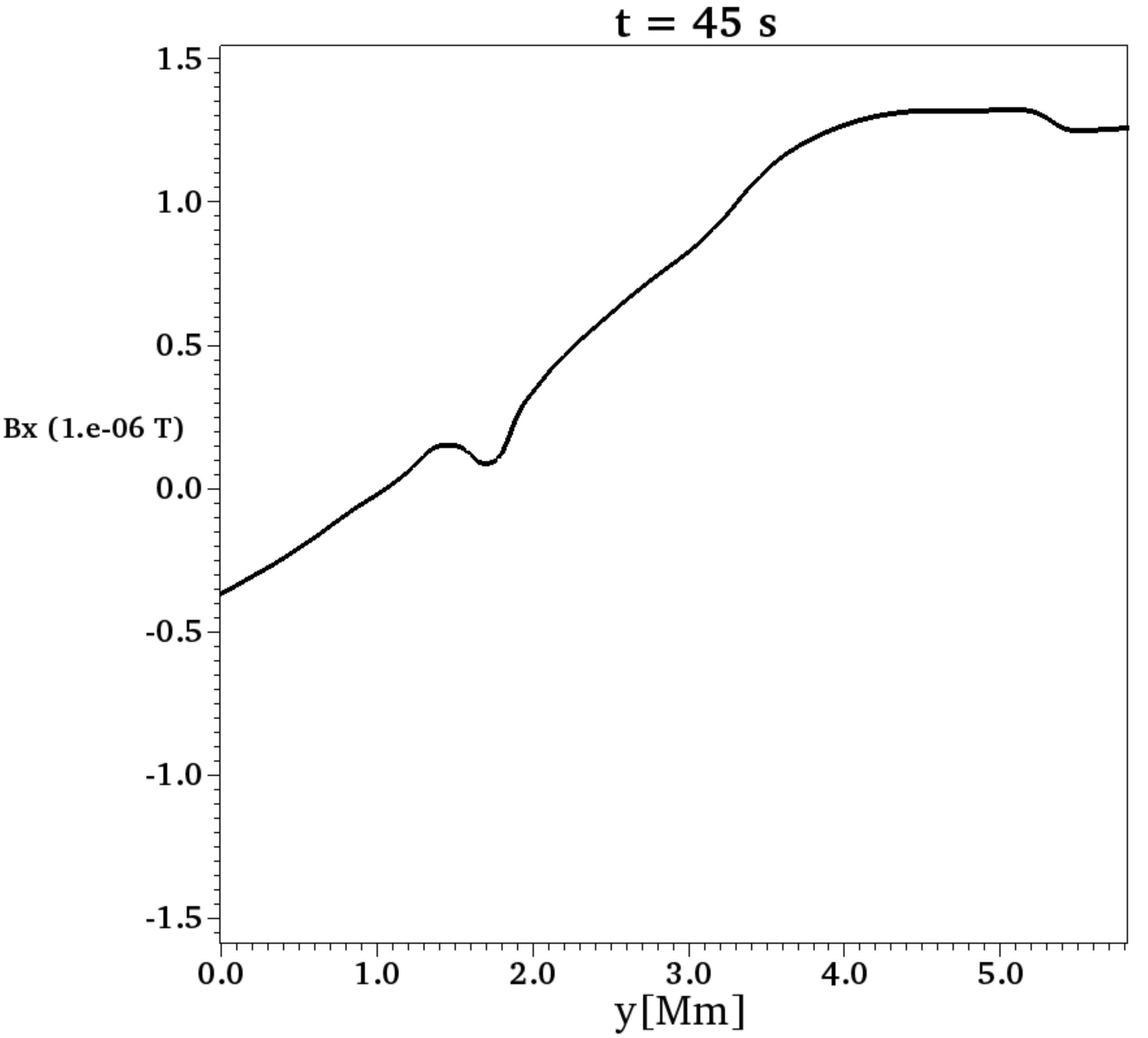}
\includegraphics[scale=0.18]{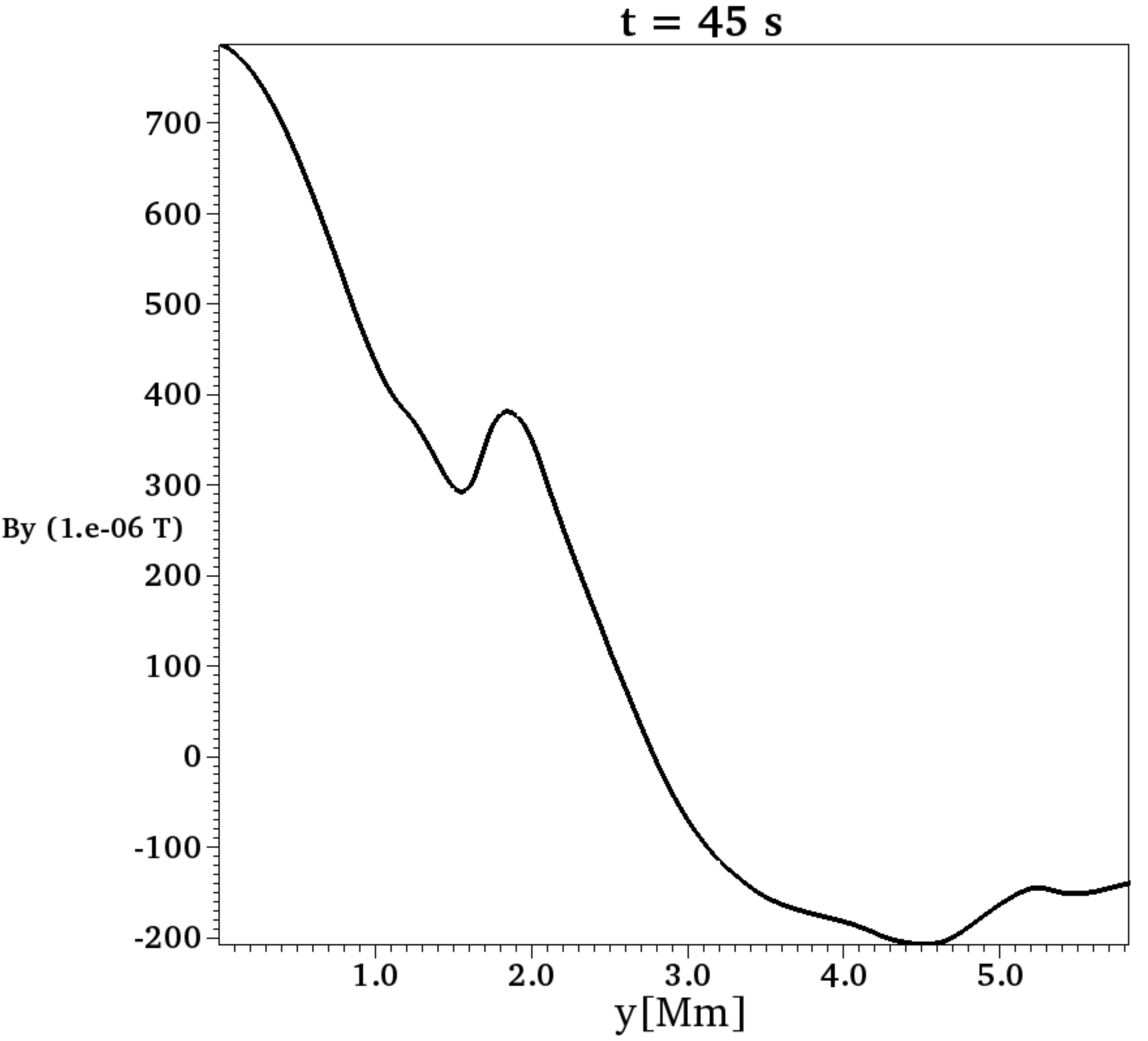}
\includegraphics[scale=0.18]{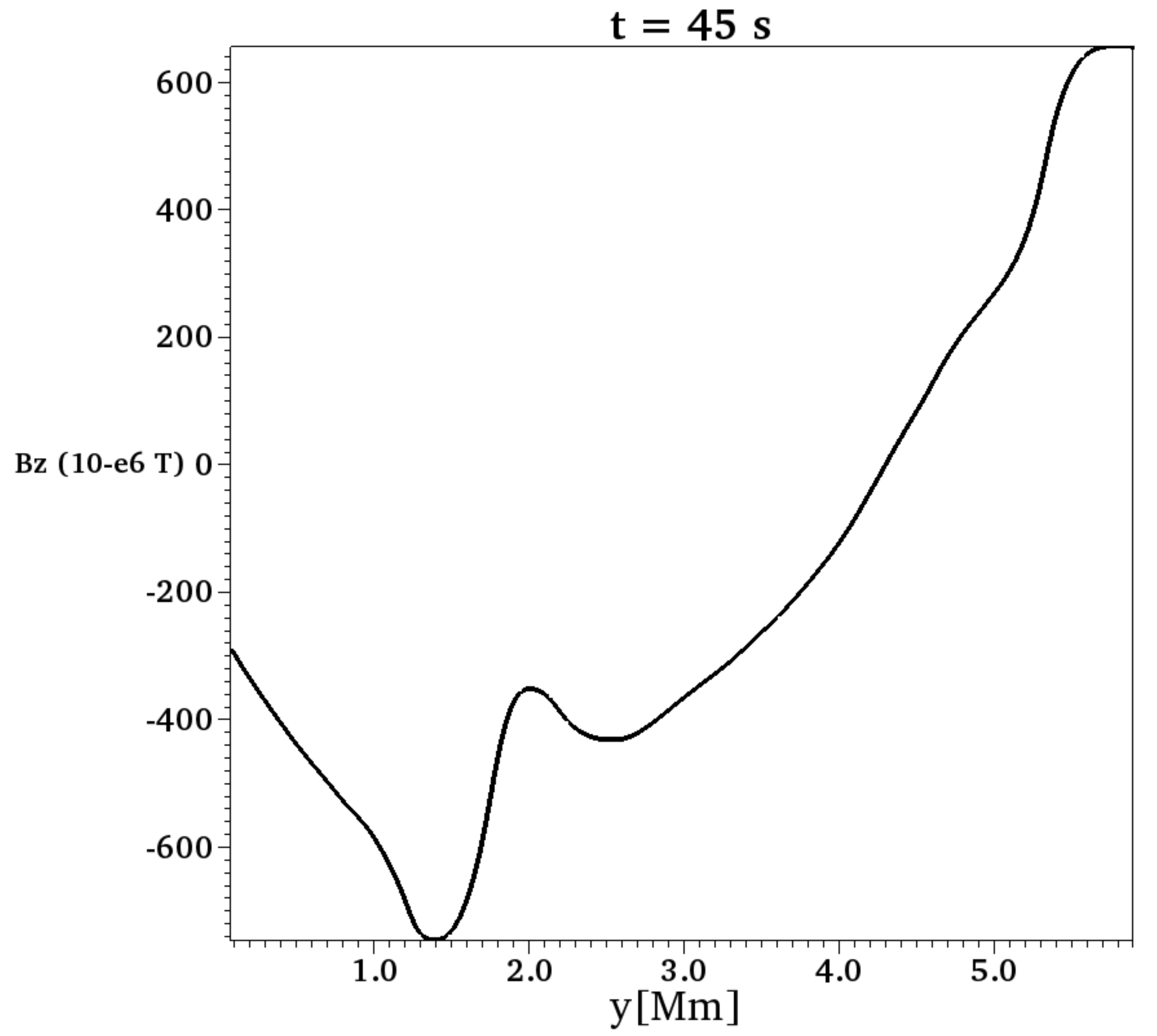}
\caption{\label{reconnection_plots} Snapshots of the $v_x$, $v_y$, $v_z$ in km $s^{-1}$ and $B_x$, $B_y$, $B_z$ in Tesla as a function of $y$ at time $t=45$ s measured at the line $z=$2.1 Mm of Figure \ref{current_density_line}.}
\end{figure*}

\begin{figure*}
\centering
\includegraphics[scale=0.18]{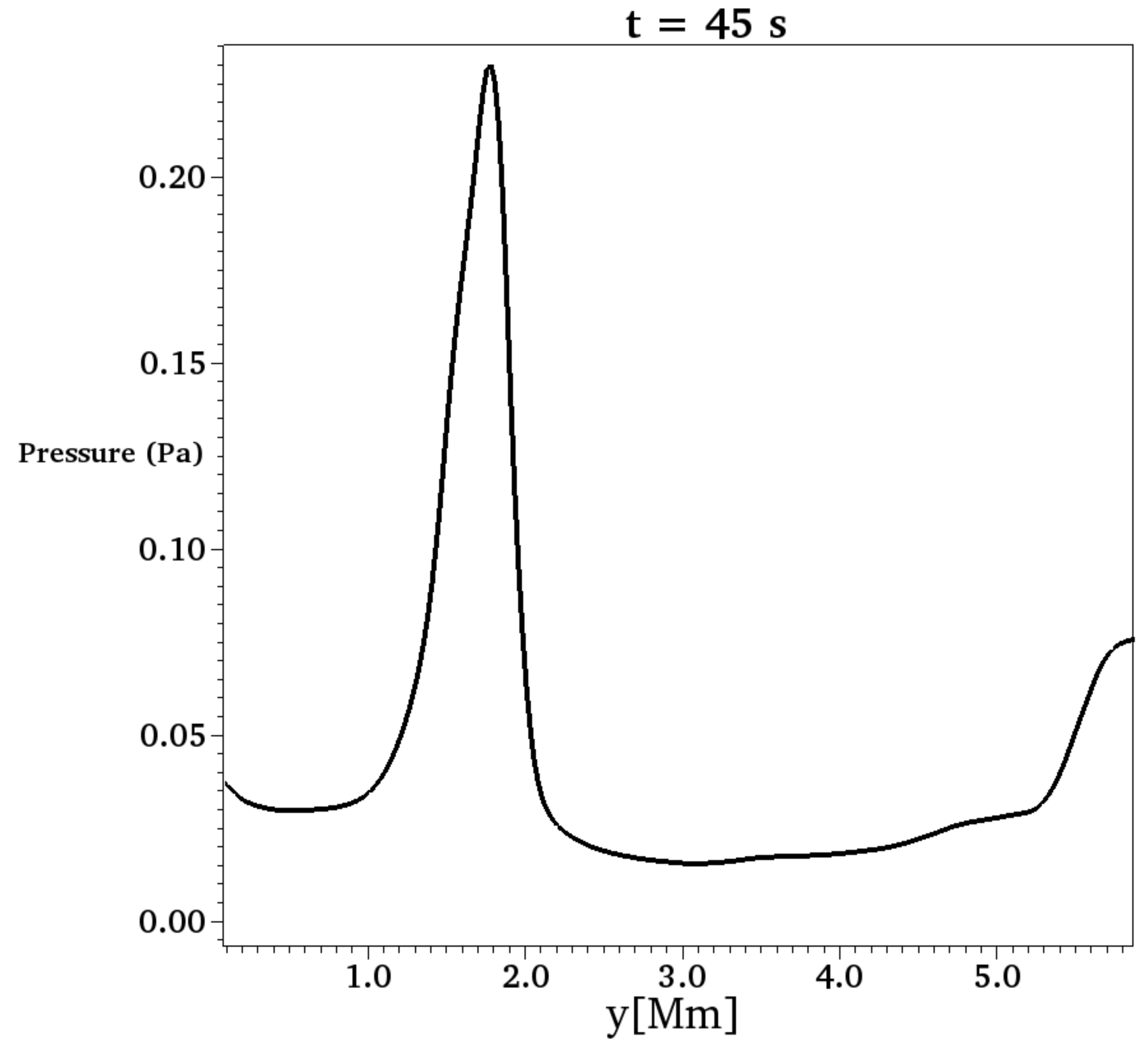}
\includegraphics[scale=0.18]{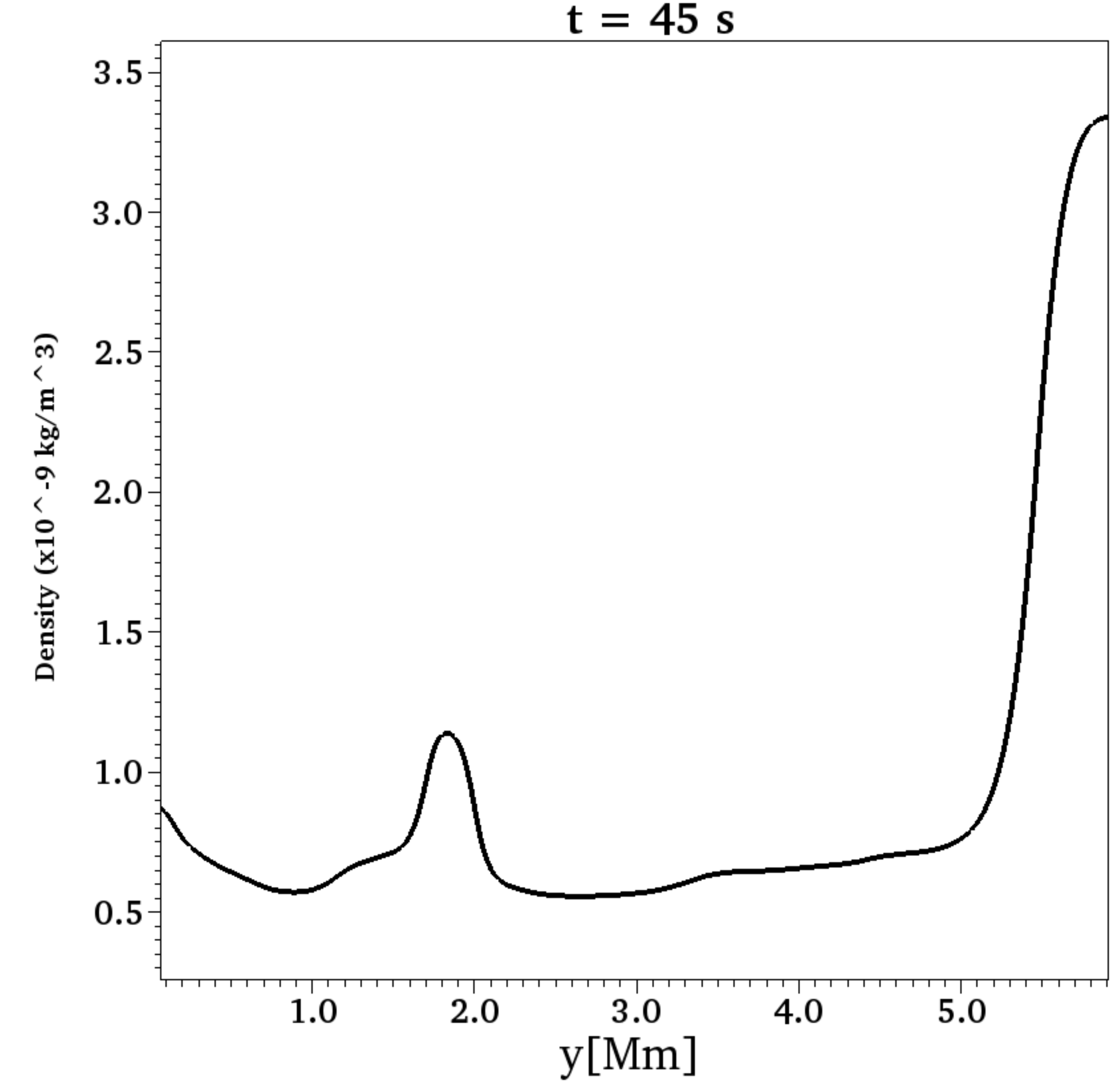}
\caption{\label{pressure_mass_rey_reconnection} Gas pressure (Pa) and mass density $\rho$ (kg m$^{-3}$) as a function of $y$ at time $45$ s measured a the line $z=$2.1 Mm of Figure \ref{current_density_line}.}
\end{figure*}

In addition, we estimate the ratio between magnetic energy density $E_{mag} =\frac{|B|^{2}}{2\mu_0}$ and kinetic energy density $E_{kin} = \frac{\rho v^{2}}{2}$ at the point A = (0.1, 1.75, 2.1) Mm shown in the right panel of Figure  \ref{current_density_line}, which is located in a region where the reconnection can be triggered. From the estimation we obtain that magnetic energy density is being converted into kinetic energy during the evolution, which is an indication of a reconnection process.

\begin{figure*}
\centering
\includegraphics[scale=0.25]{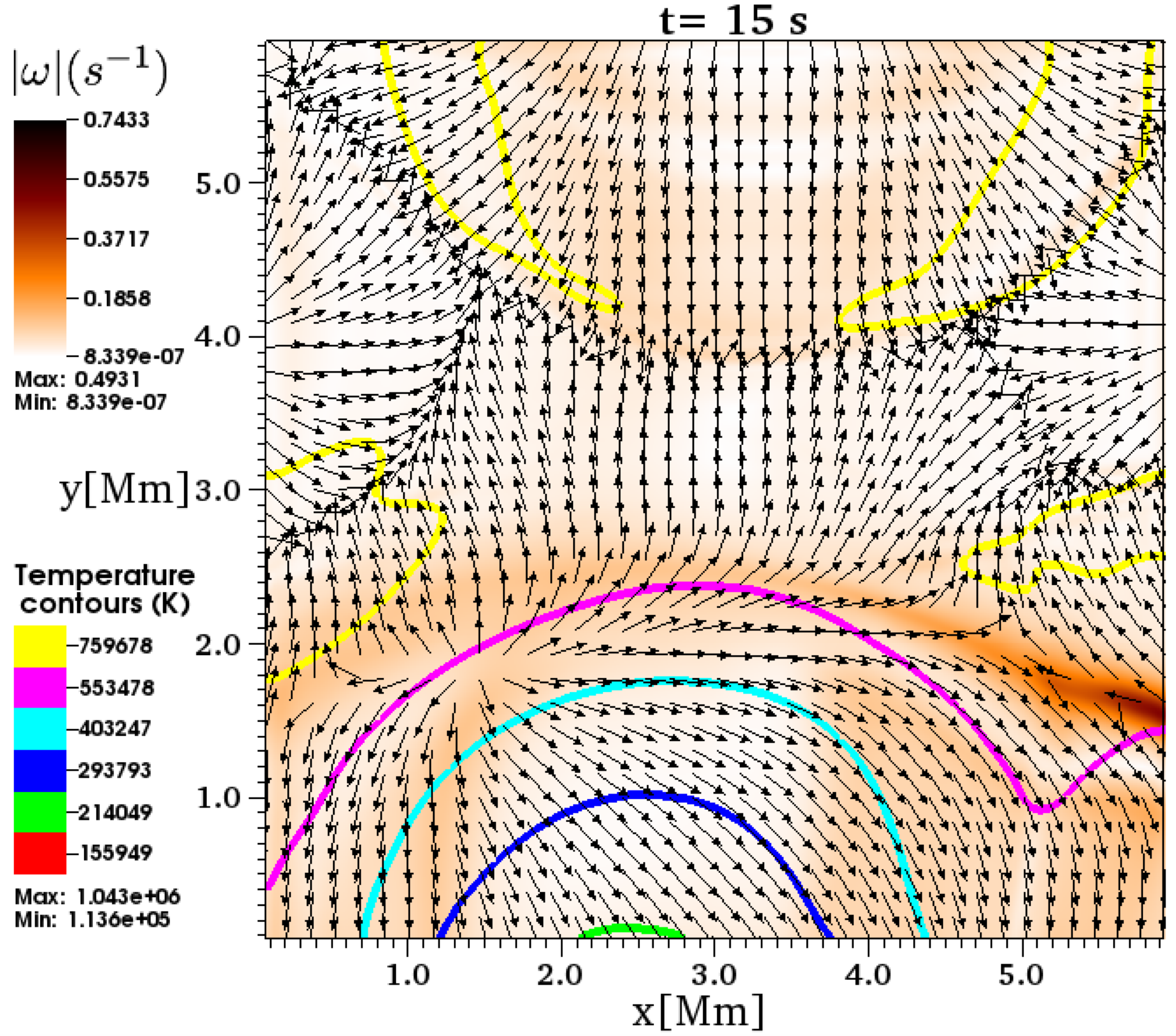}
\includegraphics[scale=0.25]{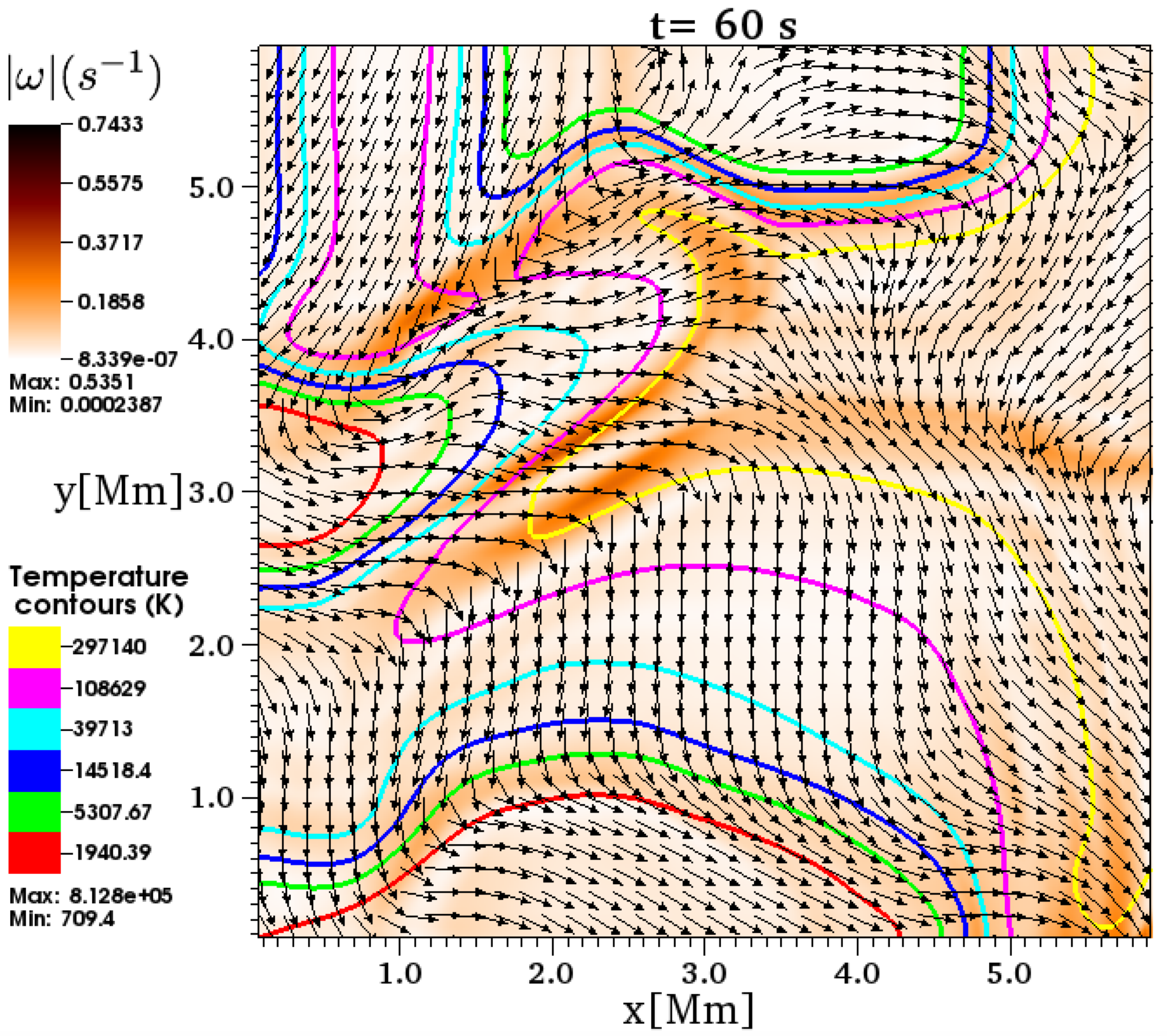}
\caption{\label{vorticity_vectorfield_xy_z=5Mm} Snapshots of the magnitude of the vorticity  $|{\bf\omega}|$ s$^{-1}$, vector velocity field and temperature contours (K) in the plane $z=5$ Mm at times 15 and 60 s.}
\end{figure*}

Another important diagnostics of Type II spicules is whether they are twisted, rotate or show torsional flows. Observations on the Doppler shift of various emission lines in the limb suggest that Type II spicules are rotating \citep{De_Pontieu_et_al_2012, Sekse2013,Sharma_2017}. Thus we calculate the vorticity ${\bf \omega}=\nabla\times{\bf v}$ and the vector velocity field in order to look for rotational motion in the spicule region. For this we consider the plane at $z=5$ Mm located approximately to the middle of the spicule. We show the magnitude of ${\bf \omega}$, velocity field and temperature contours (K) in Figure \ref{vorticity_vectorfield_xy_z=5Mm}. By $t=15$ s we can see regions where the magnitude of vorticity is high, the vector velocity field starts to circulate and the temperature is low. At time $t=60$ s we can see a region with a high value of the vorticity and a low value of the temperature. This vortex is related to the motion of the spicule structure.

\begin{figure*}
\centering
\includegraphics[width=6.0cm]{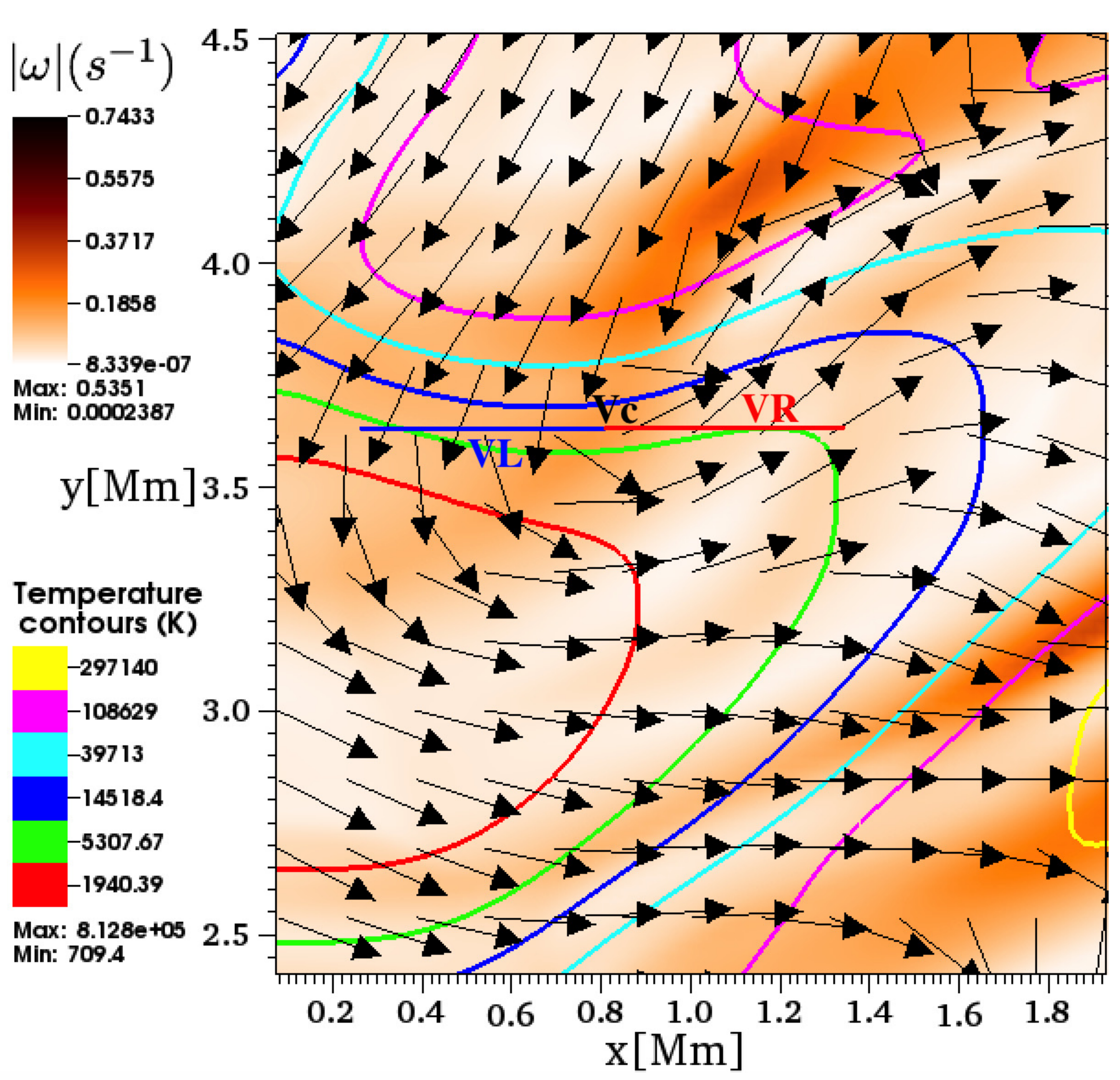}
\includegraphics[width=6.0cm]{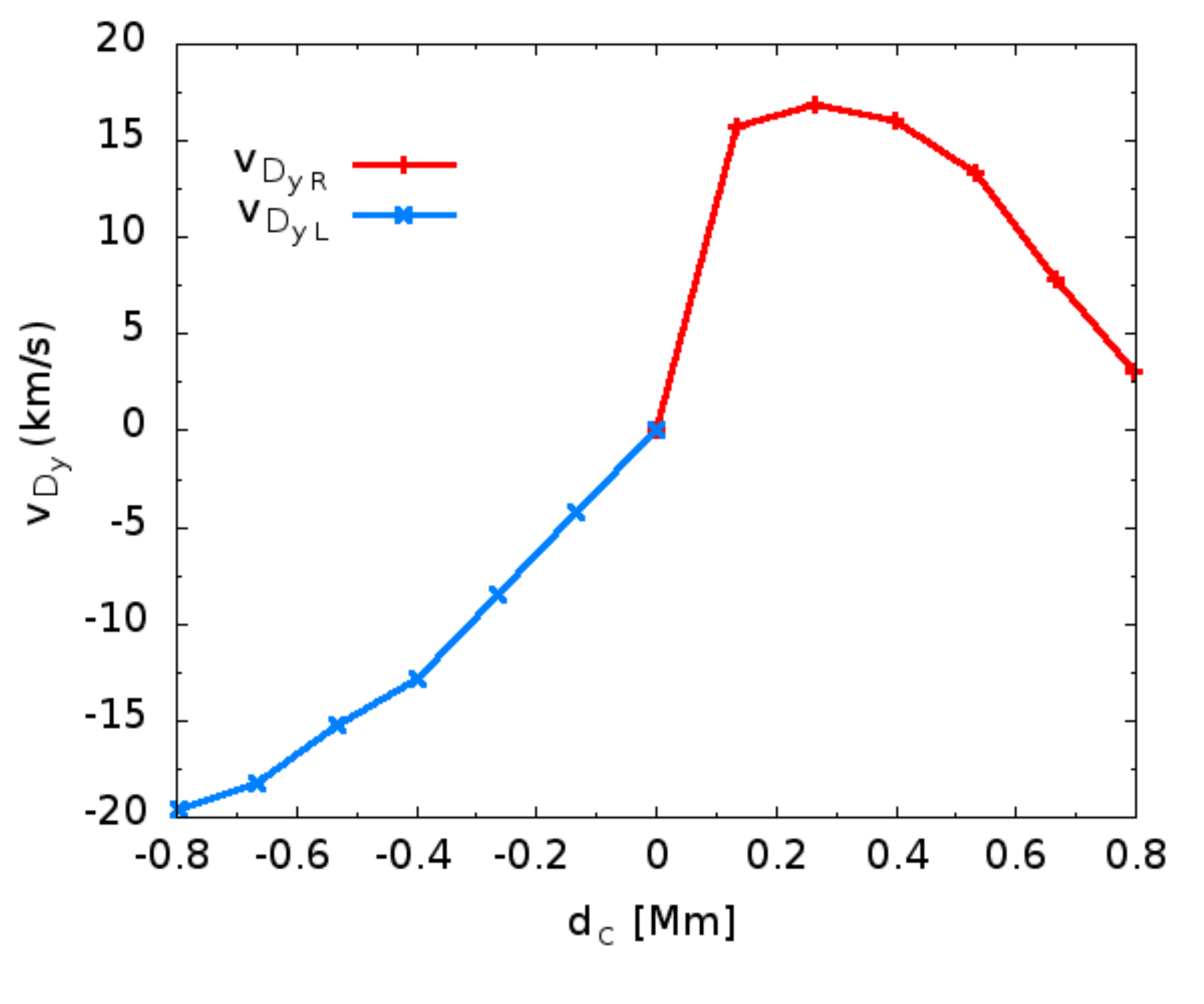}\\
\caption{\label{Doppler_effect_region_z=5Mm} (Left) Region where the Doppler shift $\Delta{\bf v}_{D}$ is estimated. This is a zoom of a vortex region of the Figure \ref{vorticity_vectorfield_xy_z=5Mm} at time $t=60$ s. (Right) Value of $\Delta{\bf v}_{D}$ of the $y$ component of velocity $v_y$ or equivalently the tangent velocity as a function of distance from the center.}
\end{figure*}

{\it Vorticity and Doppler.}  We estimate in the plane defined above the Doppler effect related to the dynamic of the spicule in a simple way. We specifically estimate this effect in a small region where the vorticity is high, the velocity vector field is circulating and the temperature is low. In order to estimate the Doppler effect we define a center in the region mentioned above where the velocity is ${\bf v}_c$. Then we chose a set of points to the left and to the right along the $x$ direction from the center (it could have been any other), with velocities ${\bf v}_L$ and ${\bf v}_R$, respectively. Then we calculate the difference in the $y$ component of these velocities with respect to that of the center, specifically ${\Delta{\bf v}_{D}}={\bf v}_{L,R}-{\bf v}_{c}$, where ${\bf v}_{L,R}$, which is an estimate of the tangent velocity of the points around the center and therefore a measure of a red and blue shift. This method is illustrated in Figure \ref{Doppler_effect_region_z=5Mm}. We show a zoom in of the vortex where the circulation of the vector velocity field is more evident. In this particular case we calculate a plot of $\Delta{\bf v}_{D}$ for the $y$ component of the velocity $v_y$ as a function of the distance $d_c$ from the center to the right or left, along the blue or red line. The amplitude of the red shift is of the order of 15 km s$^{-1}$, whereas the blue shift has an amplitude of the order 25 km $s^{-1}$. The results of the estimation of the Doppler effect due to tangent motion $\Delta v_{D_y}$ are shown also in Figure \ref{Doppler_effect_region_z=5Mm}.


\section{Conclusions}
\label{sec:conclusions}

In this paper we have presented a 3D numerical simulation on a small region of the solar atmosphere, showing the formation of a jet structure with characteristics of a Type II spicule, specifically the morphology, upward velocity range and time-scale formation. This result provides a simple explanation and is in contrast with that in \cite{Martinez-Sykora_et_al_2017}, where out of 2D simulations the formation of spicules is explained in terms of the amplification of the magnetic tension and the interaction between ions and neutrals. In our simulation we show that even if magnetic tension might be important, the magnetic pressure, which is a part of full Lorentz force is important as well, which is consistent with the results obtained in the simulations of vortex tubes \citep{Kitiashvili_2013} and in the formation of solar chromospheric jets \citep{Iijima_2017}. A quantitative distinction between the components of the different forces involved, would require the development of new analysis tools for time-dependent structures.

For this, we solve the equations of the resistive MHD submitted to the solar constant gravitational field. We use a 3D magnetic field configuration extrapolated up to the solar corona region from a simulated quiet-Sun photospheric field. This magnetic field configuration contains bipolar regions with a strong magnetic field strength at the bottom, which helps the development of the magnetic reconnection process from the photospheric level. 

A key result of our analysis is that the Lorentz force dominates over the pressure gradient in the region where the spicule takes place and helps accelerating the structure upwards.
It is also expected that the pressure gradient at the transition region contributes at accelerating the plasma upwards. 

This 3D model, reveals the complexity, since a solar atmosphere containing the transition region in combination with a magnetic field with a complex topology sketch better the complexity of the solar atmosphere.

Our findings include also that the vorticity near the spicule is important. By looking at the velocity field in a specific cross-cut of the spicule we can track the circular displacement of plasma that eventually can be identified with blue-red shifts. A detailed analysis on the torsional properties of the spicule, generated waves, rotational and radial displacements will be presented in a separate paper \citep{Gonzalez-Aviles_et_al_2017b}.

In order to contrast our simulations with other similar analyses, we mention that our simulations are limited in the sense that we do not consider thermal conductivity, radiation and partial ionization as in \cite{Martinez-Sykora_et_al_2017}, however our simulation uses a topologically complex magnetic field in full 3D. 

\acknowledgments{Acknowledgments. This research is partly supported by the following grants: Royal Society-Newton Mobility Grant NI160149, CIC-UMSNH 4.9, and CONACyT 258726 (Fondo Sectorial de Investigaci\'on para la Educaci\'on). The simulations were carried out in the facilities of  the CESCE-UNAM, Iceberg HPC Cluster and the Big Mamma cluster at the LIASC-IFM. VF and GV would like to thank the STFC for their financial support.   

\bibliographystyle{yahapj}

\end{document}